\documentclass[paper]{JHEP3}
\usepackage{cite}
\usepackage{amssymb,amsmath}
\usepackage{graphicx}
\usepackage{booktabs}

\def\kt{\tilde{k}}

\def\eps{\epsilon}

\def\Dcz{{\cal D}_0}

\def\JET{J}
\def\e{\epsilon}
\def\Flavour{F}
\def\d{\hbox{d}}

\def\l({\left(}
\def\r){\right)}

\newcommand{\order}[1]{\ensuremath{O\left(#1\right)}}
\newcommand{\en}{E_{cm}}
\newcommand{\enen}{E_{cm}^2}
\newcommand{\splitqgq}{\ensuremath{P_{qg\rightarrow Q}}}
\newcommand{\splitqqbarg}{\ensuremath{P_{q\bar{q}\rightarrow G}}}

\newcommand{\softeikonal}{\ensuremath{\mathcal{S}}}
\newcommand{\intphizerozerom}{\ensuremath{I^{(0,0,m)}_1}}
\newcommand{\intszerozerom}{\ensuremath{I^{(0,0,m)}_2}}

\newcommand{\gaussf}[4]{\ensuremath{\, _2F_1 \left(#1,#2,#3;#4\right)}}

\title{
Antenna subtraction for the production of heavy particles 
at hadron colliders
}
\author{
G.~Abelof, A.~Gehrmann--De Ridder\\
Institute for Theoretical Physics, ETH, CH-8093 Z\"urich,
Switzerland}


\abstract{
The antenna subtraction method developed originally for the computation 
of higher order corrections to jet observables from a colourless
initial state is extended for hadron collider processes involving  
a pair of massive particles and jets in the final state at 
the next-to-leading order (NLO) level.
Due to the presence of coloured initial states, the subtraction terms
need to be divided into three categories (final-final, initial-final
and initial-initial). In this paper, we outline their construction and
derive the necessary ingredients: phase space factorisation,
antenna functions and also integrated antennae, including the effects 
of massive final states in all of those building parts.
As a first application, we explicitly construct the colour-ordered
real radiation and the corresponding  antenna subtraction terms 
required at NLO for the production of a  top quark pair 
and for the production of a top quark pair in association with a
hard jet. The latter constitutes an essential
ingredient for the computation of the hadronic production of a
top-antitop pair at NNLO.}

\preprint{\today}

\allowdisplaybreaks[4]

\begin{document}

\section{Introduction}

At LHC, physics beyond the Standard Model 
will almost inevitably manifest itself by the creation of massive particles 
which decay instantaneously into multiparticle final states.
Searches for supersymmetric particles often involve 
final states with four or more jets. 
Top quarks~\cite{cdftop,d0top} are measured through their decay 
into a bottom quark and a subsequently decaying $W$-boson, yielding 
up to six-jet final states for top quark pair production.
Meaningful searches for these signals require not only a very good 
anticipation of the expected signal, but also of all standard model backgrounds
yielding identical final state signatures. Since leading-order calculations 
are affected by large uncertainties in their normalisation and their 
kinematical dependence, it appears almost mandatory to include NLO corrections.
For a long time, these corrections were available only for at most 
three final state particles. 
Up to recently, 
the principle obstacle to NLO calculations of final states with 
higher multiplicities were the one-loop virtual corrections to multi-parton 
scattering amplitudes. Application of standard quantum field theory
methods to those produced extremely large and numerically unstable 
expressions.  In the recent past, several new algorithms have emerged 
to circumvent these 
numerical instabilities~\cite{oneltech}.
The recently released packages 
CutTools~\cite{cuttools}, BlackHat~\cite{blackhat},
Golem~\cite{golem}, Rocket~\cite{rocket} and Samurai~\cite{samurai}
provide automated implementations of these 
new methods. They were already applied to
in a  number of pioneering calculations~\cite{wieders,v3jet1,v3jet2,v4jet,ttbb1,ttbb2,ttjj,ttfull1,ttfull2}.

With a mass $m_{t}=173 \pm 1.3$  GeV, the top quark is the
heaviest particle produced at colliders and due to its very large
mass it decays before it hadronises.
By studying its properties in detail, it is hoped to elucidate the origin 
of particle masses and the mechanism of electroweak symmetry breaking. 
Since its discovery at the Fermilab Tevatron, 
a number of its properties (mass, couplings) 
have been determined to an accuracy of ten to twenty per cent. With the 
large number of top quark pairs expected to be produced at the LHC, the 
study of its properties will become precision physics. 
This large production rate  will allow precise measurements of their
properties and their production cross sections with an expected 
experimental accuracy of five per cent.

Current theoretical predictions for the top quark pair production
cross section include NLO corrections ~\cite{ttfull1,ttfull2,ttnlo}  and
next-to-leading-logarithmic resummation (NLL) ~\cite{newresum}.
More recently  even the NNLL resummation effects have been completed in 
\cite{resumnnll}. These predictions lead to a theoretical uncertainty 
of the order of ten per cent.  
The same precision is available 
for single top quark production~\cite{singtop},
top-pair-plus-jets production~\cite{ttjj,ttj} and for 
top-pair-plus-bottom-pair  production~\cite{ttbb1,ttbb2}.

The top quark appears as virtual particle in hadron collider processes
and due to the small ratio between the top quark width and its mass, 
it is possible to factor the cross section of processes involving top quarks 
into the product of the production cross section for on-shell top quarks   
and the top quark decay width. Most of the calculations mentioned above 
are performed for on-shell top quark pair production.
Only most recently, the decay of the top quark has been included  
in NLO calculations \cite {ttfull1,ttfull2,Melnikov}  leading to 
a similar theoretical accuracy. 

Even for on-shell top quark pair production, a full fixed order
calculation of the total top-antitop rate at NNLO, required to match
the experimental accuracy, is missing.
NNLO calculations involving massive quarks require 
the same ingredients as their massless counterparts.
Three classes of contributions enter: double real, mixed real-virtual 
and two-loop type virtual contributions.
However, the quark mass introduces one additional scale into the calculation.
Especially the two-loop virtual corrections become therefore more involved 
than in the massless case. On the other hand, the treatment of the 
real radiation $n+2$ parton processes is expected to be easier, since the 
heavy quark mass acts as an extra infrared regulator, thus eliminating 
part of the singularity structure. 

Recent progresses has been accomplished concerning the
two-loop contributions.
Part of these two-loop virtual corrections are built with products of
one-loop virtual amplitudes. Those corrections have been computed in 
\cite{oneloop2}.
Concerning the two-loop virtual corrections build with product of
two-loop and tree-level amplitudes, the situation is different.
The two-loop virtual corrections for the processes $q\bar q \to t\bar
t$ and $gg\to t\bar t$ are not fully available at present. A purely
numerical evaluation of the quark-initiated process~\cite{czakontt}
could be partly confirmed by analytical results~\cite{ourtt}, which
were most recently extended also to the gluon-induced subprocess.

Real radiation corrections involving heavy quarks have up to now 
been investigated only at NLO. Most recently
a semi-numerical scheme to evaluate the double real radiation
corrections at NNLO has been introduced and applied to top quark pair
production~\cite{czakonsub}. 
\vspace{2mm}
 
In this paper, towards calculating the double real contributions to
top-antitop production in hadronic collisions at NNLO,   
we extend the antenna subtraction formalism to be applicable 
to the production of massive heavy quark pairs in the presence 
of coloured initial states and present results for the real and subtraction 
contributions for $t\bar{t}$ and $t\bar{t}$ +jet production at NLO.

Generally, for hard scattering observables, the inclusive cross section 
with two incoming hadrons $H_1,H_2$ can be written as  
\begin{equation}
\d \sigma = \sum_{a,b} \int 
\frac{\d \xi_1}{\xi_1} \frac{\d \xi_2}{\xi_2}\, f_{a/1}(\xi_1) \,f_{b/2}(\xi_2)
\, \d \hat{\sigma}_{ab}(\xi_1H_1,\xi_2H_2)\;, 
\label{hadroncross}
\end{equation}
where $\xi_1$ and $\xi_2$ are the momentum fractions of the 
partons of species $a$ and $b$ in both incoming hadrons, $f$ being the 
corresponding parton distribution functions and  
$\d \hat{\sigma}_{ab}(\xi_1H_1,\xi_2H_2)$ is the parton-level 
scattering cross section for incoming partons $a$ and $b$. 

The partonic cross section ${\rm d }\hat{\sigma}_{ab}$  has a perturbative
expansion in the strong coupling $\alpha_{s}$
such that theoretical predictions for a hadronic process at a given 
order in $\alpha_{s}$ are obtained when all partonic channels 
contributing to that order to the partonic cross section 
are summed and convoluted with the appropriate parton distribution 
functions as in eq.(\ref{hadroncross}).

In general, beyond the leading order, each partonic channel 
contains both ultraviolet and infrared (soft and collinear) divergences.
The ultraviolet poles are removed by renormalisation in each channel.
Collinear poles originating from the radiation of initial state partons 
are cancelled by mass factorisation counterterms and absorbed 
in the parton distribution functions.
The remaining soft and collinear poles cancel among each other when all  
partonic contributions are summed over \cite{KLN}.      
As these observables 
depend in a non trivial manner on the experimental criteria needed 
to define them, they can only be calculated numerically.
The computation of hadronic observables including higher order corrections   
therefore requires a systematic procedure to cancel 
infrared singularities among different partonic channels 
before any numerical computation of the observable can be performed.

For the task of next-to-leading 
order (NLO) calculations~\cite{ks}, the infrared divergencies present in real 
radiation contributions can be systematically extracted by 
process-independent procedures, called subtraction methods. 

More specifically, let us consider the hadronic production of $m$-jets at NLO.
A theoretical prediction for this observable is obtained 
by summing the following massless partonic contributions: 
At LO, the tree-level contribution contains $m$ partons in the final state
which build $m$ jets. 

At NLO, 
the differential cross section for the production of $m$-jets,  
${\rm d}\hat{{\sigma}}_{NLO}$ may symbolically be written as,

\begin{equation}
{\rm d}\hat{{\sigma}}_{NLO}=\int_{{\rm d}\Phi_{m+1}}
 {\rm d}\hat{{\sigma}}_{NLO}^R \;J_{m}^{(m+1)}
+ \int_{{\rm d}\Phi_m}\left({\rm d}\hat{{\sigma}}_{NLO}^V  + 
{\rm d}\hat{{\sigma}}_{NLO}^{MF}\right) J_{m}^{(m)}\\
\label{eq:signlo}
\end{equation}
where $\int_{{\rm d}\Phi_N}$ corresponds to the integration over the $N$
parton phase space.
The cross section is built with the real radiation cross section contribution 
${\rm d}\hat{\sigma}^{R}$ which has $(m+1)$ massless partons in the final
state, the one-loop cross section ${\rm d}\hat{\sigma}^{V}$ 
and the mass factorisation counterterm ${\rm d}\sigma_{NLO}^{MF}$
which have both $m$ partons in the final state.
A jet algorithm is applied separately on each of these contributions to ensure
that out of $n$ partons, $m$ jets are built in the final state.
Symbolically this recombination procedure is denoted by $J_{m}^{(n)}$.

The purpose of any subtraction method is to provide 
a subtraction term ${\rm d}\hat{\sigma}^S_{NLO}$ which has the same 
singular behaviour as the real radiation squared matrix element and
is sufficiently simple to be integrated analytically over a factorised 
form of the $(m+1)$-phase space.

Using a subtraction method, the NLO partonic cross section given in
eq. (\ref{eq:signlo})
may then be written as,
\begin{eqnarray}
{\rm d}\hat{{\sigma}}_{NLO}&=&
\int_{{\rm d}\Phi_{m+1}}\left( {\rm d}\hat{{\sigma}}_{NLO}^R 
- {\rm d}\hat{{\sigma}}^S_{NLO} \right) J_{m}^{(m+1)}\nonumber \\
&&  +\int_{{\rm d}\Phi_{m}} \left(
\int_1{\rm d}\hat{{\sigma}}^S_{NLO} + {\rm d}\hat{{\sigma}}_{NLO}^V  
+{\rm d}\hat{{\sigma}}_{NLO}^{MF} \right)\;J_{m}^{(m)}.
\label{eq:sub}
\end{eqnarray}

With this, the first integral is finite and can be integrated
numerically in four dimensions. The integrated form of the subtraction term 
has the same number of final-state partons as the virtual contributions 
and the mass factorisation counterterms.  
It can therefore be combined with those thereby canceling analytically 
the explicit infrared divergences. 
The second integral in eq. (\ref{eq:sub}) is therefore finite as well. 

The actual form of the subtraction term ${\rm d}\hat{{\sigma}}^S_{NLO}$
depends on the subtraction formalism used. The approximations of the
matrix-element in the unresolved limits being non unique, several successful
subtraction formulations have been proposed in the literature 
\cite{cs,Frixione:1995ms,Nagy:1996bz,Frixione:1997np,Somogyi:2006cz}.
The dipole formalism of Catani and Seymour \cite{cs} and the  
FKS \cite{Frixione:1995ms} of Frixione, Kunszt and Signer  
have been implemented in an automated way in 
\cite{Gleisberg:2007md,Seymour:2008mu,Hasegawa:2008ae,
Hasegawa:2009tx,Frederix:2008hu,Czakon:2009ss}.
In its original formulation, the formalism of Catani-Seymour 
\cite{cs} deals with massless partons in final and/or initial state at NLO.

An alternative subtraction formalism is given by the the antenna subtraction 
method . This formalism \cite{k,ourant}  was originally derived for processes 
involving only (massless) final state partons in $e^+e^-$ collisions.
 It has been applied in the computation of NNLO corrections
 to three-jet production 
in electron-positron annihilation 
\cite{GehrmannDeRidder:2007jk,GehrmannDeRidder:2008ug,
Weinzierl:2008iv,Weinzierl:2009nz} and related event shapes 
\cite{GehrmannDeRidder:2007bj,GehrmannDeRidder:2007hr,
GehrmannDeRidder:2009dp,Weinzierl:2009ms,Weinzierl:2009yz}, 
which were subsequently used in precision determinations 
of the strong coupling constant
\cite{Dissertori:2007xa,Dissertori:2009ik,
Dissertori:2009qa,Bethke:2008hf,Gehrmann:2009eh}.  

For processes with initial-state partons and massless final states,
the antenna subtraction formalism has been so far fully worked out 
only to NLO in \cite{k,Daleo}. It has been extended 
to NNLO for processes 
involving one initial state parton relevant for electron-proton 
scattering in ~\cite{Gionata}  
while an extension of the formalism to include two initial state 
hadrons at NNLO is under construction \cite{Boughezalnew, Joao}. 

Subtraction formalisms which deal with massive final state particles 
have been so far only developed up to the NLO level 
\cite{cdst1,cdt2,weinzierl,Mathias}.
The kinematics is more involved due to the finite value of the parton masses.
QCD radiation from massive particles can lead to soft divergencies but 
cannot lead to strict collinear divergencies, since the mass is acting as an
infrared regulator. 
In a calculation of observables 
involving massive final state fermions, logarithmic terms of the 
form $\ln (Q^2/M^2)$, where $M$ is the parton mass 
and $Q$, the typical scale of the hard scattering process can be generated.
In kinematical configurations where $Q\gg M$, these logarithmically 
enhanced contributions can become large and can spoil 
the numerical convergence of the calculation.  
The cross section calculation of $t\bar{t}$ at LHC 
is an example where such enhanced logarithmic terms arise. 
These terms are related to a process-independent 
behaviour of the matrix elements; its singular behaviour 
in the massless limit $(M \to 0)$. This singular behaviour 
is related to the  {\it quasi-collinear} \cite{cdst1} limit 
of the matrix element. In the presence of massive particles, 
the factorisation properties of matrix element and phase space 
in collinear and soft limits need to be generalised to take the mass effects 
into account.

The dipole formalism of Catani and Seymour \cite{cs} and the FKS
formalism \cite{Frixione:1995ms}
have been extended to deal with massive particles in \cite{cdst1} 
for colourless and for coloured initial states up to the NLO.
The subtraction terms constructed within this formalism account 
for the quasi-collinear limits of the matrix-element squared.

The antenna subtraction method has so far only been extended 
to deal with the production of massive fermions from a colourless 
initial state in \cite{Mathias}.
It is the purpose of this paper to present 
an extension of this method to include radiation 
off final state massive fermions produced in hadronic processes.
In this paper we aim to derive all necessary ingredients, massive
antennae, phase space factorisation and finally integrated massive
antennae for this extension. 
As a first application, we construct the subtraction terms 
$ {\rm d}\hat{\sigma}^{S}_{NLO}$ required 
for the hadronic production of a top quark pair in association with no 
or one hadronic jet. 

These two processes, $p p \rightarrow t \bar{t}$ and 
$ p p \rightarrow t\bar{t}$ +jet  
have been calculated in \cite {newresum,ttj}  using the
dipole subtraction method \cite{cdst1}.
Our aim here is however not to redo this calculation using another 
subtraction scheme. 
Instead, we construct here the subtraction terms for $t\bar{t}$ +jet 
production in a colour-ordered form which is essential 
for the computation of top quark pair production without any jets 
at NNLO within the framework of the antenna formalism.
Those subtraction terms can be used to capture all single unresolved
radiation from the double real radiation contribution for the 
$t\bar{t}$ pair production at NNLO.

The plan of this paper is as follows. In Section 2 we outline the
construction of the subtraction terms for the hadronic production 
of a heavy quark pair in association with $(m-2)$-jets at NLO. 
We present the form of the subtraction terms in all kinematical configurations
with particular emphasis on the changes caused by the presence 
of massive final state particles in the expressions of the subtraction terms 
compared to those expressions in the massless case.
Section 3 contains a list of all massive antenna functions required. 
In Section 4 we tabulate all non-vanishing single unresolved 
limits of those massive antennae while in Section 5 
results for the integrated antennae are given.
Section 6 presents a check on one of the integrated antenna. 
In Section 7, for all partonic process involved, 
we present the colour ordered real contributions and their subtraction 
terms required to evaluate the hadronic production cross section 
of a $t \bar{t}$ pair and of $t \bar{t}$ pair and a jet at NLO.     
Finally Section 8 contains our conclusions.

\section{Antenna subtraction with massive final states}

In this section, we present the general formalism necessary to
evaluate the hadronic production of a pair of heavy quarks $Q\bar{Q}$  
in association with $(m-2)$ jets at the next-to-leading order (NLO) 
in perturbative QCD.

\subsection{Real radiation contributions to heavy quark pair 
production \\ in association with jets} 
    
The leading order (LO) $m$-parton contribution to the hadronic production 
of a pair of heavy quarks $Q\bar{Q}$  in association with $(m-2)$ jets 
may be written as,  
\begin{eqnarray}
\lefteqn{{\rm d} \hat\sigma_{LO}(p_1,p_2) =
{\cal N}\,
\sum_{{m-2}}{\rm d}\Phi_{m}(k_{Q},k_{\bar{Q}},k_{1},\ldots,k_{m-2};
p_1,p_2)} \nonumber \\ && \times 
\frac{1}{S_{{m-2}}}\,
|{\cal M}_{m}(k_{Q},k_{\bar{Q}},k_{1},\ldots,k_{m-2};p_1,p_2)|^{2}\; 
\JET_{m}^{(m)}(k_{Q},k_{\bar{Q}},k_{1},\ldots,k_{m-2}).
\label{eq:siglo}
\end{eqnarray}
The momenta $p_1$ and $p_{2}$ are the momenta of the initial state partons,
the massive partons $Q$ and $\bar{Q}$ have momenta $k_{Q}$ and
$k_{\bar{Q}}$, while the momenta of the remaining $(m-2)$ massless 
final state partons are labelled $k_{1} \ldots k_{m-2}$. 
$S_{m-2}$ is a
symmetry factor for identical massless partons 
in the final state.
$\JET_{m}^{(m)}(k_{Q},k_{\bar{Q}},k_{1},\ldots k_{m-2})$ 
is the jet function. It ensures
that out of $(m-2)$ massless partons and a pair of heavy quarks $Q$
and $\bar{Q}$ present in the final state at parton level, 
an observable with a pair of heavy quark jets in association 
with $(m-2)$ jets is built.
At this order each massless or massive parton forms a jet on its own.
The normalization factor 
${\cal N}$ includes all QCD-independent factors as well as the 
dependence on the renormalised QCD coupling constant $\alpha_s$.
$\sum_{m-2}$ denotes the sum over all configurations 
with $(m-2)$ massless partons.
${\rm d}\Phi_{m}$ is the phase space for an $m$-parton final state containing 
$(m-2)$ massless and two massive partons with total
four-momentum $p_1^{\mu}+p_2^{\mu}$. In $d=4-2\e$ space-time dimensions,
this phase space takes the form:
\begin{eqnarray}
\lefteqn{\d \Phi_m(k_{Q},k_{\bar{Q}},k_{1},\ldots,k_{m-2};p_1,p_2) = 
\frac{\d^{d-1} k_Q}{2E_Q (2\pi)^{d-1}}\; 
\frac{\d^{d-1} k_{\bar{Q}}}{2E_{\bar{Q}} (2\pi)^{d-1}}} \nonumber \\
&&\times  \frac{\d^{d-1} k_1}{2E_1 (2\pi)^{d-1}}\; \ldots \;
\frac{\d^{d-1} k_{m-2}}{2E_{m-2} (2\pi)^{d-1}}\; (2\pi)^{d} \;
\delta^d (p_1+p_2 - k_{Q}-k_{\bar{Q}}-k_1 - \ldots k_{m-2}) \,,\hspace{2mm}
\end{eqnarray}
In eq.(\ref{eq:siglo}) $|{\cal M}_{m}|^2$ denotes a 
colour-ordered tree-level $m$-parton matrix element squared for $m$ partons
out of which two are massive.
These terms only account for the leading
colour contributions 
to the squared matrix elements. 
On the other hand, colour subleading 
contributions are, in general, 
given by the
interference between two colour-ordered $n$-parton
amplitudes. However, to
keep the notation simpler we denote these interference 
contributions also as $|{\cal M}_{m}|^2$. Related to these
interference terms, it is here worth noting the
following:  As soon as more than five
coloured partons are present in a given partonic process, the
subtraction of infrared
singularities present in interference terms is more involved than for
colour ordered squared matrix elements. This particular issue will be
treated in Section 7.2.1.

At, NLO the real radiation partonic contribution to the production of 
the heavy quark pair production in association with $(m-2)$ jets involves 
$(m+1)$-final state partons with two of them being massive. 
It may be written as, 
 \begin{eqnarray}
\lefteqn{{\rm d}\hat\sigma^{R}_{NLO}(p_1,p_2)=
{\cal N}\,
\sum_{{m+1}}{\rm d}\Phi_{m+1}(k_{Q},k_{\bar{Q}},k_{1},\ldots,\,k_{m-1};
p_1,p_2) }\nonumber \\ && \times 
\frac{1}{S_{{m+1}}}\,
|{\cal M}_{m+1}(k_{Q},k_{\bar{Q}},k_{1},\ldots,\ldots,k_{m-1},;p_1,p_2)|^{2}\; 
\JET_{m}^{(m+1)}(k_{Q},k_{\bar{Q}},k_{1},\ldots,k_{m-1}).\hspace{3mm}
\label{eq:real}
\end{eqnarray}
The jet function $\JET_{m}^{(m+1)}$ ensures that out of $(m-1)$-massless
partons and a $Q\bar{Q}$ pair, an observable with a pair of 
heavy quark jets in addition to $(m-2)$ jets, are built.
In other words, an $m$-jet observable is formed.

In this contribution, when the real matrix element squared $ |{\cal M}_{m+1}|^2$
is integrated over the phase space, it develops singularities 
when one parton in the final state is unresolved. In the presence of
massive partons in the final state, a parton is called unresolved,
either when it becomes soft or collinear to another massless parton or  when 
it is {\it quasi-collinear}  to a massive parton. In this latter case, it
leads to finite logarithmic terms involving the mass of the
massive parton. The notion of {\it quasi-collinear} limit will be
explicitly presented in Section 4. 
To extract the unresolved behaviours of the real matrix element, 
subtraction terms which take both the massless and massive effects 
need to be considered.

At the next-to-leading order,
the subtraction terms derived in the antenna formalism 
\cite{k,ourant,Daleo} are constructed solely with tree-level 
three-parton antenna functions. 
Those functions encapsulate all singular limits due to the emission 
of one unresolved parton between two colour-connected hard partons, 
called radiators.

Depending where the two radiators are located, in the initial or in the final
state, we distinguish three types of configurations: 
final-final, initial-final and initial-initial.
In any of those configurations, the radiated parton is 
always in the final 
state. 

The subtraction terms in a given configuration  
are constructed from products of the corresponding antenna functions 
with reduced matrix elements.
Those can be integrated over a phase space which 
is factorised into an antenna phase space (involving all
unresolved partons and the two radiators) multiplied by a reduced phase 
space, where the momenta of radiators and unresolved radiation are replaced 
by two redefined momenta. 
These redefined momenta can be in the initial or in the final state depending
on where the corresponding radiator momenta are and are defined by
appropriate mappings. 
The full subtraction term is then obtained by summing over all
antennae required in one configuration and by summing 
over all configurations needed for the problem under consideration.

The antenna subtraction terms do not provide a strictly local subtraction of
collinear singularities
in the case of a gluon splitting to two gluons or to a quark-antiquark pair.
In these, the antenna subtraction term accounts for the singular behavior
only after integration over the azimuthal angle of the two parton system
with respect to the collinear direction. As a consequence, the numerical
integration of the difference of matrix element and antenna subtraction term
is potentially unstable. By an appropriate partitioning of the final state
phase space \cite{GehrmannDeRidder:2007jk,Weinzierl-2j}, this azimuthal integration variable
can be separated off for each limit. Once this variable is separated, the
angular terms can be averaged out by a smooth one-dimensional integration,
or by combining different phase space points.

The massless and massive three parton final-final antenna functions, besides
being fundamental entities of the antenna subtraction formalism developed
for colourless initial states and for massless partons  in \cite{ourant}
or for massive partons in  \cite{Mathias} have another fundamental role.
Those can be used as basic ingredients in parton showers.
The event generator VINCIA uses these antenna functions as evolution kernels.
In its present formulation \cite{vinciaold,vincianew}, VINCIA describes
the evolution of timelike showers based on the massless \cite{ourant} 
and the massive \cite{Mathias} antenna functions.
A study concerning the importance of quark mass effects is currently ongoing. 
The results presented in this paper concerning the massive 
initial-final antennae will become relevant for initial-state 
parton showers for observables with massive final states. 

For the NLO corrections to $Q\bar{Q}$ + jets production in hadronic collisions, 
we will need all three types of subtraction terms and therefore all
three types of antenna functions.   
Since one massive radiator is always in the final state, 
the subtraction terms will involve final-final and initial-final 
massive antennae but no initial-initial antennae with massive particles. 
Antennae involving only massless partons though, will be needed 
in all three configurations. Those have been derived and 
integrated in \cite{ourant,Daleo,Joao}.
All required massive antenna functions will be presented in Section 3
and their integrated forms will be given in Section 5.

In the following, we shall present the general form of 
the subtraction terms needed in each of the three configurations 
(final-final, initial-final and initial-initial) to account for single
unresolved radiation in processes involving a heavy quark pair  
in association with jets in the final state. 
We will in particular focus on the changes introduced 
in the subtraction terms due to the presence of massive final state
particles compared to those when only massless partons are involved 
\cite{ourant,Daleo,Joao}.

\subsection{Subtraction terms for final-final configurations}

In the final-final configuration, 
the subtraction term related to the real contributions 
to the partonic process yielding a heavy quark pair in association 
with $(m-2)$ jets given in eq.(\ref{eq:real})  has to take 
into account the presence of an unresolved parton $j$ of momentum $k_{j}$ 
emitted between two hard final-state radiators $i$ and $k$ of momenta $k_{i}$ 
and $k_{k}$ respectively. It reads,
\begin{eqnarray}
{\rm d}\hat\sigma^{S,(ff)}_{NLO} &=&
{\cal N}\,\sum_{m+1}{\rm d}\Phi_{m+1}(k_{1},\ldots,k_{i},k_j,k_k,k_{m+1};p_1,p_2)
\frac{1}{S_{{m+1}}} \nonumber \\ 
 && \times \sum_{j}\;X^0_{ijk}\,
|{\cal M}_{m}(k_{1},\ldots,{K}_{I},{K}_{K},\ldots,k_{m+1};p_1,p_2)|^2\,
\nonumber\\ && 
\hspace{6mm}\times 
\JET_{m}^{(m)}(k_{1},\ldots,{K}_{I},{K}_{K},\ldots,k_{m+1})
\,.  
\label{eq:sub1}
\end{eqnarray}
This subtraction term involves the phase space for the production 
of $m+1$ partons,$\d \Phi_{m+1}$, the massive final-final three-parton 
antenna function  $X^{0}_{ijk}$, the reduced $m$-parton amplitude 
squared $|{\cal M}_{m}|^2$ and the jet function  $\JET^{(m)}_{m}$.
The jet function $\JET^{(m)}_{m}$  ensures 
that out of $(m-2)$ massless partons and a pair of massive partons, 
$(m-2)$ jets and a $Q\bar{Q}$ jet pair is build. 
The jet function and the reduced $m$-parton amplitude 
do not depend on the individual momenta ${k}_{i}$, $k_j$ and ${k}_{k}$, 
but will only depend on the redefined momenta $k_{I}$ and $k_{K}$   
which are linear combinations of the original momenta $k_{i},k_{j},k_{k}$.

Two cases are implicitly considered here. Either $i$ and $k$ are massive hard
final state radiators in which case the redefined partons $I$ and $K$
are massive or, $i$ is massless and $k$ is massive and 
the redefined partons $I$ and $K$  are massless and massive 
respectively.
In this case, one of the parton momenta  $k_{a}$ with $a \neq i,k$ 
is massive in order to obtain a reduced matrix element with two massive final
state partons.   

Eq.(\ref{eq:sub1}) holds strictly fot the subtraction of
  singularities of colour ordered matrix elements squared. Furthermore, as
 mentioned before, also interferences between partial amplitudes with
 different colour orderings appear in the subleading colour pieces. It
 will be seen in section 7.2.1 that the subtraction of infrared
 singularities appearing in these interferences needs a special
 treatment. 
However, to keep equations as brief and clear as possible, we still 
write our subtraction terms in the final-final configuration 
symbolically as in eq.(\ref{eq:sub1}).

Most of the massive final-final three-parton antenna functions $X^{0}_{ijk}$
needed to evaluate the NLO corrections to $Q\bar{Q}$ + jets production 
have been derived in \cite{Mathias}. Those will be listed in Section 3.
Solely, the new flavour-violating massive antennae, which are related 
to flavour violating vertices,  will be explicitly derived in that section.

The phase space $\d \Phi_{m+1}$ can be factorised as follows,
\begin{eqnarray}
\lefteqn{\d \Phi_{m+1}(k_{1},\ldots,k_i,k_j,k_k,\ldots,k_{m+1};p_1,p_2)  = }
\nonumber \\ &&
\d \Phi_{m}(k_{1},\ldots,{K}_{I},{K}_{K},\ldots,k_{m+1};p_1,p_2)
\cdot 
\d \Phi_{X_{ijk}} (k_i,k_j,k_k;{K}_{I}+{K}_{K})\;.
\label{eq:phasefinal}
\end{eqnarray}
${\rm d}\Phi_{m}$ is the $d$-dimensional
phase space for $m$ outgoing particles with momenta
$k_{1},\cdots,k_{m+1}$  with two of those momenta being massive,  
$d\Phi_{X_{ijk}}$ is the NLO final-final antenna phase space. 
It is proportional to a massive three-particle phase space relevant 
to a $1\to 3$ decay \cite{Mathias}.
Depending whether the two radiators are of equal masses or whether 
one of them is massless, different parametrisations of this antenna 
phase space are obtained. The parametrisations necessary to integrate the 
final-final massive antenna functions will be given in Section 5.

Appropriate final-final phase space mappings 
are furthermore required to define the final-state momenta $k_{I}$ and $k_{K}$ 
in the relevant kinematical configurations. Those are however not unique.
A possible mapping can be found in \cite{cdst1}.
As in the massless case, the mapping given in \cite{cdst1} 
is not symmetric under the exchange of $i$ and $k$. A possible 
symmetric version for a massive final-final definition of the mapped momenta 
will be given elsewhere. 

For the analytic integration, we can use the phase space factorisation formula
given in eq.(\ref{eq:phasefinal})  to rewrite each of the subtraction terms 
in eq.(\ref{eq:sub1}) in the form 
\begin{equation}
|{\cal M}_{m}(k_{1},\ldots,{K}_{I},{K}_{K},\ldots,k_{m+1};p_1,p_2)|^2\,
\JET_{m}^{(m)}\d \Phi_{m}\; \int d\Phi_{X_{ijk}} X_{ijk}^0.
\end{equation}
 
The integrated massive final-final antennae, 
normalised appropriately are defined by analogy to the massless case by  
\begin{equation}
{\cal X}_{ijk}^0(s_{ijk})=\frac{1}{C(\e)}\int d\Phi_{X_{ijk}} X_{ijk}^0
\label{eq:finalintegrated}
\end{equation}
with, the normalisation factor 
\begin{equation}
C(\e)=(4 \pi)^{\e}\frac{e^{-\e\gamma_{E}}}{8 \pi^2}.
\label{eq:Cepsilon}
\end{equation}
The integration is performed analytically in $d$-dimensions such that
the integrated subtraction terms can be combined with the one-loop
$m$-parton contribution.

\subsection{Subtraction terms for initial-final configurations}

In the initial-final configuration, the subtraction term 
related to the real contributions given in eq.(\ref{eq:real})  
has to take into account the presence of an unresolved parton $j$ 
of momentum 
 $k_{j}$, (which can be massive), emitted between a massless 
initial state radiator $i$ of momentum $p_{i}$  and a massive 
final state radiator $k$ of momentum $k_{k}$ and mass $m_{k}$.  
It is given by,
\begin{eqnarray}
{\rm d}\hat\sigma^{S,(if)}_{NLO} &=&
{\cal N}\,\sum_{m+1}{\rm d}\Phi_{m+1}(k_{1},\ldots,k_j,k_k,\ldots, k_{m+1};p_i,p_2)
\frac{1}{S_{{m+1}}} \nonumber \\ 
 && \sum_{j}\;X^0_{i,jk}\,
|{\cal M}_{m}(k_{1},\ldots,{K}_{K},\ldots,k_{m};p_i,p_2)|^2\,
\JET_{m}^{(m)}(k_{1},\ldots,{K}_{K},\ldots,k_{m}). \nonumber \\
\label{eq:sub2}
\end{eqnarray}

This subtraction term involves the phase space for the production 
of $(m+1)$ partons, $\d \Phi_{m+1}$, the massive three parton
initial-final antenna function  denoted by $X^{0}_{i,jk}$, the reduced
$m$-parton amplitude squared $|{\cal M}_{m}|^2$ and the jet function 
 $\JET^{(m)}_{m}$.
The reduced $m$-parton matrix element squared $|{\cal M}_{m}|^2$
does not contain any explicit dependence on the original final state 
momenta $k_{j}$ and $k_{k}$, 
but only depends on them through the redefined momentum $K_K$. 
The same holds for the jet function $\JET_m^{(m)}$.
The three parton massive antenna functions $X^{0}_{i,jk}$ depend only
on the momenta $p_i$, $k_j$ and $k_{k}$ and on the masses of the final
state partons $m_{j}$ and $m_{k}$. These 
will be presented in Section 3.

In this configuration, the initial state radiator $i$ is always massless
while the final state radiator $k$ is always massive. 
The unresolved parton $j$ can be either massless or massive  
such that two situations are covered in eq.(\ref{eq:sub2}).
In case it is massless, one of the partons with $k_{a}$
where $a \neq j,k$ has to be taken massive in order to build an event with a
pair of massive partons in addition to $(m-2)$ jets. 
Depending whether the unresolved parton is massless or massive, different
phase space parametrisations will be needed. Those will be presented
in Section 5 when the massive initial-final antennae are integrated
over the unresolved phase space.  

As it was mentioned for the final-final case, the subtraction
 terms as presented in eq.(\ref{eq:sub2}) are strictly valid for the 
 subtraction of infrared singularities of colour ordered matrix
 elements squared. Interferences of different partial amplitudes which
 also generate infrared singularities need
 a special treatment which we shall present in section 7.2.1.
To keep our equations as brief and clear as possible, we still 
write our subtraction terms in the initial-final configuration 
symbolically as in eq.(\ref{eq:sub2}).

\subsubsection{Phase space factorisation and mappings}
For the subtraction term to be integrated 
and then added to the virtual contributions and mass factorisation
counterterms, we need the phase space 
to factorise adequately. As the presence of massive final states lead to
differences in this factorisation with respect 
to the massless case considered in \cite{Daleo}  we derive 
this phase space factorisation explicitly below.

We start from the $2\rightarrow (m+1)$-particle phase space 
with a priori two massive final state particles of momenta $k_{j}$ and $k_{k}$ 
and the remaining $(m-1)$ particles staying massless. The momenta of the
initial state partons are named $p_i$ and $p_{2}$.
\begin{equation}
{\rm d}\Phi_{m+1}\left( k_1,...,k_{m+1};p_i,p_{2}\right)=
(2\pi)^d\:\delta\left( p_i+p_2-\sum_l k_l\right)\prod_l \left[{\rm d}k_l\right],
\label{eq:phif}
\end{equation}
where the phase space measure for massless and massive partons 
are respectively given by: $\left[{\rm d}k_l\right]={\rm d}^dk\:\delta^+\left(k_l^2\right)/\left(
2\pi\right)^{d-1}$ for $l\neq j,k$ and 
$\left[{\rm d}k_l\right]={\rm d}^dk_l\:\delta^+\left(k_l^2 -m^2_l\right)/\left(
2\pi\right)^{d-1}$ for $l=j,k$.
 We insert
\begin{equation}
1=\int {\rm d}^d q\:\delta(q+p_{i}-k_j-k_k),
\end{equation}
and, to take into account the masses of the final state particles $j$ and $k$ 
we insert
\begin{equation}
1=\frac{Q^2+m_K^2}{2\pi}\int\frac{{\rm d}x_i}{x_i}\int [{\rm d}K_K](2\pi)^d\:
\delta(q+p_I-K_K)
\end{equation}
with
\begin{equation}
Q^2=-q^2\:\:\:\:\:\:\:\:\:\:\:\:\:\:\:\:\:\:\:
x_i=\frac{Q^2+m_{K}^2}{2p_{i}\cdot q} 
\hspace{1cm} {\rm and} \hspace{1cm} p_{I}=x_i\,p_i.
\end{equation}
We find that the original phase space for $(m+1)$ partons 
given in eq.(\ref{eq:phif}) can be written as the product of 
an $m$-parton (${\rm d}\Phi_m$ ) and a two-to-two particle phase space 
(${\rm d}\Phi_2$) as follows,
\begin{eqnarray}
{\rm d}\Phi_{m+1}(k_1,...,k_{m+1};p_{i},p_2)
={\rm d}\Phi_m(k_1,...,K_K,...,k_{m+1};x_{i}p_{i},p_2)\nonumber\\
\times\frac{(Q^2+m_K^2)}{2\pi}{\rm d}\Phi_2(k_j,k_k;p_{i},q)
\frac{{\rm d}x_{i}}{x_{i}}.
\label{factorisation}
\end{eqnarray}
Replacing this phase space in the subtraction term given in 
eq.(\ref{eq:sub2}) 
we can explicitly carry out the integration of the antenna functions 
over the two-to-two particle phase space and 
get the integrated expression for the subtraction term to be added to the virtual contributions and the mass
factorisation counterterms. For this
purpose it is convenient to define the integrated massive initial-final 
antenna as,
\begin{equation}
{\cal X}_{i,,jk}=\frac{1}{C(\epsilon)}
\int {\rm d}\Phi_2\frac{(Q^2+m_K^2)}{2\pi}X_{i,jk},
\end{equation}
where
$C(\epsilon)$ is given in eq.(\ref{eq:Cepsilon}) and the initial-final
massive antenna phase space denoted by ${\rm d}\Phi_{X_{i,jk}}$ is given by,
\begin{equation}
{\rm d}\Phi_{X_{i,jk}}= {\rm d}\Phi_2\frac{(Q^2+m_K^2)}{2\pi}
\label{eq:initialfinalphase}.
\end{equation}

Due to the presence of massive particles, the phase space mapping for
initial-final configurations derived in the massless case in
\cite{Daleo}  needs to be generalized as well:
In a process of the form $ p_i +q \to k_j +k_k$ with $(q^2<0)$ ,
to find a mapping from the original momenta $p_i$ in the initial state, $k_{j}$
and $k_{k}$ in the final state, to the redefined momenta $p_{I}$ and 
$K_K$ that ensures phase space factorisation, the following conditions
need to be fulfilled: 
The remapped final state momentum $K_K$ must be on-shell, 
and momentum must be conserved.
This implies,
\begin{equation}
p_i-k_j-k_k=x_{i}p_{i}-K_K=p_{I}-K_{K}
\end{equation}
for the phase space to factorise as in eq.(\ref{factorisation}).

In addition, concerning the single unresolved behaviour of
the remapped momenta, the following  requirements have to be satisfied:
\begin{eqnarray}
&& x_ip_{i}\rightarrow p_i \:\:\:\:\:\:\:\:\:\:\:\:\:\:\:\:\: K_K\rightarrow k_k \:\:\:\:\:\:\:\:\:\:\:\:\:\:\:\:\:\: \mbox{when $j$ becomes soft}\nonumber\\
&& x_ip_i\rightarrow p_i \:\:\:\:\:\:\:\:\:\:\:\:\:\:\:\:\: K_K\rightarrow k_j+k_k
  \:\:\:\:\:\:\:\:\: \mbox{when $j$ becomes collinear or quasi-collinear with
    $k$}\nonumber \\
&& x_ip_i\rightarrow p_i-k_j \:\:\:\:\:\:\:\: K_K\rightarrow k_k
  \:\:\:\:\:\:\:\:\:\:\:\:\:\:\:\:\:\: \mbox{when $j$ becomes collinear or
    quasi-collinear with $i$}.\nonumber\\ 
\label{conditions}
\end{eqnarray}
It can be seen that when parton $j$ 
becomes soft, collinear or quasi-collinear to parton $k$ 
we have $x_{i}\rightarrow 1$, 
while for the case in which partons $i$ and $j$ are collinear or
quasi-collinear, $x_i\rightarrow 1-z$, with $z$ being the fraction 
of the initial state momentum $p_{i}$ carried by the unresolved momentum $j$. 
The notion of quasi-collinear limit can be viewed as an  
extension of the collinear limit when at least one of the partons
becoming collinear to another is massive. 
Its explicit definition in this initial-final configuration 
will be given in
Section 4. 

All these conditions given above are interpolated by,
\begin{eqnarray}
& K_K=k_j+k_k-(1-x_i)p_i &\nonumber\\
& x_i=\dfrac{s_{ij}+s_{ik}-s_{jk}}{s_{ij}+s_{ik}}.
\label{eq:mappingi}
\end{eqnarray}
With this choice, the mass relation for momenta before 
and after mapping reads,
\begin{equation} 
m_{K}^2=m_{k}^2+m_{j}^2. 
\label{eq:mK2}
\end{equation}
Note also that in eq.(\ref{eq:mappingi}), $s_{ab}$ stand for $2p_a\cdot p_b$.
as everywhere in this paper. In terms of those, massless \cite{Daleo}
and massive definitions of $x_i$ are the same. 
The massless phase space factorisation and mappings given in
\cite{Daleo} can be recovered by setting the masses of the final state 
particles to zero.

\subsection{Subtraction terms for initial-initial configurations}
Additional divergent contributions may finally also 
occur in the real matrix element squared given in eq.(\ref{eq:real})  when 
a massless final state parton becomes unresolved 
with respect to two initial state partons. 
In this case,  the subtraction terms are constructed exclusively 
with massless three parton initial-initial antennae.
In those, the initial state partons are the hard
radiators and this situation was studied in detail in \cite{Daleo}.

The subtraction term associated to an unresolved massless parton $j$ 
and two hard initial state radiators $i$ and $k$ with momenta 
$p_{i}$ and $p_{k}$ in the partonic process 
$ p_{i}+p_{k} \to k_{j} + k_{Q} +k_{\bar{Q}}+ (m-2)$ partons takes the form  
\begin{eqnarray}
{\rm d}\sigma^{S,(ii)}_{NLO} &=&
{\cal N}\,\sum_{m+1}{\rm d}\Phi_{m+1}(k_{Q},k_{\bar{Q}},k_{1},
\ldots,k_j,\ldots, k_{m-1};p_i,p_k)
\frac{1}{S_{{m+1}}} \nonumber \\ 
 && \times \sum_{j}\;X^0_{ik,j}\,
|{\cal M}_{m}(\tilde{k}_{Q},\tilde{k}_{\bar{Q}},
\tilde{k}_{1},\ldots,\tilde{k}_{j-1},\tilde{k}_{j+1}\ldots,k_{m-1};x_ip_i,x_kp_k)|^2\,
\nonumber\\ && 
\hspace{6mm}\times \JET_{m}^{(m)} (\tilde{k}_{Q},\tilde{k}_{\bar{Q}},\tilde{k}_{1},
\ldots,\tilde{k}_{j-1},\tilde{k}_{j+1}\ldots,\,
\ldots,\tilde{k}_{m-1}). \nonumber \\ && 
\label{eq:subii}
\end{eqnarray}
This subtraction term involves the phase space for the production 
of $(m+1)$ partons, $\d \Phi_{m+1}$, the massless three parton
initial-initial antenna function  denoted by $X^{0}_{ik,j}$, the reduced
$m$-parton amplitude squared $|{\cal M}_{m}|^2$ and the jet function 
 $\JET^{(m)}_{m}$.
As  mentioned for the final-final and initial-final cases, 
the subtraction terms as presented in eq.(\ref{eq:subii}) are strictly 
valid for the subtraction of infrared singularities of colour ordered
matrix elements squared only. Interferences of different partial
amplitudes need a special treatment which we shall present in Section 7.2.1.
However, to keep our equations as brief and clear as possible, we still 
write our subtraction terms in the initial-initial configuration symbolically as in eq.(\ref{eq:subii}).

All momenta (massless and massive) 
in the arguments of the reduced
matrix elements $|{\cal M}_{m}|^2$
and the jet function $\JET_{m}^{(m)} $ 
have to be redefined. They are denoted with tildes in eq. (\ref{eq:subii}).
The two hard radiators are simply rescaled by factors 
$x_i$ and $x_k$ respectively. 
The other momenta present in the reduced matrix element squared are
boosted by a Lorentz transformation onto the new set of 
momenta $\{\kt_l,\,l\neq j\}$ as described for massless partons 
in \cite{Daleo}. 
We have checked that the same boost \cite{cs} 
required to redefine the momenta of the massless partons 
can be used to redefine the momenta of the massive partons.
Consequently, the presence of massive partons 
in the final state does not influence the way the phase space 
factorises and how the mapping is defined so
that the massless factorisation \cite{Daleo} can be used in this context.

The phase space ${\rm d}\Phi_{m+1}$ 
factorises into the convolution of a massive $m$-particle phase space 
involving only redefined momenta and a massless initial-initial
antenna phase space related to  the phase space 
of the unresolved parton $j$, as in the massless case.

All  required initial-initial massless antennae $X^{0}_{ik,j}$ 
needed for the construction of the subtraction terms for $t\bar{t}$ 
production in association with jets have been derived and
integrated in \cite{Daleo,Joao}.

\section{Massive antenna functions}
In this section, we aim to list all massive antenna functions which
are needed to construct the subtraction terms for the hadronic
production of a heavy quark pair and for the production of a
heavy quark pair in association with one additional jet at NLO. 

\subsection{General features of antenna functions}

In Section 2, we saw that the 
subtraction terms defined in the three configurations 
(final-final, initial-final and initial-initial) are 
built with products of the corresponding antenna 
functions (denoted by $X$) with reduced matrix element squared.
At this order (NLO), only tree level three-particle antenna functions 
are required. 
Those describe all unresolved (soft, collinear and quasi-collinear) radiation
emitted between a pair of colour-ordered hard partons, the radiators.
Originally, in \cite{ourant}, the antenna functions were derived 
for massless final-final configurations.
Those are defined by the pair of hard partons they collapse to in the
unresolved limits and in all cases, the antenna functions are derived 
from physical matrix elements.
Generally, the quark-antiquark antenna functions are obtained
from $\gamma \to q\bar{q} +(partons) $, the quark-gluon antenna 
functions from $\tilde{\chi} \to \tilde{g}+(partons)$
\cite{gluino} and the gluon-gluon antenna functions from $H \to (partons)$
\cite{higgs}.

The three parton final-final antenna functions were obtained by 
normalising the colour-ordered three-parton tree level 
matrix elements squared to the matrix element squared 
for the basic two-parton process, omitting all couplings and colour factors.
As such the tree-level three parton massless or massive  
final-final antenna functions {\it are scalars in colour space},  
have mass dimension $(-2)$ and are defined by 
\cite{ourant,Mathias} as,
\begin{eqnarray}
X_{ijk}^0 = S_{ijk,IK}\, \frac{|{\cal M}^0_{ijk}|^2}{|{\cal
    M}^0_{IK}|^2}\; .
\end{eqnarray}
$S$ denotes the symmetry factor associated to the antenna, 
which accounts both for potential identical particle symmetries and 
for the presence of more than one antenna in the basic two-parton
process. It is chosen such that the antenna function reproduces 
the unresolved limits of a matrix element with identified particles.

At NLO the existing three parton tree level massless final-final $X_{ijk}^0 $ 
  antennae are:
\begin{itemize}
\item Quark-antiquark: the only antenna functions of this kind at NLO
  are the A-Type antennae, and they are obtained from the ratio
  $|{\cal M}(\gamma^*\rightarrow q\bar{q}g)|^2/|{\cal
    M}(\gamma^*\rightarrow q\bar{q})|^2$. Since the quark and the
  antiquark are of the same flavour, in the following these antennae
  will be referred as  flavour conserving A-Type antennae.
\item Quark-gluon: there are two different antenna functions of this kind: D-Type and E-Type. The D-Type antennae are obtained from the ratio $|{\cal M}(\tilde{\chi}\rightarrow\tilde{g}gg)|^2/|{\cal M}(\tilde{\chi}\rightarrow\tilde{g}g)|^2$, while the E-Type are computed from $|{\cal M}(\tilde{\chi}\rightarrow\tilde{g}q\bar{q})|^2/|{\cal M}(\tilde{\chi}\rightarrow\tilde{g}g)|^2$  \cite{gluino}.
\item Gluon-gluon: there are also two different antennae of this kind: F-Type and G-Type. The F-Type antenna functions are obtained from $|{\cal M}(H\rightarrow ggg)|^2/|{\cal M}(H\rightarrow gg)|^2$, while the G-Type are calculated from the ratio $|{\cal M}(H\rightarrow gq\bar{q})|^2/|{\cal M}(H\rightarrow gg)|^2$ \cite{higgs}.
\end{itemize}

The initial-final antennae and the initial-initial antennae denoted
respectively by $X_{i,jk}^0 $  and $X_{ik,j}^0 $  are 
in principle defined by crossing one or two {\it massless} partons  
from the final to the initial state in the final-final antennae $X_{ijk}^0$.
However, this crossing may not be unambiguous for initial-final 
configurations\cite{Daleo}. 

All these types of antenna functions are needed for the construction 
of the subtraction terms for heavy quark pair production 
except the G-Type antennae. Those antennae are required  
in all three (final-final, initial-final or initial-initial) 
configurations and in their massless or massive versions. 
In the latter case, the final state quark ($q$) or gluino
($\tilde{g}$) is taken massive. Note that if both of those partons are
present in the final state, the corresponding massive antenna is
obtained by taking only the gluino $\tilde{g}$ to be massive. The
antenna obtained taking both of these partons massive is not required
for the construction of the subtraction terms for heavy quark pair
production +jets at NLO, considered here. 

In addition to these flavour conserving antennae, we will 
also need flavour violating massive quark-antiquark A-Type antennae in both 
final-final and initial-final configurations. Those involve flavour
violating vertices and will be defined below.

In all these antennae denoted by $X_{ijk}$ in the final-final configuration 
and crossings of those, the labels $i,j,k$ can either 
stand for massless or massive partons. The particle with label $j$ 
always denotes the unresolved parton. 
The massive antennae will depend on the invariants 
$s_{ij},s_{ik}$ and $s_{jk}$ (with $s_{ab}=2p_ap_b$ throughout this paper) 
and on the masses of the final states heavy partons.  

In the following, when presenting the expressions of the required
massive antennae we will use a clear labeling of those antennae.  
We shall specify which partons in them are taken massless, 
those will be indexed with $q$, or taken massive and then indexed with $Q$.
For conciseness, the ${\cal O}(\epsilon)$ 
in the expression of the unintegrated antennae will be omitted.

\subsubsection{Flavour-violating antennae}
For the construction of the subtraction terms for $t \bar{t}$ +jet 
production at NLO, we encounter two types of
flavour violating antennae. Both are  A-Type quark-antiquark antennae and are 
labelled : $A^0_{qg\bar{Q}}$ and $A^0_{qgQ}$. In those, $q$ represents a massless
quark and $Q$ and $\bar{Q}$ represent a massive quark or antiquark of
different flavour than the massless quark $q$.
Note that for symmetry reasons of the A-Type antenna, 
the role of $q$
and $\bar{Q}$ in   $A^0_{qg\bar{Q}}$  or of $q$ and $Q$ in
$A^0_{qgQ}$  are interchangeable resulting in the same two flavour-violating 
antennae. We encounter these two types of flavour violating antennae 
in final-final and initial-final configurations.
In the latter case, a massless quark is always in the initial state
and plays the role of the initial state radiator.
It is worth noting that the limits covered by flavour violating antennae 
with a gluon in the initial state can be covered by flavour conserving 
initial-final antennae, such that these gluon initiated flavour
violating antennae are not required here.

The massive final-final flavour violating antennae have their massless
counterparts given by $A_{qg\bar{q}'}$ and $A_{qgq'}$ where $q'$ and
$\bar{q}'$ are massless quarks or antiquarks of different flavour than $q$.
These massless counterparts have been used
as essential ingredients for the construction of the subtraction terms 
required for the computation of $e^+e^- \to 3$ jets at NNLO 
\cite{GehrmannDeRidder:2007jk}.
Those massless A-Type flavour violating antennae have exactly 
the same properties and singular structure as a massless
flavour-conserving A-Type quark-antiquark antenna given by 
 $A^{0}_{q g\bar{q}}$ \cite{ourant}. Those have also the same unintegrated 
form as this antenna.

In the massive case,
the final-final flavour violating A-Type antennae  
$A^0_{qg\bar{Q}}$ and $A^0_{qgQ}$, having one massive particle in the
final state,  have the same properties 
and singular structure as a massive flavour-conserving A-Type  
antenna of the form $A^{0}_{Qg\bar{Q}}$  where  
the massive quark $Q$ or the massive antiquark $\bar{Q}$ (by symmetry
arguments)  present in this flavour conserving antenna 
is taken massless.

By construction, within the antenna formalism,
in the unresolved (collinear, soft or quasi-collinear) limits, 
the three parton antennae yield massless or
massive universal unresolved factors given either by splitting
functions or eikonal factors in their massless or massive
forms. The spin properties of the final states are determinant 
to obtain these unresolved factors.
Consequently, all flavour-violating antennae (massless or massive) can
be generated by processes resulting from the decay of a charged
W-boson into massless or massive final state fermions. 
When the particle in the final state is a (massless or massive) antiquark  
such that the flavour violating antenna 
contains a {\it quark and an antiquark} of different flavours, 
those final states can be Dirac fermions. 
When the particles in the antennae are {\it two
quarks} of different flavours, those final states can be produced 
in extended versions of the Standard Model \cite{Rosiek} 
and will involve Dirac and Majorana fermions.   

\subsection{Massive NLO final-final antenna functions}
For the construction of the subtraction terms relevant for 
heavy quark pair production at hadron colliders at NLO, 
we will need three-parton massive final-final flavour conserving and  
flavour violating A-Type antennae and flavour conserving 
quark-gluon antenna functions.
All flavour-conserving massive final-final antennae required have been
derived in \cite{Mathias}. They are given below for completeness. 
The flavour violating massive final-final antennae are new.  

The generic process considered to define the massive 
three-parton final-final antenna $X^{0}_{ijk}$ is given 
by $ q \to k_{i}+k_{j}+k_{k}$  where $q$ is the virtuality 
of the colourless initial state, with $q^2>0$.
In the following, the center of mass energy  of the decaying particle $E_{cm}$  
will be used instead of $q$ with $q^2=E_{cm}^2$
The final state radiator partons $i$ and $k$ 
are either both massive or only one of them is such. Parton $j$ 
defines always the unresolved massless final state parton. 

The massive three-parton final-final antenna $X^{0}_{ijk}$ 
will depend on on the invariants $s_{ij}$, $s_{jk}$ and $s_{ik}$, 
on the masses $m_{i}$ and $m_{k}$ of the final state partons $i$ and $k$
and on $E_{cm}^2$.

\subsubsection{Quark-Antiquark antennae}
The flavour conserving quark-antiquark massive antenna function denoted by
$A^0_{Qg\bar{Q}}$ has a massive quark $Q$ and a massive antiquark $\bar{Q}$ 
as radiators and reads,
\begin{equation}
\begin{split}
 A^{0}_{3}(1_Q,3_g,2_{\bar{Q}}) &= \frac{1}{ 4\left(\enen + 2 m_{Q}^2\right)} 4 \left(\frac{2 s_{12}^2}{s_{13} s_{23}}+\frac{2
   s_{12}}{s_{13}}+\frac{2
   s_{12}}{s_{23}}+\frac{s_{23}}{s_{13}}+
   \frac{s_{13}}{s_{23}}\right.\\
& \quad \left.+ m_{Q}^2 \left(\frac{8 s_{12}}{s_{13} s_{23}}-\frac{2
   s_{12}}{s_{13}^2}-\frac{2
   s_{12}}{s_{23}^2}-\frac{2
   s_{23}}{s_{13}^2}-\frac{2}{s_{13}}-\frac{2
   }{s_{23}}-\frac{2 s_{13}}{s_{23}^2}\right)\right.\\
& \quad \left. + m_{Q}^4 \left(-\frac{8}{s_{23}^2}-\frac{8}{s_{13}^2}\right)\right)
  	+\order{\epsilon}.
\end{split}
\label{eq:A03mff}
\end{equation}
This function is  normalised to the two-particle process
$\gamma^{*} \to Q\bar{Q}$, whose matrix element squared (omitting 
couplings) is given by 
\begin{equation}
A^{0}_{2}(1_{Q},2_{\bar{Q}})=4 \left[(1-\eps)\,E_{cm}^2 +2\,m_{Q}^2\right].
\end{equation}

\subsubsection{Quark-Gluon antennae}
For the subtraction terms required for the production of 
heavy particles in addition to jets, only  
massive quark-gluon antennae with one massive radiator 
in the final state are needed.

The quark-gluon massive antennae with either two gluons or a
massless quark-antiquark pair in the final state are normalised by the 
two-particle matrix element squared 
relevant for the process $\tilde{\chi}  \to \tilde{g} g$, with the gluino $\tilde{g}$
being massive with mass $m_{Q}$ and the gluon $g$ being massless. 
This two-particle matrix element squared omitting couplings
reads,
\begin{equation}
X^{0}_{2}(1_{Q},2_{g})=4\,(1-\eps)\,\left(E_{cm}^2-m_{Q}^2 \right)^2,
\end{equation}
where $X$ can stand for the $E$ or $D$-Type antennae here.

The quark-gluon massive antenna $ E^{0}_{Q q'\bar{q}'}$  
with a pair of massless quarks $q'\bar{q}'$ and a
massive radiator quark $Q$  in the final state reads, 
\begin{equation}
 E^{0}_{3}(1_{Q},3_{q'},4_{\bar{q}'}) = \frac{1}{4 (\enen -
   m_{Q}^2)^2} 4 \left( s_{13} + s_{14} + \frac{s_{13}^2}{s_{34}} +
   \frac{s_{14}^2}{s_{34}} - 2 \en m_{Q}\right) + \order{\epsilon}.
\label{eq:ffetype}
\end{equation}

The quark-gluon massive antenna $D^{0}_{Qgg}$ 
with two gluons and a massive radiator $Q$ in the final state reads, 
\begin{equation}
\begin{split}
D_{3}^{0}(1_{Q},3_{g},4_{g})&= \frac{1}{4 (\enen -m_{Q}^2)^2}
4 \left(
	\left( 9 s_{13} + 9 s_{14} + \frac{4 s_{13}^2}{s_{14}} + \frac{4 s_{14}^2}{s_{13}} + \frac{4 s_{13}^2}{s_{34}} + \frac{2 s_{13}^3}{s_{14} s_{34}} \right. \right. \\
	& \quad \quad \left. \left. + \frac{6 s_{13} s_{14}}{s_{34}} + \frac{4 s_{14}^2}{s_{34}} + \frac{2 s_{14}^3}{s_{13} s_{34}} + 6 s_{34} + \frac{3 s_{13} s_{34}}{s_{14}} + \frac{3 s_{14} s_{34} }{s_{13}} + \frac{s_{34}^2}{s_{13}} + \frac{s_{34}^2}{s_{14}} \right)
	\right. \\
	& \quad \left.
	- m_{Q}^2 \left( 6 + \frac{2 s_{13}^2}{s_{14}^2} + \frac{4 s_{13}}{s_{14}} + \frac{4 s_{14}}{s_{13}} + \frac{2 s_{14}^2}{s_{13}^2} + \frac{6 s_{34}}{s_{13}} + \frac{4 s_{13} s_{34}}{s_{14}^2} \right. \right.\\
	& \quad \quad \left. \left. + \frac{6 s_{34}}{s_{14}} + \frac{4 s_{14} s_{34}}{s_{13}^2} + \frac{2 s_{34}^2}{s_{13}^2} + \frac{2 s_{34}^2}{s_{14}^2} + \frac{2 s_{34}^2}{s_{13} s_{14}} \right)
	\right. \\
	& \quad \left.
	+ 2 \en m_{Q} - \en m_{Q}^3\frac{2  s_{34}}{s_{13} s_{14}} 
+ m_{Q}^4\frac{2 s_{34}}{s_{13} s_{14}} 
\right) + \order{\epsilon}.
 \end{split}
\end{equation}

This tree-level antenna function $D_{3}^{0}(1_{Q},3_{g},4_{g})$ 
contains two antennae, corresponding to the following configurations: 
gluon ($3_g$) radiated between the massive quark and gluon ($4_g$) 
and also the configuration where gluon ($4_g$) is radiated between
 the quark and gluon ($3_g$). 
The separation between these two configurations is not free from ambiguity, 
since the collinear limit of the two gluons has to be split between 
the two configurations. We consider the following decomposition
\begin{equation}
D_3^0(1_Q,3_g,4_g)=d_3^0(1_Q,3_g,4_g)+d_3^0(1_Q,4_g,3_g)
\end{equation}
where the sub-antenna denoted by $d_{3}^{0}$ is given by
\begin{eqnarray}
&& d_3^0(1_Q,3_g,4_g)=\frac{1}{(E_{cm}^2-m_Q^2)^2}\nonumber\\
&&\:\:\:\:\:\:\times \left[ \frac{9s_{13}+9s_{14}+6s_{34}}{2}+\frac{4s_{14}^2}{s_{13}}+\frac{4s_{14}^2}{s_{34}}+\frac{2s_{14}^3}{s_{13}s_{34}}+\frac{3s_{13}s_{14}}{s_{34}}+\frac{3s_{14}s_{34}}{s_{13}}+\frac{s_{34}^2}{s_{13}}\right. \nonumber\\
&&\:\:\:\:\:\:\:\:\:\: \left. -m_Q^2\left( 3+\frac{2s_{14}^2}{s_{13}^2}+\frac{4s_{14}}{s_{13}}+\frac{6s_{34}}{s_{13}}+\frac{6s_{14}}{s_{13}}+\frac{4s_{14}s_{34}}{s_{13}^2}+\frac{2s_{34}^2}{s_{13}^2}+\frac{s_{34}^2}{s_{13}s_{14}}\right)\right. \\
&&\:\:\:\:\:\:\:\:\:\: \left.+E_{cm}m_Q-(E_{cm}-m_Q)m_Q^3\frac{s_{34}}{s_{13}s_{14}}\right]+O(\epsilon)\nonumber
\end{eqnarray}

Both of these sub-antennae will be needed individually 
to construct the subtraction term for $t\bar{t}$ +jet at NLO. 
Those will not need to be integrated separately though. 

\subsubsection{Flavour violating antennae}
In addition to the massive flavour-conserving final-final antennae
given above, for the construction of the subtraction terms for  
$t \bar{t}$ +jet production at NLO,
we also need two types of massive flavour violating final-final antennae.
We will require an antenna involving as radiators: one massive quark $Q$
and one massless antiquark $\bar{q}$ , denoted by $A^{0}_{Qg\bar{q}}$ and 
one antenna with a massive quark $Q$ and a massless quark $q$ 
denoted by $A^{0}_{Qg\bar{q}}$ .
The expressions of these two antennae are related by the exchange of 
a massless quark versus an antiquark in the final state. 
Therefore, we here present only one of those namely $A^0_{Qg\bar{q}}$.
 It is given by,
\begin{eqnarray}
&& A_3^0(1_Q,3_g,2_{\bar{q}})=\frac{1}{(E_{cm}^2+m_Q^2)}\left[ \frac{2s_{12}}{s_{13}}+\frac{2s_{12}}{s_{23}}+\frac{2s_{12}^2}{s_{13}s_{23}}+\frac{s_{13}}{s_{23}}+\frac{s_{23}}{s_{13}}\right.\nonumber \\
&&\:\:\:\:\:\:\:\:\:\:\:\:\:\:\:\:\:\:\:\:\:\:\:\:\:\:\:\:\:\:\:\:\:\:\:\:\:\:\:\:\:\:\:\:\:\:\:\:\:\:\:\:\:\:\:\:\left. -m_Q^2\left(
    \frac{2s_{12}}{s_{13}^2}+\frac{2s_{23}}{s_{13}^2}+\frac{2}{s_{13}}\right)\right]+O(\epsilon)
\label{eq:fvff}
\end{eqnarray}

This antenna function is normalised with the following two-parton
 matrix element squared:
\begin{equation} 
A_2^0(1_Q,2_{\bar{q}})=4\,(1-\epsilon)\left[E_{cm}^2+m^2_{Q}\right].
\end{equation}
It is worth noting that the expression of this final-final flavour
violating antenna $A^{0}_{Qg\bar{q}}$ given in eq.(\ref{eq:fvff}) 
differs significantly from the flavour conserving antenna $A^0_{Qg\bar{Q}}$ 
given above in eq.(\ref{eq:A03mff}) as expected.

\subsection{Massive NLO initial-final antenna functions}
To construct our subtraction terms for heavy quark pair production in
association with jets at NLO, we
will also need massive initial-final antennae of different types.  
Flavour-conserving and flavour-violating massive quark-antiquark A-Type antennae 
and flavour conserving massive quark-gluon antennae are required. 

In principle, the massive initial-final massive antennae 
can be obtained from the corresponding expressions for the massive
final-final antennae given above by appropriately crossing 
one massless parton from the final to the initial state. 
By this crossing procedure, the presence of an overall uneven number 
of fermions crossed to define the initial-final antenna leads to an overall 
minus sign in the definition of the antennae.
Furthermore, this crossing procedure is not non ambiguous
for the quark-gluon D-Type antennae initiated by a gluon. 

In general, the initial-final antennae $X_{i,jk}$ are normalised to 
the reduced two-parton matrix element squared $|{\cal M}^0_{I,K}|^2$ 
to which the three-parton matrix element squared $|{\cal M}^0_{i,jk}|^2$ 
tends in the limits. For the D-Type antennae initiated by a gluon, 
depending which limit is considered, 
the reduced two-parton matrix element to which 
the three-parton matrix element collapses to, can be different.
In particular, the nature of the initial state parton in the three and 
two-parton matrix elements squared needed as ratio to define 
the antennae may change.

The generic process necessary to define the massive initial-final 
three parton antenna $X^{0}_{i,jk}$ is given by $q +p_{i}\to k_{j}+k_{k}$ 
with $q^2<0$ and $q^2=-Q^2$. 
In this process, $p_{i}$ is the momentum of the initial 
state radiator $i$ whereas $k_{j}$ and $k_{k}$ are the momenta 
of the unresolved parton $j$ and the final state radiator $k$
respectively. Depending on the situation considered, the 
one massive parton present in the final state can either be the
unresolved parton or the final state radiator.

In any case, the initial-final antennae $X_{i,jk}$ will depend 
on the invariants $s_{ij}$, $s_{jk}$ and $s_{ik}$, 
on the masses $m_{j}$ and $m_{k}$ of the final state partons $j$ and $k$
and on $Q^2$.

\subsubsection{Quark-Antiquark antennae}
By crossing a gluon to the initial state in the expression of the final-final
massive quark-antiquark antenna $A^{0}_{Qg\bar{Q}}$ one obtains the  
initial-final massive quark-antiquark antenna $A^{0}_{g;Q\bar{Q}}$.
In it, the gluon plays the role of the initial state radiator, 
the final state radiator is a massive quark or an antiquark (by
symmetry arguments). 
The unresolved parton is correspondingly either a massive antiquark or 
a quark which can become quasi-collinear to the initial state gluon. 
This antenna is given by,

\begin{eqnarray}
A_3^0(3_g;1_Q , 2_{\bar{Q}})&=&-\frac{1}{[Q^2-2m_Q^2]}
\left( -\frac{2s_{12}^2}{s_{13} s_{23}}+\frac{2s_{12}}{s_{13}}
+\frac{2s_{12}}{s_{23}}-\frac{s_{13}}{s_{23}}-\frac{s_{23}}{s_{13}}\right.\nonumber \\
&&\left. -m_Q^2 \left( \frac{8s_{12}}{s_{13}s_{23}}-\frac{2s_{12}}{s_{13}^2}-\frac{2s_{12}}{s_{23}^2}+\frac{2s_{13}}{s_{23}^2}+\frac{2s_{23}}{s_{13}^2}+\frac{2}{s_{13}}+\frac{2}{s_{23}}\right)\right. \label{eq:ifAg} \\
&&\left. -m_Q^4\left(-\frac{8}{s_{13}^2}-\frac{8}{s_{23}^2}\right)\right)
+O(\epsilon). \nonumber
\end{eqnarray}

This function has been normalised to the two-particle matrix element
related to the process $\gamma^{*} Q\to Q$. It corresponds to the matrix 
element in which the process $\gamma^{*} g\to Q \bar{Q}$ 
reduces to in all its limits.
Omitting couplings, this normalisation factor is given by 
\begin{equation}
A^{0}_{2}(1_{Q};2_{Q})=4 \left[(1-\eps) Q^2 -2\,m_{Q}^2\right].
\label{eq:Aifnorm}
\end{equation}

Note that the resulting antenna has an overall minus sign made
explicit in eq.(\ref{eq:ifAg}) due to the uneven
number of fermions crossed to define it.

\subsubsection {Quark-gluon antennae}
As in the massless case,
the quark-gluon final-final massive antennae are separated into two
categories depending if the final state radiator gluon splits into a
quark-antiquark pair (E-Type) or into two gluons (D-Type).
These initial-final massive antennae depend furthermore 
on the mass  $m_{\chi}$ of the decaying
neutralino with momentum $q$.
Following our definition of $Q^2$, it is given by $m_{\chi}=\sqrt{-Q^2}$.
\\

\parindent 0em
A){\bf E-Type antennae}\\
\parindent 1.5em
Only one case of E-Type initial-final massive antenna functions 
is required:
$E_3^0(4_q;3_q,1_{Q})$ which has a massless initial state radiator $4_{q}$,
a massive final state radiator denoted by $1_{Q}$ and an 
unresolved parton which is the final state quark $3_{q}$  of the same flavour 
as the initial state quark $4_{q}$.
This antenna accounts
for the massless initial state collinear behaviour and is given by, 

\begin{eqnarray}
E_3^0(4_q;3_q,1_Q)=-\frac{1}{\left(Q^2+m_Q^2\right)^2}
\left(-s_{14}+s_{13}-\frac{s_{13}^2}{s_{34}}-\frac{s_{14}^2}{s_{34}}
-2m_{Q}m_{\chi}\right)+O(\epsilon).
\label{eq:E}
\end{eqnarray}
It is normalised to a gluon initiated process, namely 
$\tilde{\chi} g  \to Q $. 
The corresponding two-particle matrix element squared omitting couplings
reads,
\begin{equation}
E^{0}_{2}(2_g;1_Q)=4\,(1-\eps)\,\left(Q^2+m_{Q}^2 \right)^2.
\end{equation}

As a consequence of the crossing procedure, the antenna defined in
eq.(\ref{eq:E}) has an overall minus sign made explicit in that expression.\\

\parindent 0em
B){\bf D-Type antennae}\\
\parindent 1.5em
The massive final-final D-Type antenna denoted by $D_3^0(1_Q,3_g,4_g)$ 
has only one massive final state particle, the quark denoted by $1_{Q}$.
The massive initial-final D-Type antenna 
is in principle obtained by crossing one of the two gluons in this function 
to the inital state. However, since the initial state gluon can split either 
into a quark-antiquark pair or into two gluons, 
two possible reduced matrix elements can serve as normalization 
for this antenna. 
A simple crossing of a gluon from the 
final-final quark-gluon antenna is not sufficient to define it unambiguously.
This ambiguity requires the 
decomposition of the gluon-initiated D-type antenna function 
into sub-antennae according to the reduced matrix elements it factorises to 
in the different limits. 
According to the limit considered, those reduced matrix elements are 
related to the process
$\tilde{\chi} +g \to Q$ or to the process $\tilde{\chi} +Q \to g$.
These two different reduced matrix elements 
define the two different limiting behaviours in which the antenna 
needs to be decomposed.
The decomposition 
can be achieved by separating the terms in the crossed function 
according to their contributions to a given limiting behaviour. 
For those terms which give contributions to more than one limiting 
behaviour, partial fractioning is applied. 
The antenna corresponding to a reduced matrix element
initiated by a quark will be denoted by  $D^0_{g,Qg}$  
and the antenna corresponding to  a reduced
matrix element initiated by a gluon by $D^0_{g;gQ}$.

For $D^0_{g;Qg}$,  the final state hard radiator
is a gluon and the parton becoming unresolved is a massive quark $Q$ which 
can become quasi-collinear to the initial state gluon. This
antenna  is given by,
{\allowdisplaybreaks
\begin{eqnarray}
D_3^0(4_g;1_Q,3_g)&=&-\frac{1}{\left(Q^2+m_Q^2\right)^2}\nonumber\\
&&\:\:\:\bigg(-\frac{4s_{13}^2}{s_{14}}+\frac{2s_{13}^3}{\left(Q^2+s_{13}+m_Q^2\right)s_{14}}+\frac{3s_{13}s_{34}}{s_{14}}-\frac{s_{34}^2}{s_{14}}\nonumber \\
&&\:\:\:\:\:\ -m_Q^2\left(\frac{2s_{13}^2}{s_{14}^2}-\frac{4s_{13}}{s_{14}}+\frac{6s_{34}}{s_{14}}-\frac{4s_{13}s_{34}}{s_{14}^2}+\frac{2s_{34}^2}{s_{14}^2}-\frac{2s_{34}^2}{s_{13}s_{14}}\right)\\
&&\:\:\:\:\:\:\:+2m_Q^4\frac{s_{34}}{s_{13}s_{14}}-2m_Q^3m_{\chi}\frac{s_{34}}{s_{13}s_{14}}\bigg)+O(\epsilon).\nonumber\\ 
\label{eq:ifdtypeq}
\end{eqnarray}}
This antenna is normalised to the matrix element associated to the process 
$\tilde{\chi} Q  \to g $ with a massive quark in the initial
state. Omitting couplings it reads,
\begin{equation}
D^{0}_{2}(1_Q;2_g)=4\,(1-\eps)\,\left(Q^2+m_{Q}^2 \right)^2.
\end{equation}

As this D-Type  antenna is initiated by a gluon 
but normalised by a quark-initiated process, it has an overall minus
sign made explicit in eq.(\ref{eq:ifdtypeq}). 

For $D^0_{g,gQ}$  , the final state hard radiator is a massive quark 
and the parton becoming unresolved is a gluon which can become soft,
collinear to the initial state gluon or quasi-collinear to the final-state 
massive quark. This three-parton
antenna reads, 
\begin{eqnarray}
D_3^0(4_g;3_g,1_Q)&=&\frac{1}{\left(Q^2+m_{Q}^2\right)^2}\nonumber\\
&&\bigg(9s_{13}-9s_{14}-6s_{34}+\frac{4s_{14}^2}{s_{13}}-\frac{4s_{13}^2}{s_{34}}-\frac{4s_{14}^2}{s_{34}}+\frac{s_{34}^2}{s_{13}}\nonumber\\
&&+\frac{2s_{13}^3}{\left(Q^2+s_{13}+m_Q^2\right)s_{34}}+\frac{2s_{14}^3}{s_{13}s_{34}}+\frac{6s_{13}s_{14}}{s_{34}}+\frac{3s_{14}s_{34}}{s_{13}}\\
&&-m_Q^2\left(6+\frac{2s_{14}^2}{s_{13}^2}-\frac{4s_{14}}{s_{13}}-\frac{6s_{34}}{s_{13}}+\frac{4s_{14}s_{34}}{s_{13}^2}+\frac{2s_{34}^2}{s_{13}^2}\right)\nonumber\\
&&+2m_Q m_{\chi}\bigg)+O(\epsilon).\nonumber\\
\end{eqnarray}

This function is normalised to the matrix element associated to the
gluon-initiated process $\tilde{\chi} g  \to Q $ with  
the corresponding two-particle matrix element squared omitting couplings
reading,
\begin{equation}
D^{0}_2(1_g;2_g)=4\,(1-\eps)\,\left(Q^2+m_{Q}^2 \right)^2.
\end{equation}
Being initiated and normalised by a gluon, this antenna 
has no overall minus sign. 

\subsubsection{ Flavour violating antennae}
Finally, 
as for the final-final case, in addition to the flavour conserving 
antennae given above,
for the construction of the subtraction terms for  
$Q\bar{Q}$ +jets production at NLO, we also need two types of massive
flavour violating initial-final antennae. 
Those are such that in both cases a massless quark is 
playing the role of the initial state radiator;
the final state radiators can either be a massive 
quark or a massive antiquark. The expressions of these two antennae 
are related by the exchange of a 
massive quark versus a massive antiquark in the final state such that
only one expression is given below.
The unintegrated form of the antenna having a massive quark and a
gluon in the final state given by $A^0_{q,gQ}$ is,

\begin{eqnarray}
A_3^0(2_q;3_g,1_Q)=\frac{1}{Q^2-m_Q^2}\left[\frac{2s_{12}^2}{s_{13}s_{23}}+\frac{2s_{12}}{s_{13}}-\frac{2s_{12}}{s_{23}}+\frac{s_{13}}{s_{23}}+\frac{s_{23}}{s_{13}}\right.\nonumber\\
\left.-m_Q^2\left( \frac{2s_{12}}{s_{13}^2}+\frac{2s_{23}}{s_{13}^2}-\frac{2}{s_{13}}\right)\right]+{\cal O}(\epsilon).
\label{eq:Afvif}
\end{eqnarray}

This antenna function is normalised with the following two-parton
 matrix element squared:
\begin{equation} 
A_2^0(2_q;1_Q)=4\,(1-\epsilon)\left[Q^2-m^2_{Q}\right]
\end{equation}
and has therefore no overall minus sign.

\section{Singular limits of the massive antennae}

The factorisation properties of tree-level QCD squared matrix 
elements for massless partons are well known \cite{fac1}. 
An $(m+1)$-parton squared amplitude factorises into a product 
of a reduced $m$-parton squared matrix element and a soft eikonal  
factor in the soft limit, or an Altarelli-Parisi splitting function 
in the collinear case.
If massive partons are involved the factorisation still takes place but the
collinear and soft behaviour have to be generalised to take the mass
effects into account.
Those generalised limits will be described
both in final-final and initial-final configurations below. 

In the following, we shall here first recall 
all unresolved factors (massless and
massive)  in the final-final and initial-final configurations before
tabulating the limits of all antenna functions encountered in Section 3.

\subsection{Single unresolved massless factors}

When only massless partons are involved, when 
a gluon $j$ emitted between two massless hard
radiators $i$ and $k$ becomes soft, the squared matrix element factorises
and the eikonal factor that factorises off the squared matrix element is
\begin{equation}
S_{ijk}=\frac{2s_{ik}}{s_{ij}s_{jk}}.
\end{equation}

When two massless partons become collinear,
the matrix element factorises yielding specific Altarelli-Parisi splitting
functions corresponding to a particular parton-parton splitting.
Those functions depend on $z$, the fractional momentum carried 
by the unresolved parton. 
Depending whether the unresolved parton is collinear to an initial or to a
final state parton, the definition of $z$ will be different.
For two final state particles $i$ and $j$ of momenta $p_i$ and $p_j$ 
becoming collinear, we have, in the limit,
\begin{equation}
p_i\to z p_{ij},\quad p_j\to (1-z) p_{ij} ,\quad s_{ik}\to z s_{ijk},
\quad s_{jk}\to (1-z) s_{ijk}\,,
\label{eq:zdefinal}
\end{equation}
whereas for a final state particle $j$ of momentum $p_j$ 
becoming collinear with an initial state parton $i$ of momentum $p_i$ 
we have
\begin{equation}
p_j\to z p_{i},\quad p_{ij}\to (1-z) p_{i} ,
\quad s_{ik}\to \frac{s_{ijk}}{1-z},\quad s_{jk}\to \frac{z s_{ijk}}{1-z}\;.
\label{eq:zdefinitial}
\end{equation}

The splitting functions denoted by $P_{ij \rightarrow (ij)}(z)$
corresponding to the collinear limit of two final state partons $i$
and $j$ are given in \cite{AP}  by,
\begin{eqnarray}
&& P_{qg\rightarrow Q}(z)=\frac{1+(1-z)^2-\epsilon z^2}{z}\\
&& P_{q\bar{q}\rightarrow G}(z)=\frac{z^2+(1-z)^2-\epsilon}{1-\epsilon}\\
&& P_{gg\rightarrow G}(z)=2\left[\frac{z}{1-z}+\frac{1-z}{z}+z(1-z)\right].
\end{eqnarray}

When the collinearity arises between an initial  $i$
and a final state parton $j$, the splitting functions  
denoted by $P_{ij \leftarrow (ij)}(z)$ are given in \cite{AP}  by,
\begin{eqnarray}
&& P_{gq\leftarrow Q}(z)=\frac{1+z^2-\epsilon(1-z)^2}{(1-\epsilon)(1-z)^2}
=\frac{1}{1-z}\frac{1}{1-\epsilon}P_{qg\rightarrow Q}(1-z)\\
&& P_{qg\leftarrow Q}(z)=\frac{1+(1-z)^2-\epsilon z^2}{z(1-z)}
=\frac{1}{1-z}P_{qg\rightarrow Q}(z)\\
&& P_{q\bar{q}\leftarrow G}(z)=\frac{z^2+(1-z)^2-\epsilon}{1-z}
=\frac{1-\epsilon}{1-z}P_{q\bar{q}\rightarrow G}(z)\\
&& P_{gg\leftarrow G}(z)=\frac{2(1-z+z^2)^2}{z(1-z)^2}
=\frac{1}{1-z}P_{gg\rightarrow G}(z).
\end{eqnarray}
The additional factors $(1-\epsilon)$ and $1/(1-\epsilon)$  account for 
the different number of polarizations of quark and gluons in the cases
in which the particle entering the hard processes changes its type.

In all splitting functions defined above,
the label $q$ can stand for a massless quark or an antiquark 
since charge conjugation implies that 
$P_{qg\rightarrow Q}=P_{\bar{q}g\rightarrow\bar{Q}}$ and 
$P_{qg\leftarrow Q}=P_{\bar{q}g\leftarrow\bar{Q}}$ . The labels $Q$ 
and $G$ in those denote the parent parton 
of the two collinear partons,  which is massless.

\subsection{Single unresolved massive factors}

When the final state partons are massive, the emission of extra
radiation from those can still lead to soft
divergences, but not to collinear singularities 
since the mass of the final state parton regulates those. 
The relation between matrix element squared and 
the splitting functions needs to be extended from massless to
massive. Similar factorisation formulae for matrix elements 
as in the massless case hold provided the collinear limit
 is generalized \cite{cdst1} to the {\it quasi-collinear} limit.
 
{\subsubsection{Quasi-collinear limit in final-final and initial-final 
configurations}

In the final-final configuration, two final state massive partons 
can become quasi-collinear to each other resulting in a parent parton 
which is massive. 
The limit when a massive parton $(ij)$ of momentum
$p_{(ij)}$ and mass $m_{(ij)}$
decays quasi-collinearly into two massive partons $i$ and $j$ of
masses $m_i$ and $m_j$ is defined by,
\begin{eqnarray}
&& p_{j}^{\mu}\rightarrow z p_{(ij)}^{\mu},\:\:\:\:\:\:\:\:\:\: p_{i}^{\mu}\rightarrow 
(1-z)p_{(ij)}^{\mu}\\
&& p_{(ij)}^2=m_{(ij)}^2
\end{eqnarray}
with the constraints
\begin{equation}
p_i\cdot p_j,\:m_i,\:m_j,\:m_{(ij)}\rightarrow 0,
\end{equation}
at fixed ratios
\begin{equation}
\frac{m_i^2}{p_i\cdot p_j},\:\frac{m_j^2}{p_i\cdot p_j},\:
\frac{m_{(ij)}^2}{p_i\cdot p_j}.
\end{equation}

The difference obtained between taking the {\it quasi-collinear limit} 
between two final state particles or taking this limit when one 
initial and one final state particles are involved is closely
related to the difference obtained in these two situations 
for the {\it massless collinear limit}.
When a massive parton of momentum $p_{j}$ becomes
quasi-collinear to an initial state massless parton $p_{i}$ we have, 
\begin{equation}
p_j\to z p_{i},\quad p_{(ij)}\to (1-z) p_{i} ,
\label{eq:zdeinitial}
\end{equation}
with the constraints,
\begin{equation}
p_i\cdot p_j,\:m_j,\:m_{(ij)}\rightarrow 0,
\end{equation}
at fixed ratios
\begin{equation}
\frac{m_j^2}{p_i\cdot p_j},\;\frac{m_{(ij)}^2}{p_i\cdot p_j}.
\end{equation}

Only the fractional momentum $z$ carried by the unresolved
parton needs to be defined accordingly in both final-final and 
initial-final situations.
For the quasi-collinear limits, it is defined exactly as for the
collinear limits. 
For the final-final case, $z$ is defined as in eq.(\ref{eq:zdefinal})
whereas in the initial-final case it is defined by
eq.(\ref{eq:zdeinitial}).

The key difference between the massless collinear limit and the
quasi-collinear limit is given by the constraint that the on-shell masses
squared of the final state partons have to be kept of the same order 
as the invariant $s_{ij}=2 p_i\cdot p_j$, with the latter becoming small. 

\subsubsection{Factorisation in the quasi-collinear limits}
In these quasi-collinear limits (final-final or initial-final),
the $(m+1)$-parton matrix element squared factorises into a reduced
$m$-parton matrix element and unresolved massive factors.
These single unresolved massive factors are generalizations
of the massless unresolved factors defined above.

The generalized soft eikonal factor $S_{ijk}(m_i,m_k)$ for a massless
gluon $j$ emitted between two massive partons $i$ and $k$ depends on
the invariants $s_{lm}=2p_l\cdot p_m$ built with the partons $i$, $j$
and $k$ but also on the masses $m_i$ and $m_k$ of partons $i$ and $k$. 
It is given by 
\cite{cdst1,Mathias}
\begin{equation}
S_{ijk}(m_i,m_k)=\frac{2s_{ik}}{s_{ij}s_{jk}}-\frac{2m_i^2}{s_{ij}^2}-
\frac{2m_k^2}{s_{jk}^2}.
\end{equation}

The massive splitting functions will depend on $z$, 
the fractional momentum carried 
by the unresolved parton $j$ and on the masses $m_{i}$ and $m_{j}$ 
of the partons $i$ and $j$ becoming quasi-collinear. 
All the mass dependence can be parametrized by $\mu_{(ij)}$ 
given by, 
\begin{equation}
\mu_{(ij)}=\frac{m_i^2+m_j^2}{(p_i+p_j)^2-m_{(ij)}^2}.
\end{equation}

The massive splitting functions 
denoted by  $P_{ij\rightarrow (ij)}(z,\mu_{ij}^2)$ for a massive parton $(ij)$ 
which splits into partons $i$ and $j$, both being in the final state, 
have been given in the appendix of \cite{cdt2}
 and in \cite{Mathias}. Those read,
 \begin{eqnarray}
&& P_{qg\rightarrow Q}(z,\mu_{qg}^2)=\frac{1+(1-z)^2-\epsilon z^2}{z}-2\mu_{qg}^2\\
&& P_{q\bar{q}\rightarrow G}(z,\mu_{q\bar{q}}^2)=\frac{z^2+(1-z)^2-\epsilon
+\mu_{q\bar{q}}^2}{1-\epsilon},
\label{eq:pqqm}
\end{eqnarray}
where
\begin{equation}
\mu_{qg}^2=\frac{m_Q^2}{s_{qg}}\:\:\:\:\:\:\:\:\:\:\:\:\:\:\:\:\mbox{and}\:\:\:\:\:\:\:\:\:\:\:\:\:\:\:\:\mu_{q\bar{q}}^2=\frac{2m_Q^2}{s_{q\bar{q}}}.
\end{equation}
Naturally, the gluon-gluon splitting function $P_{gg\rightarrow G}(z)$ is left
unchanged. 
The splitting function $P_{q\bar{q}\rightarrow G}(z,\mu_{q\bar{q}}^2)$
given in eq.(\ref{eq:pqqm}) and related to the quasi-collinear limit 
of two massive partons does not correspond to a limiting behaviour 
of the antenna functions required here and given in Section 3. It  
is given here for completeness.

So far we have treated all generalisations of soft and collinear massless 
factors needed to treat final-final configurations involving massive partons. 
For the initial-final situations,
since all the initial state partons are taken massless the only 
initial-final splitting function that changes when we allow massive partons
in the final state is $P_{q\bar{q}\leftarrow G} (z,\mu_{qg}^2)$ given by, 
\begin{equation}
P_{q\bar{q}\leftarrow G} (z,\mu_{qg}^2)=\frac{z^2+(1-z)^2-\epsilon}{1-z}+2\mu_{qg}^2.
\end{equation}
The definition of the momentum fraction $z$ present in this formula 
will be the same as in the massless case given by eq.(\ref{eq:zdefinitial}).

\subsection{Singular limits of the massive antenna functions}
In this subsection we list all the non-vanishing soft, 
collinear and quasi-collinear limits of the massive final-final and
initial-final antenna functions given in Section 3.
The limits of the massless antennae
also needed to construct the subtraction terms for $Q\bar{Q}$+ jet
production can be found in \cite{ourant,Daleo,Joao}.  

\subsubsection{Final-final antenna functions}
The limits of the massive flavour conserving 
quark-antiquark and quark-gluon antenna functions 
have been derived  in \cite{Mathias}. We give them here for
completeness. The limits of the flavour violating antennae are new.

The singular limits of the massive quark-antiquark antenna 
$A^{0}_{Qg\bar{Q}}$ are
\begin{equation}
 \begin{split}
A_3^{0}(1_Q,3_g,2_{\bar{Q}}) & \xrightarrow{3_g \rightarrow 0} 
\softeikonal_{132}(m_Q,m_Q),\\
A_3^{0}(1_Q,3_g,2_{\bar{Q}} )& \xrightarrow{3_g \parallel 1_Q} \frac{1}{s_{13}}
\splitqgq(z,\mu^2_{qg}),\\
A_3^{0}(1_Q,3_g,2_{\bar{Q}}) & \xrightarrow{3_g \parallel 2_{\bar{Q}}} \frac{1}{s_{23}}
\splitqgq(z,\mu^2_{qg}).\\
\end{split}
\end{equation}

The only non-vanishing singular limit of the quark-gluon E-Type antenna 
$E^0_{Qq'\bar{q}'}$  is the collinear massless limit of the massless 
quark-antiquark pair,
\begin{equation}
E^0_3(1_Q,3_{q'},4_{\bar{q}'})\xrightarrow{3_{q'} \parallel 4_{\bar{q}'}}
\frac{1}{s_{34}}  \splitqqbarg(z),
\end{equation}
while for the D-type $D^0_{Qgg}$ and d-type $d^0_{Qgg}$ antennae we have,
\begin{eqnarray}
&& D_3^0(1_Q,3_g,4_g)\stackrel{^{3_g\rightarrow0}}{\longrightarrow}S_{134}(m_Q,0)\nonumber\\
&& D_3^0(1_Q,3_g,4_g)\stackrel{^{4_g\rightarrow0}}{\longrightarrow}S_{143}(m_Q,0)\nonumber\\
&& d_3^0(1_Q,i_g,j_g)\stackrel{^{i_g\rightarrow0}}{\longrightarrow}S_{1ij}(m_Q,0)\nonumber\\
&& d_3^0(1_Q,i_g,j_g)\stackrel{^{j_g\rightarrow0}}{\longrightarrow}0\nonumber\\
&& D_3^0(1_Q,3_g,4_g)\stackrel{^{1_Q||3_g}}{\longrightarrow}\frac{1}{s_{13}}P_{qg\rightarrow Q}(z,\mu_{qg}^2)\nonumber\\
&& D_3^0(1_Q,3_g,4_g)\stackrel{^{1_Q||4_g}}{\longrightarrow}\frac{1}{s_{14}}P_{qg\rightarrow Q}(z,\mu_{qg}^2)\\
&& d_3^0(1_Q,i_g,j_g)\stackrel{^{1_Q||i_g}}{\longrightarrow}\frac{1}{s_{1i}}P_{qg\rightarrow Q}(z,\mu_{qg}^2)\nonumber\\
&& d_3^0(1_Q,i_g,j_g)\stackrel{^{1_Q||j_g}}{\longrightarrow}0\nonumber\\
&& D_3^0(1_Q,3_g,4_g)\stackrel{^{3_g||4_g}}{\longrightarrow}\frac{1}{s_{34}}P_{gg\rightarrow G}(z)\nonumber\\
&& d_3^0(1_Q,3_g,4_g)\stackrel{^{3_g||4_g}}{\longrightarrow}\frac{1}{s_{34}}\left( P_{gg\rightarrow G}(z)-\frac{2z}{1-z}-z(1-z)\right)\nonumber\\
&& d_3^0(1_Q,4_g,3_g)\stackrel{^{3_g||4_g}}{\longrightarrow}\frac{1}{s_{34}}\left( P_{gg\rightarrow G}(z)-\frac{2(1-z)}{z}-z(1-z)\right). \nonumber
\end{eqnarray}

For the massive A-Type flavour violating antenna $A^0_{Qg\bar{q}}$ we have 
\begin{eqnarray}
&& A_3^0(1_Q,3_g,2_{\bar{q}})\stackrel{^{3_g\rightarrow0}}{\longrightarrow}S_{132}(m_Q,0)\\
&& A_3^0(1_Q,3_g,2_{\bar{q}})\stackrel{^{1_Q||3_g}}{\longrightarrow}\frac{1}{s_{13}}P_{qg\rightarrow Q}(z,\mu_{qg}^2)\\
&& A_3^0(1_Q,3_g,2_{\bar{q}})\stackrel{^{2_{\bar{q}}||3_g}}{\longrightarrow}\frac{1}{s_{23}}P_{qg\rightarrow Q}(z).
\end{eqnarray}

Again, for symmetry reasons the role of $Q$ and $\bar{q}$ 
are interchangeable in these formulae.

\subsubsection{Initial-final antenna functions}
For the quark-antiquark initial-final massive antenna
$A^0_{g,Q\bar{Q}}$, only quasi-collinear limits are present. Those are:
\begin{eqnarray}
&& A_3^0(3_g;1_Q,2_{\bar{Q}})\stackrel{^{1_Q||3_g}}{\longrightarrow}\frac{1}{s_{13}}P_{q\bar{q}\leftarrow G}(z,\mu_{qg}^2)\\
&& A_3^0(3_g;1_Q,2_{\bar{Q}})\stackrel{^{2_{\bar{Q}}||3_g}}
{\longrightarrow}\frac{1}{s_{23}}P_{q\bar{q}\leftarrow G}(z,\mu_{qg}^2)
\end{eqnarray}

For the E-Type antenna $E^0_{q;qQ}$, the only non-vanishing singular limit is,
\begin{equation}
E_3^0(4_q;3_q,1_Q)\stackrel{^{3_q||4_q}}{\longrightarrow}\frac{1}{s_{34}} 
P_{gq\leftarrow Q}(z).
\end{equation}

The limits of the two quark-gluon D-Type antennae initiated by a gluon 
$D_{g,Qg}$ and $D_{g,gQ}$ are given by : 
\begin{equation}
D_3^0(4_g;1_Q,3_g)\stackrel{^{1_Q||4_g}}{\longrightarrow}
\frac{1}{s_{14}}P_{q\bar{q}\leftarrow G}(z,\mu_{Qg}^2),
\end{equation}
which has only a quasi-collinear limit when the unresolved quark
$1_{Q}$ is collinear to the initial-state gluon.
The limits of  $D_3^0(4_g;3_g,1_Q)$ with an unresolved gluon are
\begin{eqnarray}
&& D_3^0(4_g;3_g,1_Q)\stackrel{^{1_Q||3_g}}{\longrightarrow}
\frac{1}{s_{13}}P_{qg\rightarrow Q}(z,\mu_{qg}^2)\\
&& D_3^0(4_g;3_g,1_Q)\stackrel{^{3_g||4_g}}{\longrightarrow}
\frac{1}{s_{34}}P_{gg\leftarrow G}(z)\\
&& D_3^0(4_g;3_g,1_Q)\stackrel{3_g\rightarrow 0}{\longrightarrow}
S_{134}(m_Q,0).
\end{eqnarray}

Finally, the initial-final flavour violating antenna $A^0_{q,gQ}$ has 
its limits given by,
\begin{eqnarray}
&&A_3^0(2_q;3_g,1_Q)\stackrel{^{1_Q||3_g}}{\longrightarrow}\frac{1}{s_{13}}P_{qg\rightarrow Q}(z,\mu_{qg}^2)\\
&&A_3^0(2_q;3_g,1_Q)\stackrel{^{2_q||3_g}}{\longrightarrow}\frac{1}{s_{23}}P_{qg\leftarrow Q}(z)\\
&&A_3^0(2_q;3_g,1_Q)\stackrel{^{3_g\rightarrow0}}{\longrightarrow}S_{132}(m_Q,0).
\end{eqnarray}

\section{Integrated massive antenna functions}
\parindent 1.5em
To combine the antenna subtraction terms with the virtual corrections 
and the mass factorization counterterms in a given kinematical configuration, 
the antenna functions 
have to be integrated over the appropriate factorised antenna phase space. 
After integration, the 
implicit soft and collinear singularities present in the antenna functions 
turn into explicit poles in 
the dimensional regularization parameter $\e$, and the remaining phase space 
corresponds to the same $n$-particle kinematics as the virtual 
contributions or the mass factorisation counterterms. 
In this section, we derive the integrated forms of the massive antenna 
functions defined in Section 3. Only full antennae,  
denoted by capital letters $X_{ijk}$  (or crossing of those),
need to be integrated, while  partial antennae denoted  
$x_{ijk}$ sum up to $X_{ijk}$ prior to integration. 

\subsection{Properties of the integrated massive antennae}
The results  will be presented in two forms, in expanded and unexpanded
forms in the dimensional regularization parameter $\e$.
The unexpanded forms of the integrated antennae 
are functions of a few master integrals obtained after standard
reduction techniques \cite{IBP1,IBP2,higgsnnlo,4particle,Laporta} 
have been applied. The masters will be
given here analytically to all orders in $\e$. 
Additionally, the integrated antennae  will be presented 
after an $\e$-expansion has been performed on these all-order results up 
to finite order in $\e$. In this expanded 
form, the poles in $\e$  become explicit and can be related to process
independent infrared singularity operators and splitting kernels. 

Final-final integrated antennae have their pole part 
entirely related to colour ordered massive infrared 
singularity operators $\bf{I}^{(1)}_{ij}$, which will be defined below.
For initial-final antennae, those infrared operators are not
sufficient to capture all singularities present in the integrated
antennae. Additional pole terms can arise due to the presence 
of massless intial-final parton-parton collinear singularities which 
cancel against the mass factorisation counterterms.
These pole terms are proportional to 
universal and process-independent splitting kernels $p^{(0)}_{ij}(x)$. 
Those are defined for example in \cite{Daleo} and will be given below.

As a check on our results for the expanded forms of the 
integrated antennae we consider the following:
At NLO, a particular antenna can be regarded as the sum of two
particular dipoles in the dipole formalism of \cite{cs,cdst1}. 
The two radiators present in an given antenna play then 
both the role of emitter and spectator in the corresponding two
dipoles. Up to terms which do not 
give rise to singularities when integrated over the phase space, 
the sum of these two dipoles and the antenna are the same 
(up to coupling factors).
The pole parts of a given integrated antenna 
and those of the corresponding  integrated dipoles can therefore
be related.  For each of the integrated antenna we will specify 
how this comparison is performed.

\subsubsection{Infrared singularity structure}
Owing to the universal factorisation properties of QCD amplitudes 
in infrared singular limits, 
it is possible to describe the infrared pole structure 
of virtual loop corrections and, consequently, 
also of integrated subtraction terms, by 
the product of infrared singularity operators 
with tree level amplitudes. These infrared 
singularity operators are a priori tensors in colour space. 
In a colour ordered framework, they decompose into different 
colour-ordered infrared singularity operators.   
In massless QCD, there is solid evidence to assume that the  
infrared singularity operators consist only of combinations of two-particle 
correlations~\cite{becherneubert,gardimagnea,dixon} 
at all orders in perturbation theory. 
Explicit forms of the massless operators are known 
to three-loop order~\cite{mochvogt}. 
In massive QCD, only the one-loop infrared singularity operator 
is made up entirely of two-parton correlations~\cite{cdt2}, 
while multi-particle correlations can contribute 
at higher loop order~\cite{neubertferroglia}. 
The explicit form of the massive infrared singularity 
operators is known to two loops~\cite{neubertferroglia} 
and was used to predict the pole 
structure of the two-loop matrix elements 
for $q\bar q\to t\bar t$ and $gg\to t\bar t$.  

We are concerned only with the infrared singularity structure at 
NLO in the present study. 
Consequently, the integrated massive antenna functions will contain a pole 
structure in terms of the massive one-loop infrared singularity operators. 
Containing only two-parton correlations irrespective of the particle 
masses involved, these can be expressed straightforwardly in a colour-ordered 
form. 
We introduce the mass-dependent colour-ordered NLO 
real radiation singularity operator
\begin{equation}
{\bf I}^{(1)}_{ij} (\e,s_{ij},m_{i},m_{j},\lambda\mu^2)\,,
\end{equation}
which describes the unresolved real radiation 
between partons $i$ and $j$. It is a function 
of the invariant mass of the parton pair, of the masses of the partons and of 
a product of kinematical parameters $\lambda \mu^2$, 
which determines the logarithmic pole coefficient.
 In this form, $\mu^2$ is dependent only on the mass combination 
of the radiator partons, while 
$\lambda$ takes account of the nature of the kinematically 
allowed endpoint in the different kinematical configurations. 
$\lambda =1$ for all final-final antenna functions
and for initial-final antenna functions with equal masses, while $\lambda=
x_0^2$ with 
\begin{equation}
x_0=\frac{Q^2}{Q^2+m_Q^2}
\label{eq:x0}
\end{equation} 
for initial-final antenna functions with 
one massless and one massive radiator. 

In their most general form those 
infrared operators have their functional dependence given by,
$${\bf I}^{(1)}_{Q\bar{Q}} (\e,s_{Qg},m_{Q},m_{Q},\mu^2), \qquad
{\bf I}^{(1)}_{Qg} (\e,s_{Qg},m_{Q},m_{Q},\lambda \mu^2) \qquad \mbox{and}\qquad  
{\bf I}^{(1)}_{Qg,\Flavour} (\e,s_{ij},m_{Q},0,\mu^2).\; $$
The third operator describes contributions arising from the 
splitting of a gluon into a quark-antiquark pair which are proportional 
to the number of light quark flavours $N_F$. 

The massive infrared operators which have a non-trivial mass dependence,
required in both final-final and intial-final configurations are, 
\begin{eqnarray}
&&{\bf I}^{(1)}_{Q\bar{Q}} \left(\e,s_{Q\bar Q},m_{Q},m_{Q},
\lambda\frac{1-\sqrt{r}}{1+\sqrt{r}}\right)=\nonumber\\
&&\hspace{25mm} - \frac{e^{\e \gamma_E}}{2\Gamma(1-\e)}\,
\left[ \frac{s_{Q\bar{Q}}+2 m_Q^2}{s_{Q\bar{Q}}^2 } \right ]^{\e}
\left \{
\frac{1}{\e}\left(\frac{(1+r)}{2\sqrt{r}})
\ln\left(\frac{1-\sqrt{r}}{1+\sqrt{r}}\right)+
  1 \right) \right \} 
\nonumber \\ 
&&{\bf I}^{(1)}_{Qg}\left(\e,s_{Q g},m_{Q},0,\lambda \frac{m_{Q}^2}{s_{Qg}+m_{Q}^2}
\right)=\nonumber\\
&&\hspace{25mm} - \frac{e^{\e \gamma_E}}{2\Gamma(1-\e)}\, 
\left[\frac{s_{Q g}+m_{Q}^2}{s_{Qg}^2} \right]^{\e}
\left \{
\frac{1}{2 \e^2}+\frac{1}{2\e}\left(\frac{17}{6}\right) +\frac{1}{2 \e}
\ln\left(\frac{\lambda m_{Q}^2}{s_{Q g}+m_Q^2} \right)\right \} \, 
\nonumber\\
&&{\bf I}^{(1)}_{Q g,\Flavour} \left(\e,s_{Q g},m_{Q},0,
\frac{m_{Q}^2}{s_{Qg}+m_{Q}^2}\right) =  
\frac{e^{\e \gamma_E}}{2\Gamma(1-\e)}\, 
\left[\frac{s_{Q g}+m_{Q}^2}{s_{Qg}^2}\right]^{\e}
\left(\frac{1}{6\e} \right)\,, \nonumber \\
\label{eq:Ione}
\end{eqnarray}
with 
\begin{equation}\label{eq:r}
r=1-\frac{4 m_{Q}^2}{s_{Q \bar{Q}}+2m_{Q}^2}. 
\end{equation}
The antiquark-gluon operators are obtained by charge conjugation:
$${\bf I}^{(1)}_{g\bar Q} (\e,s_{g\bar Q},m_{Q},0,\lambda \mu^2) = 
{\bf I}^{(1)}_{Qg} (\e,s_{gQ},m_{Q},0,\lambda \mu^2),$$
$${\bf I}^{(1)}_{g\bar Q,\Flavour} (\e,s_{g\bar Q},m_{Q},0,\mu^2) = 
{\bf I}^{(1)}_{gQ,\Flavour} (\e,s_{gQ},m_{Q},0,\mu^2)\;.
$$

In addition to these infrared operators which have their massless
counterparts defined in \cite{ourant}, we also have an infrared
operator associated with the flavour-violating antennae $A_{Qg\bar{q}}$
and $A_{Qgq}$ defined in Section 3. As those antennae have the same
unintegrated form both of them are related to one infrared
operator denoted by ${\bf I}^{(1)}_{q\tilde{Q}}$, where $\tilde{Q}$ 
can stand for a massive quark or antiquark. 
As the quark-gluon operator ${\bf I}^{(1)}_{Qg}$, this operator 
has a mass-dependent logarithmic term proportional to $\lambda \mu^2$
with $\lambda=x_0^2$ in the initial-final configuration.
It is given by,
\begin{eqnarray}
&&{\bf I}^{(1)}_{q\tilde{Q}} \left(\e,s_{q \tilde{Q}},m_{\tilde{Q}},0
,\lambda\frac{m^2_{\tilde{Q}}}{s_{q\tilde{Q}}+m^2_{\tilde{Q}}} \right)=\\
&&\hspace{25mm}- \frac{e^{\e \gamma_E}}{2\Gamma(1-\e)}\, 
\left[\frac{s_{q\tilde{Q}}+m_{\tilde{Q}}^2}{s_{q\tilde{Q}}^2} \right]^{\e}
\left \{
\frac{1}{2 \e^2}+\frac{1}{2\e}\left(\frac{5}{2}\right) +\frac{1}{2
  \e}\ln \left(\frac{\lambda\: m_{\tilde{Q}}^2}{s_{q\tilde{Q}}+m^2_{\tilde{Q}}}\right) \right \}\nonumber.
\end{eqnarray}

\subsection{Integrated massive final-final antennae}
The integrated massive final-final flavour conserving antennae 
have been derived as function of a few master integrals in
\cite{Mathias}.
In this section,
all final-final integrated antennae will be presented in expanded 
form such that their pole part can be related
to the colour-ordered massive singularity operators ${\bf I}^{(1)}_{ij}$
defined above.
Before we shall derive the integrated massive flavour-violating
antenna and give its expanded and unexpanded forms.

In general, the integrated final-final antennae denoted 
by ${\cal X}_{ijk}$ are given 
as the integration over the final-final antenna phase space $\d \Phi_{X_{ijk}}$ 
of the unintegrated final-final antennae $X_{ijk}$ 
as given in  eq.(\ref{eq:finalintegrated}).
Those will depend on the masses 
of the final state particles and on $E_{cm}^2$. 

We start by giving the NLO massive final-final antenna 
phase space  $\d \Phi_{X_{ijk}}$ necessary to evaluate 
these integrated final-final antennae.
In the most general case, where the involved partons $i,j,k$ 
have three different masses $m_{i},m_{j}$ and $m_{k}$, 
the massive antenna phase space $d\Phi_{X_{ijk}}^{(m_i,m_j,m_k)}$ 
is given by \cite{weinzierl,Mathias},
\begin{equation}
\label{eq:phix}
\begin{split}
\int d \Phi_{X_{ijk}}^{(m_i,m_j,m_k)} & (s_{ij},s_{jk},s_{ik}) =\\
	&\,(2 \pi)^{1-d} \frac{2 \pi^{d/2-1}}
{\Gamma\left(\frac{d}{2}-1\right)}
 \frac{1}{4} \left( (\en^2-m_I^2-m_K^2)^2 - 4 m_I^2 m_K^2\right)^{\frac{3-d}{2}}
\\
	&\int d s_{ij}\, ds_{jk}\,ds_{ik}\, \delta( \en^2-m_i^2-m_j^2-
m_k^2-s_{ij}-s_{jk}-s_{ik})\\
	&[4\,\Delta_3(p_{i},p_{j},p_{k})]^\frac{d-4}{2} \;
\theta(\Delta_3(p_{i},p_{j},p_{k})). 
\end{split}
\end{equation}
The masses $m_{I}$ and $m_{K}$ appearing in this equation are
combinations of the masses $m_{i},m_{j}$ and $m_{k}$.
The function $\Delta_3(p_{i},p_{j},p_{k})$  
is the Gram determinant for massive particles of momenta 
$p_{i},p_{j},p_{k}$ given in terms of invariants $s_{ij}=2p_i \cdot p_j$ 
and masses $m_{i},m_{j},m_{k}$ by,
\begin{equation}
\Delta_3(p_{i},p_{j},p_{k}) = 
\frac{1}{4}\left(s_{ij}s_{ik}s_{jk} - m_i^2 s_{jk}^2-m_k^2s_{ij}^2
-m_j^2 s_{ik}^2 + 4 m_i^2 m_j^2 m_k^2 \right).
\end{equation} 

To be able to evaluate the integrated flavour-violating massive
final-final antennae we need to consider the case where only one
particle in the final state is massive while the other two are taken
massless. We consider, $m_i=m_{j}=0$ and $m_{k}=m_{Q}$, 
in which case the masses of the remapped momenta 
$p_{I}$ and $p_{K}$ have their masses 
given by $m_{I}=m_{i}=0$, $m_{K}=m_{k}$. We use the following
parametrisation of the final-final massive antenna phase space  
with one massive parton $d\Phi_{X_{ijk}}^{(0,0,m)}$ given by, 
\begin{eqnarray}\label{eq:Phix2}
d \Phi_{X_{ijk}}^{(0,0,m)}&=& \frac{(4 \pi)^{\epsilon-2}}{\Gamma(1-\epsilon)} \left( \enen \right)^{1-\epsilon} \left(u_0\right)^{2 - 2 \eps}\nonumber\\
&& \quad \int_0^1 du u^{1-2\eps} (1-u)^{1-2\eps}(1-u_0 u)^{-1+\eps} \int_0^1 dv v^{-\eps} (1-v)^{-\eps},
\end{eqnarray}
with
\begin{eqnarray}
&& \quad u_0 = 1 - \frac{m_{Q}^2}{\enen},\nonumber\\
&& \quad s_{ij} = \enen u_0^2 \frac{u(1-u)v}{(1-u_0 u)},\nonumber\\
&& \quad s_{ik} = \enen u_0 (1-u).
\end{eqnarray}

Integrating the flavour violating antenna $A^0_3(1_Q,3_g,2_{\bar{q}})$
given in eq.(\ref{eq:fvff})  over this phase space 
$d\Phi_{X_{ijk}}^{(0,0,m)}$  we obtain,
\begin{eqnarray}\label{eq:flavourviolint}
&&{\cal A}^0_3(1_Q,3_g,2_{\bar{q}})=\frac{8\pi^2 (4\pi)^{-\epsilon} e^{\epsilon\gamma_E}}{2E_{cm}^4 u_0^2 (1-u_0)\epsilon^2 (1-2\epsilon)}\nonumber\\
&&\times\bigg\{E_{cm}^2\left[ 12(1-u_0)^2-2\epsilon(29-54u_0+24u_0^2)+\epsilon^2(98-176u_0+73u_0^2)\right.\nonumber\\
&&\:\:\:\:\:\:\left. +\epsilon^3(-68+120u_0-47u_0^2)+2\epsilon^4(8-12u_0+3u_0^2)\right] I_{1}^{(0,0,m)}\\
&&\:\:\:+3(1-\epsilon)\left[ 4(1-u_0)+2\epsilon(-7+6u_0)+\epsilon^2(14-11u_0)+2\epsilon^3(-2+u_0)\right]I_{2}^{(0,0,m)}\bigg\}\nonumber .
\end{eqnarray}
In this expression, the master integrals required are $I_{1}^{(0,0,m)}$
and $I_{2}^{(0,0,m)}$ were derived in \cite{Mathias}.
The master integral $I_{1}^{(0,0,m)}$ corresponding to the 
integrated $ 1\to 3$ phase space measure with one massive final state 
is given by,
\begin{eqnarray}
\intphizerozerom &=& \int d \Phi_{X_{ijk}}^{(0,0,m)}  \nonumber \\
&=& (\enen)^{1-\eps} u_0^{2-2\eps} 2^{-4+2\eps} \pi^{-2+\eps}
\frac{\Gamma(2-2\eps)
  \Gamma(1-\eps)}{\Gamma(4-4\eps)}\gaussf{1-\eps}{2-2\eps}{4-4\eps}{u_0}\,,
\nonumber \\
\end{eqnarray}
while $I_{2}^{(0,0,m)}$ is given by,
\begin{eqnarray}
\intszerozerom &=& \int d \Phi_{X_{ijk}}^{(0,0,m)} (s_{ik}) \nonumber \\
&=& (\enen)^{2-\eps} u_0^{3-2\eps} 2^{-5+2\eps} \pi^{-2+\eps} 
\frac{\Gamma(2-2\eps) \Gamma(1-\eps)}{\Gamma(4-4\eps)}
 \gaussf{1-\eps}{2-2\eps}{5-4\eps}{u_0}\,, \nonumber \\
\end{eqnarray}

Expanding the integrated flavour-violating A-Type antenna 
${\cal A}^0_{Qg\bar{q}}$ given in eq. (\ref{eq:flavourviolint}) 
in powers of $\epsilon$ up to finite order, we obtain
\begin{eqnarray}
&&{\cal A}^0_{Qg\bar{q}}(1_Q,3_g,2_{\bar{q}})=
-2{\bf I}^{(1)}_{\tilde{Q}\bar{q}}\left(\e,s_{Q g\bar{q}},m_Q,0,
\frac{m_{Q}^2}{m_{Q}^2+s_{Q g\bar{q}}}\right)-\left[ \frac{1}{1-\mu^2}-\frac{19}{4}+\frac{5\pi^2}{12}\right.\nonumber\\
&&\hspace{20mm}\left. +\left( \frac{1}{(1-\mu^2)^2}-\frac{5}{4} \right){\rm ln}(\mu^2)+\frac{1}{4}{\rm ln}^2(\mu^2)+{\rm Li}_2(1-\mu^2)\right]+{\cal O}(\epsilon),
\end{eqnarray}
where
\begin{equation}\label{eq:mu}
\mu^2=\frac{m_Q^2}{E_{cm}^2}.
\end{equation}

All other integrated flavour conserving final-final antennae 
given in terms of masters in \cite{Mathias} can be written in terms 
of the real infrared singularity operators ${\bf I}^{(1)}_{ij}$ 
defined above up to finite order as follows,
\begin{eqnarray}
\lefteqn{{\cal A}^{0}_3(1_Q,3_g,2_{\bar{Q}})=-2 {\bf I}^{(1)}_{Q \bar{Q}}
\left(\e,s_{Q g\bar{Q}},m_{Q},m_{Q},\frac{1-\sqrt{r}}{1+\sqrt{r}}\right)} 
\nonumber\\
&& +\frac{1}{8\sqrt{r}}\bigg\{ \frac{1}{r-3}\bigg[-40\ln(2)(r-3)(1+r)\tanh^{-1}(\sqrt{r})\nonumber\\
&&\hspace{20mm}+2\sqrt{r}(-39+17r-16\ln(2)(r-3))\nonumber\\
&&\hspace{20mm}+(-33+10r-r^2+16\ln(2)(r-3)(1+r))\ln\left(\frac{1+\sqrt{r}}
{1-\sqrt{r}}\right)\nonumber\\
&&\hspace{20mm}-16(r-3)\tanh^{-1}(1-2r)\left(-2\sqrt{r}+(1+r)
\ln\left(\frac{1+\sqrt{r}}{1-\sqrt{r}}\right)\right)\\
&&\hspace{20mm}+10(r-3)(1+r)\left( \ln^2\left(1+\sqrt{r}\right) -\ln^2\left(1-\sqrt{r}\right)\right)\bigg]\nonumber\\
&&\hspace{7mm}+4(1+r)\bigg[ 5{\rm Li}_2\left( \frac{1+\sqrt{r}}{2}\right) -5{\rm Li}_2\left( \frac{1-\sqrt{r}}{2}\right)+3{\rm Li}_2(r)-12{\rm Li}_2 \left( \sqrt{r}\right)\nonumber\\
&&\hspace{24mm}+2\ln\left(\frac{1+r}{2}\right)\left(1+\ln\left( \frac{1-\sqrt{r}}{1+\sqrt{r}} \right) \right)\bigg]\bigg\} +{\cal O}(\epsilon),\nonumber 
\end{eqnarray}
\begin{eqnarray}
\lefteqn{{\cal E}^{0}_{3}(1_{Q}, 3_{q'},4_{\bar{q}'})=-4 {\bf I}^{(1)}_{Qg,\Flavour}
\left(\e,s_{Q q'\bar{q}'},m_{Q},0,\frac{m_{Q}^2}{m_{Q}^2+s_{Qq'\bar{q}'}}\right)}
\nonumber\\
&&\hspace{15mm}-\frac{1}{6(1-\mu^2)^3}\bigg[ 6+ 3\mu-14\mu^2+14\mu^4-3\mu^5 
-14\mu^2+14\mu^4-3\mu^5-6\mu^6
\nonumber\\
&&\hspace{15mm}-2\mu^3(-3-3\mu+\mu^3)\ln(\mu^2)\bigg]+{\cal O}(\epsilon),
\end{eqnarray}
\begin{eqnarray}
\lefteqn{{\cal D}^{0}_3(1_{Q},3_g,4_g)=-4 {\bf I}^{(1)}_{Qg}\left(\e,s_{Q gg},m_{Q},0,
\frac{m_{Q}^2}{m_{Q}^2+s_{Q gg}}\right)+\frac{1}{12(1-\mu^2)^3}}\nonumber\\
&&\times\bigg[-2(1-\mu^2)(4(-14+\pi^2)-3\mu+\mu^2(126-8\pi^2)+9\mu^3+\mu^4(-74+4\pi^2)) \nonumber\\
&&\hspace{6mm} -2(-3+33\mu^2+6\mu^3-51\mu^4+17\mu^6)\ln(\mu^2)+3(1-\mu)^2(-2-4\mu-\mu^2+\mu^3)\ln^2(\mu^2)\nonumber\\
&&\hspace{6mm}+12(1-\mu)^2(-2-4\mu-\mu^2+\mu^3){\rm Li}_2(1-\mu^2)\bigg]+{\cal O}(\epsilon),\nonumber\\
\end{eqnarray}
with $r$ and $\mu$ defined as in eq. (\ref{eq:r}) and eq. (\ref{eq:mu}) respectively. 

Let us notice that the last term present in the expansion of 
${\cal A}_{Qg \bar{Q}}$ arises through the expansion at finite order of 
\begin{displaymath}
\frac{1}{\e}\left(\frac{(1+r)}{2\sqrt{r}}\right)
\ln\left(\frac{1-\sqrt{r}}{1+\sqrt{r}}\right)\;\left[1-\frac{2m_{Q}^2}{E_{cm}^2}\right]^{+2\e}
\end{displaymath}
with $E_{cm}^2=s_{Q g\bar{Q}}+2 m_{Q}^2$.
This term arises since we have chosen to factor 
\begin{equation*}
 \left(\frac{s_{Q\bar{Q}}+2 m_{Q}^2}{s_{Q\bar{Q}}^{2}} \right)^{+\e}
\end{equation*} 
in ${\bf I}^{(1)}_{Q\bar{Q}}$ as unexpanded overall factor 
whereas the integrated antenna ${\cal A}_{Qg \bar{Q}}$
is naturally proportional to $\left[s_{Q g\bar{Q}}+2 m_Q^2\right]^{-\e}$. 
We have chosen to define the ${\bf I}^{(1)}_{Q\bar{Q}}$ operator such, 
in order to have similar overall factors in all ${\bf I}^{(1)}_{ij}$-type 
operators.

From these results on the expanded integrated final-final antennae, 
we see that the pole parts of all of those can be captured 
solely with poles present in the massive colour-orderered ${\bf I}^{(1)}_{ij}$  
operators defined above.   

Finally, for all integrated final-final antennae, 
we have compared the pure pole parts of those
with the pure pole parts of the corresponding sums of
the integrated dipoles presented in \cite{weinzierl}. Setting couplings 
and colour factors to one we found full agreement.

More explicitly, the poles of  ${\cal A}_{Qg\bar{Q}}$ can be compared
with the poles of two dipoles having both two massive final states,
a spectator $Q$ and an emitter $\bar{Q}$. The poles of ${\cal E}_{Qq'\bar{q}'}$ 
can be compared with those of two dipoles with a massless final state
emitter $q$ and a massive final-state spectator $Q$ while the poles of 
${\cal D}_{Qg g}$ have to be compared with the sum of two 
dipoles where the massive quark and the gluon in the final state can
both be emitter or spectator.

 \subsection{Integrated massive initial-final antennae}
In order to obtain the integrated massive initial-final antennae
${\cal X}_{i,jk}$,
the massive unintegrated initial-final antennae $X_{i,jk}$   
defined in Section 3 need to be integrated over the initial-final 
massive antenna phase space ${\rm d}\Phi_{X_{i,jk}}$ 
as given in eq.(\ref{eq:initialfinalphase}). As a result,
the integrated antennae will all depend on $Q^2=-q^2$, on $x$,
the momentum fraction carried by the initial state parton $p_{i}$, and on 
the masses of the final states present in a given antenna.
As we saw in Section 2, the momentum fraction $x$ depends on the number of 
massive particles present in the final state in a given antenna.
For two massive final states $j$ and $k$ of masses 
$m_{j}$ and $m_{k}$  $x$ is given by, 
\begin{displaymath}
x=\frac{Q^2 +m_{j}^2+m_{k}^2}{2 p_{i}\cdot q}.
\label{eq:xif2}
\end{displaymath}  

Generally, the pole parts of the  
integrated massive initial-final antennae are related to the
massive $\bf{I}^{(1)}_{ij}$ operators defined above and to the $x$-dependent 
colour ordered splitting kernels $p^{(0)}_{ij}(x)$ . 
The splitting kernels describe the initial-final
massless collinear singularities and are given by \cite{Daleo},
\begin{eqnarray}
&&p^{(0)}_{qq}(x)=\frac{3}{2}\,\delta(1-x)+2\Dcz(x)-1-x\,,
\nonumber \\
&&p^{(0)}_{qg}(x)=1-2x+2x^2\,,\nonumber \\
&&p^{(0)}_{gq}(x)=\frac{2}{x}-2+x\,,\nonumber \\
&&p^{(0)}_{gg}(x)=\frac{11}{6}\,\delta(1-x)+2\Dcz(x)+\frac{2}{x}-4+2x-2x^2\,,
\nonumber \\
&&p^{(0)}_{gg,F}(x)=-\frac{1}{3}\,\delta(1-x)\,,
\end{eqnarray}
with the distributions
\begin{displaymath}
{\cal D}_n(x) = \left(\frac{\ln^n(1-x)}{1-x}\right)_+\;.
\end{displaymath} 

In the following, we shall specify the phase space 
parametrisations for the initial-final antenna phase space 
${\rm d}\Phi_{X_{i,jk}}$  
needed to integrate the different antenna types 
before giving the results for the integrated initial-final 
massive antennae in unexpanded and expanded forms.

\subsubsection{Phase space parametrisations for initial-final configurations}
In the most general case, the massive two-to-two phase space 
${\rm d}\Phi_2$ to which the initial-final 
antenna phase space ${\rm d}\Phi_{X_{i,jk}}$  defined in eq.
(\ref{eq:initialfinalphase}) is proportional to,  
is related to the process $ q +p_i \to p_j +p_k$
with $p_{j}$ and $p_{k}$ the momenta of the final state partons with masses
$m_{j}$ and $m_{k}$. It can be written as 
\begin{equation}
{\rm d}\phi_2(m_j,m_k)=\frac{(4\pi)^{\epsilon}}{8\pi\Gamma(1-\epsilon)}
E_{cm}^{2\epsilon-2}\left[ \left( E_{cm}^2-m_j^2-m_k^2 \right)^2
-4m_j^2 m_k^2 \right]^{\frac{1-2\epsilon}{2}}
{\rm d}y\:y^{-\epsilon}(1-y)^{-\epsilon},
\label{phaseifAtype}
\end{equation}
where $y$ runs from $0$ to $1$. In this parametrisation the invariants 
take the following form
\begin{eqnarray}
2p_i\cdot p_j&=&\frac{Q^2+m_j^2+m_k^2}{2x E_{cm}^2}\times\nonumber\\ 
&&\left( E_{cm}^2+m_j^2-m_k^2-(2y-1)\sqrt{\left( E_{cm}^2-m_j^2-m_k^2
  \right)^2
-4m_j^2m_k^2}\right)\\
2p_i\cdot p_k&=&\frac{Q^2+m_j^2+m_k^2}{2x E_{cm}^2}\times\nonumber\\
&&\left( E_{cm}^2-m_j^2+m_k^2+(2y-1)\sqrt{\left( E_{cm}^2-m_j^2-m_k^2
  \right)^2
-4m_j^2m_k^2}\right),
\end{eqnarray}
and the partonic center of mass energy reads
\begin{equation}
E_{cm}=\sqrt{(p+q)^2}=\sqrt{\frac{Q^2(1-x)+m_j^2+m_k^2}{x}}.
\end{equation}
In this case also $x$ is defined as,
\begin{displaymath}
x=\frac{Q^2 +m_{j}^2+m_{k}^2}{2 p_{i}.q}.
\end{displaymath}  

\parindent 1.5em

In our 
application to $t\bar{t}$ +jet production,
we are only interested in the cases where $m_j=m_k=m_Q$ 
or in the case where $m_{j}\neq 0$ and $m_k=0$. 
For the first case, the parametrisation given above simplifies to
\begin{equation}\label{phasesp2m}
 {\rm d}\phi_2(m_Q,m_Q)=\frac{(4\pi)^{\epsilon-1}}{2\Gamma(1-\epsilon)}E_{cm}^{2\epsilon-2}\left[ \left( E_{cm}^2-2m_Q^2\right)^2-4m_Q^4\right]^{\frac{1-2\epsilon}{2}}{\rm d}y\:y^{-\epsilon}(1-y)^{-\epsilon},
\end{equation}
while the invariants read
\begin{eqnarray}
2p_i\cdot p_j &=&\frac{Q^2+2m_Q^2}{2x E_{cm}^2}\left( E_{cm}^2-(2y-1)
\sqrt{\left( E_{cm}^2-2m_Q^2\right)^2-4m_Q^4}\right)\\
2p_i\cdot p_k &=&\frac{Q^2+2m_Q^2}{2x E_{cm}^2}\left( E_{cm}^2+(2y-1)
\sqrt{\left( E_{cm}^2-2m_Q^2\right)^2-4m_Q^4}\right),
\label{eq:sij2m}
\end{eqnarray}
and the center of mass energy is
\begin{equation}\label{eq:xa03}
E_{cm}=\sqrt{\frac{Q^2(1-x)+2m_Q^2}{x}} \hspace{1.5cm} {\rm with} \hspace{1.5cm}
x=\frac{Q^2 +2\,m_{Q}^2}{2 p_{i}.q}.
\end{equation}

\parindent 1.5em

In the case in which $m_k=0$ and $m_j=m_Q$ the phase space reduces to
 \begin{equation}\label{phasesp1m}
 {\rm d}\phi_2(m_Q,0)=\frac{(4\pi)^{\epsilon-1}}{2\Gamma(1-\epsilon)}
E_{cm}^{2\epsilon-2}\left( E_{cm}^2-m_Q^2 \right)^{1-2\epsilon}{\rm d}y\:y^{-\epsilon}(1-y)^{-\epsilon},
\end{equation}
the invariants are
\begin{eqnarray}
2p_i\cdot p_j &=&\frac{Q^2+m_Q^2}{xE_{cm}^2}\left[ E_{cm}^2-y
\left( E_{cm}^2-m_Q^2\right)\right]\\
2p_i\cdot p_k &=&\frac{Q^2+m_Q^2}{xE_{cm}^2}\left[y\left( E_{cm}^2-m_Q^2\right)
\right]
\label{eq:sij1m}
\end{eqnarray}
and the center of mass energy is
\begin{equation}
E_{cm}=\sqrt{\frac{Q^2(1-x)+m_Q^2}{x}}\hspace{1cm} {\rm with}
\hspace{1cm}
x=\frac{Q^2 +m_{Q}^2}{2 p_{i}\cdot q}.
\end{equation}

\subsubsection{Integrated forms of the initial-final massive antennae}

The integrated forms of the antennae can be obtained using reduction
techniques, using the extension of the
integration by parts method \cite{IBP1, IBP2} 
in \cite{higgsnnlo,4particle}
to reduce the phase space integrals to master integrals. 
In this task, we express all the invariants
in the antenna functions as massive propagators and write the three 
on-shell conditions $p_a^2=m_a^2$ ($a=j,k$) as cut propagators. 
Since the invariants $s_{ab}$ appearing in the
antennae are not all independent from each other, 
the integrated antennae are written as a one-loop diagrams with on-shell
conditions and not as two-loop diagrams as in the final-final case.    
The reduction to master integrals is therefore easier. It was 
done using the Laporta algorithm \cite{Laporta} with two
independent implementations: the Mathematica package 
FIRE \cite{firepaper}, and an in-house implementation 
in FORM \cite{form}. For all integrated
 initial-final antennae required and defined in Section 3, we find four master
integrals. Those can be evaluated analytically 
in terms of gamma functions and hypergeometric functions. 
The all order expressions as well as the expanded expressions of all
initial-final massive antenna functions 
will be presented below. \\

\parindent 0em
A) {\bf Quark-Antiquark antennae}\\
\parindent 1.5em
The integrated form of the massive three-parton initial-final 
quark antiquark antenna given by ${\cal A}_{g,Q{\bar{Q}}}$ 
is obtained by integrating 
its unintegrated form $A_{g,Q,\bar{Q}}$ defined in eq.(\ref{eq:ifAg})
over the initial-final massive antenna phase ${\rm d \Phi}_{X_{i,jk}}$ 
given in eq.(\ref{eq:initialfinalphase})
using the parametrisation of the two-to-two parton phase space  
${\rm d}\Phi_2(m_Q,m_Q)$ given in eq.(\ref{phasesp2m}).
It depends on the virtuality of the incoming boson $Q^2$, 
the masses of the final states both being $m_{Q}$ and on $x$ given by 
$x=\frac{Q^2 +2 m_{Q}^2}{2 p_{i}\cdot q}$. It reads, 
\begin{eqnarray}
\lefteqn{{\cal A}_3^0(3_g;1_Q,2_{\bar{Q}})= \frac{2(4\pi)^{1-\epsilon} 
e^{\epsilon\gamma_E}}{Q^2+2m_Q^2}}
\nonumber\\ 
&&\:\:\:\:\times \Bigg\{ \left[ -1-\frac{2m_Q^2(2-\epsilon)}
{Q^2(1-\epsilon)-2m_Q^2}+2x(1-2\epsilon)\left(1-\frac{Q^2x}{Q^2+2m_Q^2}\right)
\right] I_2(m_Q,m_Q)\nonumber\\
&&\:\:\:\:\:\:\:\:\:\:\:\:\:\:\: +\frac{1}{x}
\Bigg[ \frac{4(2-\epsilon)^2 m_Q^4}{Q^2(1-\epsilon)-2m_Q^2}
+2m_Q^2(2-\epsilon+2x^2\epsilon)\nonumber\\
&&\:\:\:\:\:\:\:\:\:\:\:\:\:\:\:  +(1-2x+2x^2)(1-\epsilon)(Q^2+2m_Q^2)\Bigg]
  I_3(m_Q,m_Q,s_{13}) \Bigg \}.
\label{eq:Aint}
\end{eqnarray}
$I_2(m_Q,m_Q)$ is the integrated phase space 
measure with two massive final states. It is given by
\begin{equation}
I_2(m_Q,m_Q)=(4\pi)^{\epsilon-1}\frac{\Gamma(1-\epsilon)}{2\Gamma(2-2\epsilon)}\left( \frac{x}{Q^2(1-x)+2m_Q^2} \right)^{\epsilon} \left( 1-\frac{4m_Q^2x}{Q^2(1-x)+2m_Q^2} \right)^{\frac{1-2\epsilon}{2}},
\end{equation}
while the remaining master integral reads
\begin{eqnarray}
I_3(m_Q, m_Q, s_{ij})&=&\int {\rm d}\Phi_2(m_Q,m_Q)\frac{1}{s_{ij}}\nonumber\\
&=&(4\pi)^{\epsilon-1}\frac{\Gamma(1-\epsilon)}{\Gamma(2-2\epsilon)}\left( \frac{x}{Q^2+2m_Q^2}\right) \left( \frac{x}{Q^2(1-x)+2m_Q^2} \right)^{\epsilon}\nonumber\\ &&\times\frac{v^{1-2\epsilon}}{1+v} \gaussf{1}{1-\epsilon}{2-2\epsilon}{\frac{2v}{1+v}}
\end{eqnarray}
where, 
\begin{equation}\label{eqv}
v=\left(1-\frac{4m_Q^2}{E_{cm}^2}\right)^{\frac{1}{2}}=\left( 1-\frac{4m_Q^2 x}{Q^2(1-x)+2m_Q^2}\right)^{\frac{1}{2}}.
\end{equation}

The integrated form of the gluon-initiated quark-antiquark massive antenna $A^0_{g;Q\bar{Q}}$ is finite, since it has only quasi-collinear limits.
Expanding the all order result given in eq.(\ref{eq:Aint}) 
in powers of $\e$ up to finite order, we find
\begin{eqnarray}
&&{\cal A}_3^0(3_g;1_Q,2_{\bar{Q}})=\frac{1}{(Q^2+2m_Q^2)^2(Q^2-2m_Q^2)}\nonumber\\
&&\times\bigg\{ \left[Q^4(1-2x+2x^2)+4m_Q^2 Q^2+m_Q^4(4+8x-8x^2)\right]{\rm ln}\left( \frac{1-v}{1+v} \right)\\
&&\:\:+v\left[ Q^4(1-2x+2x^2)+4m_Q Q^2(1-x)+m_Q^4(4-8x) \right]\bigg\}+{\cal O}(\e).\nonumber
\end{eqnarray}

\parindent 0em
B){\bf Quark-Gluon antennae}\\
\parindent 1.5em
As only one massive parton is present in the final state,
the fractional momentum $x$ carried by the initial state momentum $p_{i}$ 
is defined as $x=\frac{Q^2 +m_{Q}^2}{2p_i \cdot q}$. 
For the initial-final antenna phase space ${\rm d \Phi}_{X_{i,jk}}$  
over which the antennae are
integrated, we use  the parametrisation of the two-by-two parton phase space  
${\rm d}\Phi_2(m_Q,0)$ given in eq.(\ref{phasesp1m}). 
After reduction, we found that two master integrals are needed. 

The master integral $I_2(m_Q,0)$ corresponds to the integrated phase 
space in eq.(\ref{phasesp1m}) and reads,
\begin{equation}
I_2(m_Q,0)=(4\pi)^{\epsilon-1}\frac{\Gamma(1-\epsilon)}{2\Gamma(2-2\epsilon)}\left( \frac{x}{Q^2(1-x)+m_Q^2} \right)^{\epsilon} \left( 1-\frac{m_Q^2x}{Q^2(1-x)+m_Q^2} \right)^{1-2\epsilon},
\end{equation}
while the remaining integral is given by
\begin{eqnarray}
I_3(m_Q,0,s_{ik})&=&\int {\rm d}\Phi_2(m_Q,0)\frac{1}{s_{ik}}\nonumber\\
&=&(4\pi)^{1-\epsilon}\frac{\Gamma(1-\epsilon)}{2\Gamma(2-2\epsilon)}\left( \frac{x}{Q^2+m_Q^2} \right) \left( \frac{x}{Q^2(1-x)+m_Q^2} \right)^{\epsilon}\\ 
&&\times \:\:\:u^{1-2\epsilon} \gaussf{1}{1-\epsilon}{2-2\epsilon}{u}\nonumber
\end{eqnarray}
with $u$ being given by
\begin{displaymath}
u=1-\frac{x m_{Q}^2}{Q^2(1-x) +m_{Q}^2}.
\end{displaymath}
\\
\parindent 0em
B.1) {\bf E-Type antennae}\\
\parindent 1.5em
The integrated form of $E^0_{q,qQ}$ is only proportional
to the integrated phase space measure $I_2(m_Q,0)$ and reads, 
\begin{eqnarray}
\lefteqn{{\cal E}_3^0(4_q;3_q,1_{Q})=\frac{(4\pi)^{1-\epsilon}e^{\epsilon\gamma_E}}
{[Q^2+m_Q^2]}}\nonumber\\
&&\times\Bigg\{ 2m_Q m_{\chi}-\frac{( Q^2(1-x)+m_Q^2 )( 1+(1-x)^2 )}{\epsilon x(1-x)}+m_Q^2\left( \frac{1+x}{1-x} \right)-Q^2(1-2x)\Bigg\} I_2(m_Q,0)\nonumber\\
\end{eqnarray} 
Its $\e$ expansion is given by,
\begin{eqnarray}
&&{\cal E}_3^0(4_q;3_q,1_{Q})=\frac{e^{\epsilon\gamma_E}}{\Gamma(1-\e)}
\left[ Q^2+m_{Q}^2 \right]^{-\e}
\times \Bigg \{ -\frac{1}{2\e}\;p_{qg}^{(0)}(x) \nonumber \\
&&\hspace{10mm} +\frac{1}{2x}(2-2x+x^2) \left[-2+2\ln(1-x)-\ln(x) -\ln(1-x_0x) \right]\\
&&\hspace{10mm}+\frac{Q^2(-1+3x-2x^2)+m_Q^2(1+x)+2m_Qm_{\chi}(1-x)}{2(Q^2(1-x)+m_Q^2)}+{\cal O}(\e) \Bigg \}, \nonumber
\end{eqnarray}
with $x_0$ being given in eq.(\ref{eq:x0}),
and where we have kept the natural phase space factor 
for initial-final quark-gluon integrated antennae, 
$$\frac{e^{\epsilon\gamma_E}}{\Gamma(1-\e)}\left[ Q^2+m_{Q}^2 \right]^{-\e}$$
unexpanded.
Since this antenna has only one massless initial-final collinear limit, 
the pole part in its integrated form is only related to the $x$-dependent 
splitting kernel $p^{(0)}_{qg}(x)$.
This expression can be directly compared with the expression 
obtained in \cite{weinzierl} for the integrated dipole 
involving a massless initial emitter $q$ and a massive final-state spectator 
$Q$ and full agreement is found. \\

\parindent 0em
B.2) {\bf D-Type antennae}\\
\parindent 1.5em
The integrated form of the gluon-initiated initial-final antenna 
$D^0_{g,Qg}$, which is such that the reduced matrix element 
is induced by a quark and which has a massive quark $Q$ as unresolved parton, is
\begin{eqnarray}
&&{\cal D}_3^0(4_g;1_Q,3_g)=\frac{(4\pi)^{1-\epsilon}e^{\epsilon\gamma_E}}{(Q^2+m_Q^2)}\frac{1}{(1-\epsilon)x(1-x)}\nonumber\\
&&\times\Bigg\{ \frac{1}{2(Q^2+m_Q^2)(Q^2(1-x)+m_Q^2)}\nonumber\\
&&\hspace{4mm}\:\:\:\:\times \bigg[ Q^6(1-x)^2 \left[ -3+6x-4x^2+4x^3+2\epsilon (1-4x+6x^2-6x^3)\right.\nonumber\\
&&\hspace{4mm}\:\:\:\:\:\:\:\:\:\:\:\:\:\:\:\:\:\:\:\:\:\:\:\:\:\:\:\:\:\:\:\:\:\left. +\epsilon^2(1+2x-8x^2+8x^3) \right]\nonumber\\
&&\hspace{4mm}\:\:\:\:\:\:\: -Q^4 m_Q^2 (1-x) \left[ 9-25x+32x^2-16x^3+4x^4-2\epsilon (3-14x+30x^2\right.\nonumber\\
&&\hspace{4mm}\:\:\:\:\:\:\:\:\:\:\:\:\:\:\:\:\:\:\:\:\:\:\:\:\:\:\:\:\:\:\:\:\:\:\:\:\:\:\left. -24x^3+6x^4)-\epsilon^2 (3+5x-24x^2+32x^3-8x^4) \right]\nonumber\\
&&\hspace{4mm}\:\:\:\:\:\:\: +Q^2 m_Q^4 \left[ -9+32x-57x^2+38x^3-8x^4 + \epsilon (6-32x+82x^2-76x^3+24x^4)\right.\nonumber\\
&&\hspace{4mm}\:\:\:\:\:\:\:\:\:\:\:\:\:\:\:\:\:\:\:\:\:\:\:\:\:\:\left.+\epsilon^2 (3+4x-25x^2+38x^3-16x^4) \right] \nonumber\\
&&\hspace{4mm}\:\:\:\:\:\:\: +m_Q^6 \left[ -3+10x-19x^2+4x^3+2\epsilon (1-5x+12x^2-6x^3)\right.\nonumber\\
&&\hspace{4mm}\:\:\:\:\:\:\:\:\:\:\:\:\:\:\:\:\:\:\:\:\left. +\epsilon^2 (1+2x-7x^2+8x^3) \right]\\
&&\hspace{4mm}\:\:\:\:\:\:\: +2\epsilon (1-\epsilon) Q^4 m_Q m_{\chi}(-1-x+2x^2)\nonumber\\
&&\hspace{4mm}\:\:\:\:\:\:\:-4Q^2 m_Q^3 m_{\chi}x \left[ -x(1-x)+\epsilon (1-\epsilon)(1+x-x^2) \right]\nonumber\\
&&\hspace{4mm}\:\:\:\:\:\:\: +2m_Q^5 m_{\chi}x \left[ 2x-\epsilon (1+x) +\epsilon^2 (1+x)\right] \bigg] I_2(m_Q,0)\nonumber\\
&&\hspace{3mm}-\bigg[(1-\epsilon)^2 Q^4 x (1-3x+4x^2-2x^3)\nonumber\\
&&\hspace{6mm}+2Q^2m_Q^2 \left[1-4x+4x^2-2x^3+\epsilon(-2+7x-6x^2+2x^3)+\epsilon^2(1-2x+2x^2)\right]\nonumber\\
&&\hspace{6mm}+m_Q^4\left[1-7x+4x^2-2x^3+\epsilon(-2+8x-4x^2)+\epsilon^2(1-x+2x^3)\right]\nonumber\\
&&\hspace{6mm}-2\epsilon(1-\epsilon)Q^2m_Qm_{\chi}x+2(1-\epsilon+\epsilon^2)m_Q^3m_{\chi}x\bigg]I_3(m_Q,0,s_{ik})\Bigg\}\nonumber
\end{eqnarray}
Its expansion up to finite order in $\e$ reads,
\begin{eqnarray}
\lefteqn{{\cal D}_3^0(4_g;1_Q,3_g)=\frac{1}{4 (Q^2+m_{Q}^2)^2\,(1-x)}}\nonumber\\
&&\times \bigg\{ \frac{(Q^2+m_Q^2)(1-x)}{(Q^2(1-x)+m_Q^2)^2 x}\bigg[ Q^6(1-x)^2(-3+6x-4x^2+4x^3)-Q^4m_Q^2(1-x)(9-25x\nonumber\\
&&\hspace{20mm}+32x^2-16x^3+4x^4)+Q^2m_Q^4(-9+32x-57x^2+38x^3-8x^4)\nonumber\\
&&\hspace{20mm}+m_Q^6(-3+10x-19x^2+4x^3)+4Q^2 m_Q^3 m_{\chi}x^2(1-x)+4m_Q^5 m_{\chi}x^2\bigg]\nonumber\\
&&\hspace{5mm}-2\ln \left( \frac{m_Q^2 x}{Q^2(1-x)+m_Q^2} \right)\bigg[ Q^4(1-x)(1-2x+2x^2)-2Q^2m_Q^2(-1+4x-4x^2+2x^3)\nonumber\\
&&\hspace{20mm}-m_Q^4(-1+7x-4x^2+x^3)+2m_Q^3m_{\chi}x\bigg]\bigg\}
+{\cal O}(\epsilon)
\nonumber\\
\end{eqnarray}
This integrated antenna is finite since its unintegrated form 
has only quasi-collinear limits.
 
Finally, the integrated form of the initial-final massive D-Type quark-gluon 
antenna $D^0_{g,gQ}$ which is initiated by a gluon, which has  
a reduced matrix element induced by a gluon and 
where the final state gluon is the unresolved parton, is given by, 
\begin{eqnarray}
&&{\cal D}_3^0(4_g;3_g,1_Q)=\frac{(4\pi)^{1-\epsilon}e^{\epsilon\gamma_E}}{2}\nonumber\\
&&\:\:\:\:\times \Bigg\{ \frac{Q^2(1-\epsilon)^2+2m_Qm_{\chi} \epsilon(1-\epsilon) + m_Q^2(3-4\epsilon-\epsilon^2) }{(1-\epsilon)[ Q^2(1-x)+m_Q^2]}\nonumber\\
&&\hspace{10mm} -\frac{4(1-2\epsilon)[ Q^2(1-x)+m_Q^2] \left( 1-x+x^2 \right)^2}{\epsilon x(1-x)^2 (Q^2+m_Q^2)} \nonumber\\
&&\hspace{10mm} +\frac{2}{(1-\epsilon)(1-x)^2(Q^2+m_Q^2)}\bigg[ 2m_Q m_{\chi}(1-\epsilon+\epsilon^2)(1-x)^2 \\
&&\hspace{10mm}+Q^2(1-x) [ -5+3x+\epsilon(8-6x)-\epsilon^2(1-x) ]  \nonumber\\ 
&&\hspace{10mm}+m_Q^2 [ -9+16x-9x^2+2\epsilon (4-7x+4x^2)+\epsilon^2(1-x)^2 ] \bigg]\Bigg\} I_2(m_Q,0)\nonumber
\end{eqnarray}
The $\e$ expansion of this integrated antenna reads
\begin{eqnarray}\label{expd03g}
\lefteqn{{\cal D}_3^0(4_g;3_g,1_Q)= 
-2{\bf I}^{(1)}_{Qg}\left(\e,Q^2,m_Q,0,x_0^2 \frac{m_Q^2}{Q^2 +m_{Q}^2}\right)\;
\delta(1-x)-\frac{1}{2\e}p_{gg}^{(0)}(x)}\nonumber\\
&&-\frac{3}{2}+\left( 1-\frac{\pi^2}{12} \right)\delta(1-x)
-{\cal D}_0(x)+2{\cal D}_1(x)\nonumber\\
&&+\frac{1}{4(Q^2(1-x)+m^2)^2}\bigg[ Q^4(1-x)-2Q^2m_Q^2(1-x)(2-9x)
-5m_Q^4(1-3x)\nonumber\\
&&\hspace{15mm}+4Q^2m_Qm_{\chi}(1-x)^2+4m_Q^3m_{\chi}(1-x)\bigg]\\
&&+2\left( -2+\frac{1}{x}+x-x^2\right)\ln (1-x)+\left(\frac{1}{2}\delta(1-x)+\frac{1}{1-x}-{\cal D}_0(x) \right)\ln( 1-x_0)\nonumber\\
&&+\frac{1}{2}\delta(1-x)\left(\frac{17}{6}\ln(x_0^2)+\ln(x_0^2)\ln(1-x_0)+\frac{1}{2}\ln^2(x_0^2)+\frac{1}{2}\ln^2(1-x_0)\right) \nonumber\\
&&-\frac{(1-x+x^2)^2}{x(1-x)}\left[ \ln(x)+\ln(1-x_0x) \right]- \frac{1}{2} p_{gg}^{(0)}(x)\ln\left( \frac{x_0^2(Q^2+m^2)}{Q^4}\right) + {\cal {O}}(\e),\nonumber
\end{eqnarray}
where $x_0$ is given in eq. (\ref{eq:x0}). 

From this expansion it can be seen that all the pole parts 
are contained in the infrared singularity 
operator $${\bf I}^{(1)}_{Qg}
\left(\e,Q^2,m_Q,0,x_0^2\frac{m_Q^2}{Q^2 +m_{Q}^2}\right)$$ 
as well as in the splitting kernel $p_{gg}^{(0)}(x)$.
A factor $x_{0}^2$ appears explicitly in 
the mass-dependent logarithmic term in the infrared operator 
${\bf I}^{(1)}_{Qg}$. It is   
necessary in order to capture all poles parts proportional 
to $\delta(1-x)$ of the integrated antenna, ${\cal D}_3^0(4_g;3_g,1_Q)$.

Furthermore, we can compare this expanded 
result for ${\cal D}^0_{g,gQ}$ with the corresponding 
integrated dipoles in \cite{weinzierl}.
The pole parts of our expression given in eq.(\ref{expd03g}) corresponds 
to the pole part of the sum of two dipoles which have  
either a massless initial spectator and a massive final-state emitter 
or a massless initial state emitter and a massive final state spectator.
Full agreement is found providing us with a strong check on our result
for ${\cal D}^0_{g,gQ}$.\\

\parindent 0em

C){\bf Flavour violating antennae}\\
\parindent 1.5em
In addition to the flavour conserving  integrated inital-final antennae, 
we also need to consider the integration over the massive 
initial-final antenna phase space of the unintegrated initial-final 
flavour-violating antenna $A^0_{q;gQ}$.
The phase space parametrisation of the initial-final antenna phase required 
here is the same parametrisation as the one used for 
integrating the initial-final quark-gluon antennae. One  uses the
parametrisation of the two-to-two particle phase space   
given by ${\rm d}\Phi_2(m_Q,0)$  in eq.(\ref{phasesp1m}). 
The integrated flavour violating antenna ${\cal A}^0_{q;gQ}$ 
can be written in term of the 
phase space measure  $I_2(m_Q,0)$ only.
In term of this master integral, its integrated form reads,
\begin{eqnarray}
&&{\cal A}_3^0(2_q;3_g,1_Q)=-\frac{(4\pi)^{1-\epsilon}e^{\epsilon\gamma_E}}
{2\epsilon(Q^2+m_Q^2)(Q^2(1-x)+m_Q^2)(1-x)^2}\nonumber\\
&&\:\:\:\:\:\:\:\left\{ Q^4(1-x)^2\left[ 2(1+x^2)-\epsilon(7-8x+6x^2)
+\epsilon^2(1-2x)^2 \right]\right.\\
&&\:\:\:\:\:\:\:\left. +2m_Q^2Q^2(1-x)\left[ 2(1+x^2)-\epsilon(7-9x+6x^2)
+\epsilon^2(1-3x+2x^2) \right]\right.\nonumber\\
&&\:\:\:\:\:\:\:\left. +m_Q^4 \left[ 2(1+x^2)-\epsilon(7-10x+7x^2)
+\epsilon^2(1-x)^2\right]\right\}I_2(m_Q,0)\nonumber
\end{eqnarray}
Expanding in powers of $\e$ we obtain
\begin{eqnarray}
\lefteqn{{\cal A}_3^0(2_q;1_Q,3_g)=-2{\bf I}^{(1)}_{q\tilde{Q}}
\left(\e,Q^2,m_Q,0,x_0^2 \frac{m_{Q}^2}{Q^2+m_{Q}^2}\right)\delta(1-x)-\frac{1}{2\e}p_{qq}^{(0)}(x)}\nonumber\\
&&+\frac{3}{2}+\left(1-\frac{\pi^2}{12}\right)\delta(1-x)-{\cal D}_0(x)+2{\cal D}_1(x)-\frac{x}{2}-\frac{m_Q^2}{4Q^2(1-x_0x)^2}\nonumber\\
&&+\frac{1}{4(1-x_0)(1-x_0x)}+\left(\frac{1}{2}\delta(1-x)+\frac{1}{1-x}-{\cal D}_0(x) \right)\ln(1-x_0)\\
&&+\frac{1}{4}\delta(1-x)\left(5\ln(x_0^2)+2\ln(x_0^2)\ln(1-x_0)+\ln^2(x_0^2)+\ln^2(1-x_0)\right)\nonumber\\
&&-(1+x)\ln(1-x)-\frac{6(1+x^2)}{1-x}\left[\ln(x)-\ln(1-x_0x) \right]\nonumber\\
&&-\frac{1}{2}p_{qq}^{(0)}(x)\ln\left( \frac{x_0^2(Q^2+m^2)}{Q^4}\right)+{\cal{O}}(\e),\nonumber
\label{expA03fvif}
\end{eqnarray}
All the pole pieces are contained in the following infrared 
singularity operator 
$${\bf I}^{(1)}_{q\tilde{Q}}\left(\e,Q^2,m_Q,0,x_0^2 
\frac{m_{Q}^2}{Q^2+m_{Q}^2}\right)$$  and the splitting kernel
$p_{qq}^{(0)}(x)$.
As for the integrated gluon-initiated quark-gluon 
antenna ${\cal D}^0_{g,gQ}$, a factor $x_{0}^2$ appears in the mass-dependent 
logarithmic term in the infrared operator ${\bf I}^{(1)}_{q\tilde{Q}}$.  
It is necessary in order to capture all poles parts proportional 
to $\delta(1-x)$ of the integrated flavour violating 
antenna, ${\cal A}_3^0(2_q;3_g,1_Q)$. 

In summary, in this subsection 5.3 we have shown that all integrated massive 
initial-final antennae have their pole parts related either 
to massive infrared operators or to splitting kernels or both.
\section{Check of ${\cal A}_{g;Q\bar{Q}}^0$}

A strong check can be performed on the integrated quark-antiquark
antenna ${\cal A}_{g;Q\bar{Q}}^0$ by comparing its expression given in 
eq.(\ref{eq:Aint}) and known results from the literature on   
the leading order heavy-quark coefficient functions.

To compare our results with $\gamma$ induced deep inelastic scattering we
consider the contraction of the hadronic tensor  $W^{\mu\nu}$    
 with the metric tensor $-g_{\mu\nu}$. 
This corresponds to the trace of the hadronic tensor, which in terms of 
the structure functions $\mathcal{F}_{2}$ and $\mathcal{F}_{L}$ 
is given by,
\begin{equation}
-W^{\mu}_{\mu}=-\frac{d-1}{2}\mathcal{F}_{L}\l(z,Q^{2}\r)+\frac{d-2}{2}\mathcal{F}_{2}\l(z,Q^2\r)\,,
\end{equation}
where the structure functions can be expanded in powers 
of the strong coupling constant.

To zeroth order in $\alpha_{s}$ these structure functions 
are given by the simple parton model result
\begin{equation}
\mathcal{F}_{L,q}^{\l(0\r)}\,=\,\mathcal{F}_{L,g}^{\l(0\r)}\,=\,0\,,\qquad\qquad\mathcal{F}_{2,q}^{\l(0\r)}\,=\,\delta\l(1-z\r)\,,\qquad\qquad\mathcal{F}_{2,g}^{\l(0\r)}\,=\,0\,.
\end{equation}
We find that the correct normalisation of $W^{\mu}_{\mu}$ 
to be checked against the antenna ${\cal A}_{g;Q\bar{Q}}^0$  
is as in the massless case \cite{Gionata}
given by
\begin{equation}\label{eq:haronictensortocheck}
-\frac{2}{{d-2}} W^{\mu}_{\mu}=\mathcal{F}_{2}-\frac{d-1}{d-2}\mathcal{F}_{L}\,.
\end{equation}
such that the following relation should hold
\begin{equation}
\label{acoeff}
{\cal A}_{g;Q\bar{Q}}^{0}\times \left[Q^2(1-\e) -2 m_{Q}^2 \right]=
-\frac{1-\epsilon}{2}
\left[ {\cal F}^{(1)}_{2,g}(x,Q^2)
-\frac{3-2\epsilon}{2-2\epsilon}{\cal F}^{(1)}_{L,g}(x,Q^2) \right]
\end{equation}
with  ${\cal F}^{(1)}_{2,g}$ and ${\cal F}^{(1)}_{L,g}$ being the
leading order heavy quark
coefficient functions.

The factor on the left-hand side of this equation which 
multiplies the integrated antenna corresponds to the massive 
two-parton antenna $A_{Q;Q}$ serving to normalise the three parton
antenna and given in eq.(\ref{eq:Aifnorm}).

The coefficient functions are given up to finite order in $\eps$ 
in \cite{klein} with a different choice of variable $x$.
$x \equiv x_{B}=\frac{Q^2}{2p\cdot q}$ is used
instead of our expression for $x$ given in eq.(\ref{eq:xa03}) by
$x=\frac{Q^2+2m_{Q}^2}{2p\cdot q}$, which
includes the masses $m_{Q}$ of the final states $Q$ and $\bar{Q}$.

We used this second definition of $x$ in our expression of the integrated
antenna ${\cal A}_{g;Q\bar{Q}}$
since this definition of $x$ is required in order to guarantee 
phase space factorisation in our subtraction formalism.  
Adapting the results given in \cite{klein} to this mass-dependent 
definition of $x$, the heavy quark coefficient functions read,

\begin{eqnarray}
{\cal F}^{(1)}_{2,g}(x,Q^2)=-\frac{2}{(Q^2+2m_Q^2)^2}\Bigg\{ v\left[ Q^4(1-8x+8x^2)-4Q^2 m_Q^2(-1+3x+x^2)\right. \nonumber\\
\left. +m_Q^4 (4+8x)\right] +\left[ Q^4(1-2x+2x^2)+4Q^2 m_Q^2(1-3x^2) \right. \label{F2} \\
\left. +m^4(4+8x-8x^2)\right] \log\left( \frac{1-v}{1+v}
\right)\Bigg\} +{\cal O}(\epsilon)\nonumber
\end{eqnarray}

\begin{equation}\label{FL}
{\cal F}^{(1)}_{L,g}(x,Q^2)=\frac{8Q^2 x}{(Q^2+2m_Q^2)^2}\Bigg\{v\left[ Q^2(1-x)+2m_Q ^2\right] +2m_Q^2 x \log\left( \frac{1-v}{1+v} \right) \Bigg\} +{\cal O}(\epsilon).
\end{equation}
To the finite order in $\eps$, one can show that 
\begin{eqnarray}
&& -\frac{1}{2}{\cal F}^{(1)}_{2,g} +\frac{3}{4}{\cal F}^{(1)}_{L,g} \nonumber\\
&&\:\:\:=\frac{1}{(Q^2+2m_Q^2)^2}\Bigg\{ v \left[ Q^4(1-2x+2x^2)+4Q^2m_Q^2(1-x^2)+m_Q^4(4+8x)\right]\label{eq:check}\\
&&\:\:\:\:\:\:\:+\left[ Q^4(1-2x+2x^2)+4Q^2m_Q^2+m_Q^4(4+8x-8x^2) \right]\log\left( \frac{1-v}{1+v} \right)\Bigg\},\nonumber
\end{eqnarray}
with the variable $v$ defined as in  eq. (\ref{eqv}) since 
the redefinition of $x$ in \cite{klein} does not affect
the definition of $v$.
The expression given in eq.(\ref{eq:check})  coincides 
with the $O(\epsilon^0)$ term of our expanded form for ${\cal A}_{g;Q\bar{Q}}$ 
after we undo the normalisation of this antenna function 
by multiplying it by $[Q^2-2m_Q^2]$. The relation given by
eq.(\ref{acoeff}) is therefore  herewith proven at ${\cal O}(\epsilon^0)$
giving us a strong check on the integrated antenna ${\cal
  A}_{g;Q\bar{Q}}$  itself.

\newcommand{\cm}{{\cal M}_5^0}
\newcommand{\cmf}{{\cal M}_4^0}
\newcommand{\cms}{{\cal M}_6^0}
\newcommand{\wt}{\widetilde}
\newcommand{\ds}{{\rm d}\hat{\sigma}}
\newcommand{\two}{2_{\bar{Q}}}

\section{Application to top quark pair production at LHC}
\parindent 1.5em
In this section we shall give  the colour-ordered real emission 
contributions for all partonic processes contributing to $t\bar{t}$ and 
$t\bar{t}+jet$ production at the LHC present at NLO. 
Together with these, we shall present their corresponding antenna subtraction terms which capture all single unresolved (soft, collinear and quasi-collinear)
 radiation of the real matrix-element squared 
for each partonic process involved.

The results presented here for the real contributions and subtraction terms 
for the process $p p \rightarrow t\bar{t}$ +1 jet at NLO 
are essential ingredients for the computation of the double real 
contributions and their subtraction terms to the production 
of $t \bar{t}$ at NNLO. 
Our subtraction terms enable to capture all single unresolved radiation 
present in those double real contributions. 
Furthermore, concerning the colour decomposition of the real matrix element squared for  $t\bar{t}$ +1 jet production at NLO, 
in the limit where the heavy quarks present in the final state 
are taken massless, the colour decomposition provided here 
corresponds to the colour decomposition of the real matrix elements 
to the processes involving a massless quark-antiquark pair 
for two jet production at NNLO.

For all partonic processes involved, 
starting from well-known amplitudes given in\cite{newresum,ttj}, 
the colour decomposition of the real matrix elements squared is presented and where possible, checked against results in the literature. 
Wherever possible, decoupling identities are used to reduce the size of the 
colour ordered decomposition of the real matrix-element squared and 
in some cases to eliminate the interference terms in those. 

The colour ordered subtraction terms are explicitly constructed 
as sums of terms involving the product of 
antenna functions with reduced matrix element squared 
and jet functions, as explained in Section 2. 

Concerning the notation of the matrix elements appearing in the real
contributions, those matrix elements denoted by 
${\cal M}$ represent colour-ordered amplitudes
in which the coupling constants and colour factors have been omitted. Furthermore, to explicitly visualize the colour connection between particles in these colour ordered amplitudes, a double semicolon is used in the labeling of the partons contributing to a given matrix elements. This double semi colon serves to separate strings of colour connected partons. Partons within a pair of double semicolons are colour-connected.
 
Since in the antenna framework a parton can only be unresolved with respect to its colour-connected neighbours, this notation helps to identify the unresolved limits present in a given colour ordered amplitude and therefore helps to construct the corresponding subtraction terms. Notationwise, we also denote gluons which are photon-like and only couple to quark lines,  with the index $\gamma$ instead of $g$, to manifestly separate leading from subleading contributions. In amplitudes where all gluons are photon-like no semicolons are used, since the concept of colour connection in not meaningful in those configurations.

Concerning the notation in the subtraction terms themselves, 
the reduced matrix-element squared present in those are also to be taken without coupling constants and coupling factors. 
In those matrix elements, the crossed momenta are denoted 
with a hat, the remapped final-state momenta are denoted with tildes 
and the remapped momenta of initial state hard radiators are denoted 
by a bar and a hat \footnote{This notation for the remapped initial state momenta was already used in \cite{Joao}}.

\subsection{$t\bar{t}$ production at LHC}
Following the general factorisation formula given in eq.(\ref{hadroncross}) for hadronic collision processes, the real NLO contributions to the production of a massive quark-antiquark pair in a hadronic collision can be written as
\begin{eqnarray}
{\rm d}\sigma^R=\int \frac{{\rm d}\xi_1}{\xi_1} \frac{{\rm d}\xi_2}{\xi_2} \bigg\{ \sum_{q} \bigg[ f_q(\xi_1)f_{\bar{q}}(\xi_2)\ds_{q\bar{q}\rightarrow Q\bar{Q}g}+f_{q}(\xi_1)f_{g}(\xi_2)\ds_{qg\rightarrow Q\bar{Q}q}\\
+f_{\bar{q}}(\xi_1)f_{g}(\xi_2)\ds_{\bar{q}g\rightarrow
  Q\bar{Q}\bar{q}}\bigg] +f_g(\xi_1)f_g(\xi_2)\ds_{g g\rightarrow Q\bar{Q}g}\bigg\}.\nonumber
\label{eq:ttbarhadron}
\end{eqnarray}
$f_{i}(\xi_{j})$ denotes the parton distribution function of parton $i$, 
which carries a fraction $\xi_{j}$ of one of the incoming hadron momenta 
and ${\rm d}\hat{\sigma}$ are the partonic cross sections. These process-dependent partonic cross sections ${\rm d}\hat{\sigma}$  are, in turn given by
\begin{eqnarray}
\ds_{q\bar{q}\rightarrow Q\bar{Q}g}&=& d\Phi_3(k_Q,k_{\bar{Q}},k_g;p_q,p_{\bar{q}})|M^0_{q\bar{q}\rightarrow Q\bar{Q}g}|^2 J_2^{(3)}(k_Q,k_{\bar{Q}},k_g), \label{eq:realemttbar1}\\
\ds_{qg\rightarrow Q\bar{Q}q}&=&d\Phi_3(k_Q,k_{\bar{Q}},k_q;p_q,p_g)|M^0_{qg\rightarrow Q\bar{Q}q}|^2 J_2^{(3)}(k_Q,k_{\bar{Q}},k_q),\label{eq:realemttbar2}\\
\ds_{gg\rightarrow Q\bar{Q}g}&=&d\Phi_3(k_Q,k_{\bar{Q}},k_g;p_{g1},p_{g2})|M^0_{gg\rightarrow Q\bar{Q}g}|^2 J_2^{(3)}(k_Q,k_{\bar{Q}},k_g),\label{eq:realemttbar3}
\end{eqnarray}
where the momenta labels in the matrix elements are omitted 
for the sake of conciseness. Each of these process-dependent 
partonic cross sections 
involves a massive two-to-three parton phase space, $d\Phi_3$, 
the corresponding matrix element squared $|M^0|^2$ 
and the appropriate jet function $J_{2}^{(3)}$ related to the selection 
criteria of 2-jet events.
Out of three partons, from which two are a $Q$ and $ \bar{Q}$ , 
an event with two jets having each  
a heavy quark $Q$ or a heavy anti-quark $\bar{Q}$ in them is formed. 
In the corresponding subtraction terms, the jet functions will all 
be of the type $J_{2}^{(2)}$ and will only depend   
on the hard final-state momenta appearing in the reduced 
matrix-element squared. 

In order to obtain the subtraction terms, the colour decomposition of the real  matrix elements present in the partonic cross sections given above 
in eqs. (\ref{eq:realemttbar1},\;\ref{eq:realemttbar2},\;\ref{eq:realemttbar3}) has to be performed. However, since colour ordering does not distinguish between initial and final state partons, it is sufficient 
to consider the colour decomposition for the two unphysical 
processes $0\rightarrow Q\bar{Q}q\bar{q}g$ and $0\rightarrow Q\bar{Q}ggg$, 
and obtain the matrix elements needed for $t \bar{t}$ production from two initial state partons  
by appropriate crossings. We shall follow this strategy in the following.

\subsubsection{Processes derived from $0\rightarrow Q\bar{Q}q\bar{q}g$}
At amplitude level, the colour decomposition for this unphysical process
$0\rightarrow Q\bar{Q}q\bar{q}g$ is 
\begin{eqnarray}
M_5^0(1_Q,2_{\bar{Q}},3_q,4_{\bar{q}},5_g)&=&g^3\:\sqrt{2}\bigg[ (T^{a_5})_{i_1,i_4}\delta_{i_2,i_3} \cm(1_Q,5_g,4_{\bar{q}};;2_{\bar{Q}},3_q)\nonumber\\ 
&&\:\:\:\:+(T^{a_5})_{i_3,i_2}\delta_{i_1,i_4} \cm(1_Q,4_{\bar{q}};;2_{\bar{Q}},5_g,3_q)\\
&&-\frac{1}{N_c}(T^{a_5})_{i_1,i_2}\delta_{i_3,i_4} \cm(1_Q,5_g,2_{\bar{Q}};;3_q,4_{\bar{q}})\nonumber\\ 
&&-\frac{1}{N_c}(T^{a_5})_{i_3,i_4}\delta_{i_1,i_2} \cm(1_Q,2_{\bar{Q}};;3_q,5_g,4_{\bar{q}})\bigg],\nonumber
 \end{eqnarray}
Squaring, and using the photon decoupling identities stating that
  \begin{eqnarray}
\cm(1_Q,2_{\bar{Q}},3_q,4_{\bar{q}},5_{\gamma})=\cm(1_Q,5_g,4_{\bar{q}};;2_{\bar{Q}},3_q)+\cm(1_Q,4_{\bar{q}};;2_{\bar{Q}},5_g,3_q)\nonumber\\
 =\cm(1_Q,5_g,2_{\bar{Q}};;3_q,4_{\bar{q}})+\cm(1_Q,2_{\bar{Q}};;3_q,5_g,4_{\bar{q}})
\label{eq:id1}
\end{eqnarray}
 gives
\begin{eqnarray}
\label{eq:sqorderqq}
&&|M_5^0(1_Q,2_{\bar{Q}},3_q,4_{\bar{q}},5_g)|^2=g^6(N_c^2-1)\nonumber\\
&&\:\:\:\:\:\:\:\:\:\:\:\:\:\:\:\:\:\times\bigg[N_c\left(|\cm(1_Q,5_g,4_{\bar{q}};;2_{\bar{Q}},3_q)|^2+|\cm(1_Q,4_{\bar{q}};;2_{\bar{Q}},5_g,3_q)|^2\right)\nonumber\\
&&\:\:\:\:\:\:\:\:\:\:\:\:\:\:\:\:\:\:\:\:+\frac{1}{N_c}\left( |\cm(1_Q,5_g,2_{\bar{Q}};;3_q,4_{\bar{q}})|^2+|\cm(1_Q,2_{\bar{Q}};;3_q,5_g,4_{\bar{q}})|^2\right.\\
&&\:\:\:\:\:\:\:\:\:\:\:\:\:\:\:\:\:\:\:\:\:\:\:\:\:\:\:\left.-2|\cm(1_Q,2_{\bar{Q}},3_q,4_{\bar{q}},5_{\gamma})|^2\right)\bigg],\nonumber
 \end{eqnarray}
where in $\cm(1_Q,2_{\bar{Q}},3_q,4_{\bar{q}},5_{\gamma})$ the parton with momentum label $5$ is a $U(1)$ photon-like gluon that only couples to the quark lines. 

Upon the crossing of $3_q$ and $4_{\bar{q}}$ to the initial state, an expression for the squared matrix element $|M^0_{q\bar{q}\rightarrow Q\bar{Q}g}|^2$ in terms of colour-ordered amplitudes is obtained by replacing $3_q$ and $4_{\bar{q}}$ by $\hat{3}_{\bar{q}}$ and $\hat{4}_q$ in the expression given in eq.(\ref{eq:sqorderqq}).
 
This, together with the phase space and the jet function as given 
in eq.(\ref{eq:realemttbar1}) leads to the corresponding real emission 
differential cross section 
due to this partonic process in a colour ordered way. 

Analogously, the crossing of $4_{\bar{q}}$ and $5_g$ 
in eq.(\ref{eq:sqorderqq}) together with the phase space 
and the jet function in eq.(\ref{eq:realemttbar2}) 
gives the colour ordered real emission corrections 
to the production of a top-antitop pair from  
the partonic process $qg\rightarrow Q\bar{Q}q$.
The corresponding subtraction terms for these two crossings are given below.

After the crossing of $3_q$ and $4_{\bar{q}}$ in eq.(\ref{eq:sqorderqq}),  
we find that the subtraction term for the partonic process 
$q\bar{q}\rightarrow Q\bar{Q}g$ is
\begin{eqnarray}
&&{\rm d}\hat{\sigma}_{q\bar{q}\rightarrow Q\bar{Q}g}^S=g^6(N_c^2-1){\rm d}\Phi_3(k_{1Q},k_{2\bar{Q}},k_{5g};p_{4q},p_{3\bar{q}})\nonumber\\
&&\:\:\:\:\times\bigg\{ N_c \bigg[ A_3^0(4_q;1_Q,5_g) 
|\cmf((\wt{15})_Q,2_{\bar{Q}},\hat{3}_{\bar{q}},\hat{\bar{4}}_{q})|^2 J_2^{(2)}(k_{\wt{15}},k_2)\nonumber\\
&&\:\:\:\:\:\:\:\:\:\:\:\:\:\:\:\:+A_3^0(3_{\bar{q}};2_{\bar{Q}},5_g) |\cmf(1_Q,(\wt{25})_{\bar{Q}},\hat{\bar{3}}_{\bar{q}},\hat{4}_q)|^2 J_2^{(2)}(k_1,k_{\wt{25}})\bigg]\nonumber\\
&&\:\:\:\:\:\:\:-\frac{1}{N_c}\bigg[ A_3^0(1_Q,5_g,2_{\bar{Q}}) |\cmf((\wt{15})_Q,(\wt{25})_{\bar{Q}},\hat{3}_{\bar{q}},\hat{4}_q)|^2 J_2^{(2)}(k_{\wt{15}},k_{\wt{25}})\\
&&\:\:\:\:\:\:\:\:\:\:\:\:\:\:\:\:+A_{3}^0(4_q,3_{\bar{q}};5_g) |\cmf(\tilde{1}_Q,\tilde{2}_{\bar{Q}},\hat{\bar{3}}_{\bar{q}},\hat{\bar{4}}_q)|^2 J_2^{(2)}(\tilde{k}_1,\tilde{k}_2)\bigg]\bigg\}.\nonumber
\end{eqnarray}
Only A-Type quark-antiquark antennae are required in all three configurations.
In the intial-final configuration, both flavour conserving 
and flavour violating massive antennae are needed.

\parindent 1.5em
If $4_{\bar{q}}$ and $5_g$ are crossed instead, the subtraction term for $qg\rightarrow Q\bar{Q}q$ will be constructed with massive flavour-conserving  
quark-antiquark A-Type and quark-gluon E-Type antennae in both final-final and initial-final configurations in addition to massless initial-initial 
A-type antennae. It reads,
\begin{eqnarray}
\lefteqn{{\rm d}\hat{\sigma}_{qg\rightarrow Q\bar{Q}q}^S=g^6(N_c^2-1){\rm d}\Phi_3(k_{1Q},k_{\bar{2Q}},k_{3q};p_{4q},p_{5g})}\nonumber\\ 
&&\times\bigg\{ N_c \bigg[ \frac{1}{2} A_3^0(5_g;1_Q,2_{\bar{Q}}) \bigg( |\cmf((\wt{12})_Q,\hat{\bar{5}}_Q,3_q,\hat{4}_q)|^2+|\cmf(\hat{\bar{5}}_{\bar{Q}},(\wt{12})_{\bar{Q}},3_q,\hat{4}_q)|^2  \bigg)J_2^{(2)}(k_{\wt{12}},k_3)\nonumber\\
&&\:\:\:\:\:\:\:\:\:\:\:+A_3^0(4_q,5_g;3_q) |\cmf(\tilde{1}_Q,\tilde{2}_{\bar{Q}},\hat{\bar{4}}_q,\hat{\bar{5}}_{\bar{q}})|^2 J_2^{(2)}(\tilde{k}_1,\tilde{k}_2)\nonumber\\
&&\:\:\:\:\:\:\:\:\:\:\:+\frac{1}{2} E_3^0(4_q;3_q,1_Q) \bigg( |\cmf((\wt{13})_Q,\hat{5}_g,\hat{\bar{4}}_g,2_{\bar{Q}})|^2 + |\cmf((\wt{13})_Q,\hat{\bar{4}}_g,\hat{5}_g,2_{\bar{Q}})|^2 \bigg)J_2^{(2)}(k_{\wt{13}},k_2)\nonumber\\
&&\:\:\:\:\:\:\:\:\:\:\:+\frac{1}{2} E_3^0(4_q;3_q,2_{\bar{Q}}) \bigg( |\cmf(1_Q,\hat{5}_g,\hat{\bar{4}}_g,(\wt{23})_{\bar{Q}})|^2+|\cmf(1_Q,\hat{\bar{4}}_g,\hat{5}_g,(\wt{23})_{\bar{Q}})|^2 \bigg)J_2^{(2)}(k_1,k_{\wt{23}})\bigg] \nonumber\\
&& -\frac{1}{N_c} \bigg[\frac{1}{2} A_3^0(5_g;1_Q,2_{\bar{Q}}) \bigg( |\cmf((\wt{12})_Q,\hat{\bar{5}}_Q,3_q,\hat{4}_q)|^2 +|\cmf(\hat{\bar{5}}_{\bar{Q}},(\wt{12})_{\bar{Q}},3_q,\hat{4}_q)|^2  \bigg)J_2^{(2)}(k_{\wt{12}},k_3)\nonumber\\
&&\:\:\:\:\:\:\:\:\:\:\:+A_3^0(4_q,5_g;3_q)|\cmf(\tilde{1}_Q,\tilde{2}_{\bar{Q}},\hat{\bar{4}}_q,\hat{\bar{5}}_{\bar{q}})|^2 J_2^{(2)}(\tilde{k}_1,\tilde{k}_2)\nonumber\\
&&\:\:\:\:\:\:\:\:\:\:\:+\frac{1}{2} E_3^0(4_q;3_q,1_Q) |\cmf((\wt{13})_Q,\hat{\bar{4}}_{\gamma},\hat{5}_{\gamma},2_{\bar{Q}})|^2  J_2^{(2)}(k_{\wt{13}},k_2)\\
&&\:\:\:\:\:\:\:\:\:\:\:+\frac{1}{2} E_3^0(4_q;3_q,2_{\bar{Q}}) |\cmf(1_Q,\hat{\bar{4}}_{\gamma},\hat{5}_{\gamma},(\wt{23})_{\bar{Q}})|^2  J_2^{(2)}(k_1,k_{\wt{23}})\bigg]\bigg\}. \nonumber
\end{eqnarray}

\subsubsection{Partonic process $gg\rightarrow Q\bar{Q}g$}
The colour decomposition for the unphysical process $0\rightarrow Q\bar{Q}ggg$ is
\begin{equation}
M_5^0(1_Q,2_{\bar{Q}},3_g,4_g,5_g)=(g\sqrt{2})^3\sum_{(i,j,k)\in P(3,4,5)} (T^{a_i}T^{a_j}T^{a_k})^{i_1}_{i_2}\cm(1_Q,i_g,j_g,k_g,2_{\bar{Q}}).
\end{equation} 
Squaring and crossing gluons $4_{g}$ and $5_{g}$ to the initial state gives
\begin{eqnarray}
\lefteqn{|M_5^0(1_Q,2_{\bar{Q}},3_g,\hat{4}_g,\hat{5}_g)|^2=g^6(N_c^2-1)}\nonumber\\
&&\times\bigg\{   \sum_{(i,j)\in P(4,5)}  \bigg[    N_c^2\left( |\cm(1_Q,3_g,\hat{i}_g,\hat{j}_g,2_{\bar{Q}})|^2+|\cm(1_Q,\hat{i}_g,3_g,\hat{j}_g,2_{\bar{Q}})|^2\right.\nonumber\\
&&\:\:\:\:\:\:\:\:\:\:\:\:\:\:\:\:\:\:\:\:\:\:\:\:\:\:\:\:\:\:\:\:\:\:\:\left.+|\cm(1_Q,\hat{i}_g,\hat{j}_g,3_g,2_{\bar{Q}})|^2\right)  \nonumber\\
&&\:\:\:\:\:\:\:\:\:\:\:\:\:\:\:\:\:\:\:\:\:\:\:-\left(|\cm(1_Q,3_g,\hat{i}_g,\hat{j}_{\gamma},2_{\bar{Q}})|^2+|\cm(1_Q,\hat{i}_g,3_g,\hat{j}_{\gamma},2_{\bar{Q}})|^2\right.\nonumber\\
&&\:\:\:\:\:\:\:\:\:\:\:\:\:\:\:\:\:\:\:\:\:\:\:\:\:\:\left. +|\cm(1_Q,\hat{i}_g,\hat{j}_g,3_{\gamma},2_{\bar{Q}})|^2\right) \bigg]\label{eq:ttgg} \\
&&\:\:\:\:\:\:\:\:\:\:+\left(\frac{N_c^2+1}{N_c^2}\right)|\overline{{\cal M}}_5^0(1_Q,3_g,\hat{4}_g,\hat{5}_g,2_{\bar{Q}})|^2\bigg\}.\nonumber
\end{eqnarray}
We have used the following photon decoupling identity, 
\begin{eqnarray}
&&\cm(1_Q,i_g,j_g,k_{\gamma},2_{\bar{Q}})=\label{eq:id3}\\
&&\:\:\:\:\:\: \cm(1_Q,i_g,j_g,k_g,2_{\bar{Q}})+\cm(1_Q,i_g,k_g,j_g,2_{\bar{Q}})+\cm(1_Q,k_g,i_g,j_g,2_{\bar{Q}})\nonumber
\end{eqnarray}
where gluon $k$ is a $U(1)$ boson decoupled from the other gluons, and
\begin{equation}\label{eq:id2}
\overline{{\cal M}}_5^0(1_Q,3_g,4_g,5_g,2_{\bar{Q}})= \sum_{(i,j,k)\in P(3,4,5)} \cm(1_Q,i_g,j_g,k_g,2_{\bar{Q}}),
\end{equation}
where all gluons are photon-like.

The colour decomposition of the matrix element squared with two gluons in the initial state given by eq.(\ref{eq:ttgg}) is obtained as follows: 
After squaring the amplitude for the process $0\rightarrow Q\bar{Q}ggg$ 
and expanding the sum of permutations concerning the three final state gluons,
the crossing of two gluons $4_{g}$ and $5_{g}$ is performed.
As the final state gluon labelled $3_g$ is fixed, 
the terms are then regrouped and a sum over the permutations 
for the two initial state gluons $4_g$ and $5_g$ only is performed. 

Only colour ordered squared matrix elements 
are involved in  eq.(\ref{eq:ttgg}) for which unresolved radiation
can be captured by a single antenna function. 
Massive flavour-conserving quark-antiquark A-Type antennae, 
massive quark-gluon D-type antennae
together with massless initial-initial gluon-gluon F-Type antennae are required.

The subtraction term reads,
\begin{eqnarray}\label{eq:subtermttbarggg}
\lefteqn{{\rm d}\hat{\sigma}^{S}_{gg\rightarrow Q\bar{Q}g}=g^6(N_c^2-1) {\rm d}\Phi_3(k_{1Q},k_{2\bar{Q}},k_{3g};p_{4g},p_{5g})}\nonumber\\
&&\times\Bigg\{\sum_{(i,j)\in P(4,5)} \Bigg[ N_c^2 \bigg[ \frac{1}{2}A_3^0(i_g;1_Q,2_{\bar{Q}}) \big( |\cmf((\wt{12})_Q,3_g,\hat{j}_g,\hat{\bar{i}}_Q)|^2+|\cmf(\hat{\bar{i}}_{\bar{Q}},3_g,\hat{j}_g,(\wt{12})_{\bar{Q}})|^2\nonumber\\
&&\:\:\:\:\:\:\:\:\:\:\:\:\:\:\:\:\:\:\:\:\:\:\:\:\:\:\:\:\:\:\:\:\:\:\:\:\:\:\:\:\:\:\: +|\cmf((\wt{12})_Q,\hat{j}_g,3_g\hat{\bar{i}}_Q)|^2 +|\cmf(\hat{\bar{i}}_{\bar{Q}},\hat{j}_g,3_g(\wt{12})_{\bar{Q}})|^2\big) J_2^{(2)}(k_{\wt{12}},k_3)\nonumber\\
&&\:\:\:\:\:\:\:\:\:\:\:\:\:\:\:\:\:\:\:\:\:\:\:\:\:\:\:\:\:\:\:\:\: +D_3^0(i_g;3_g,1_Q) |\cmf((\wt{13})_Q,\hat{\bar{i}}_g,\hat{j}_g,2_{\bar{Q}})|^2 J_2^{(2)}(k_{\wt{13}},k_2)\nonumber\\
&&\:\:\:\:\:\:\:\:\:\:\:\:\:\:\:\:\:\:\:\:\:\:\:\:\:\:\:\:\:\:\:\:\: +D_3^0(i_g;3_g,2_{\bar{Q}}) |\cmf(1_Q,\hat{j}_g,\hat{\bar{i}}_g,(\wt{23})_{\bar{Q}})|^2 J_2^{(2)}(k_1,k_{\wt{23}})\nonumber\\
&&\:\:\:\:\:\:\:\:\:\:\:\:\:\:\:\:\:\:\:\:\:\:\:\:\:\:\:\:\:\:\:\:\: +F_3^0(i_g,j_g;3_g)  |\cmf(\tilde{1}_Q,\hat{\bar{i}}_g,\hat{\bar{j}}_g,\tilde{2}_{2\bar{Q}})|^2 J_2^{(2)}(\tilde{k_1},\tilde{k_2}) \bigg]\nonumber\\
&&\:\:\:\:\:\:\:\:\:\:\:\:\:\:\:\:\:\:\:\:-A_3^0(1_Q,3_g,2_{\bar{Q}}) |\cmf((\wt{13})_Q,i_g,j_g,(\wt{23})_{\bar{Q}})|^2 J_2^{(2)}(k_{\wt{13}},k_{\wt{23}})\\
&&\:\:\:\:\:\:\:\:\:\:\:\:\:\:\:\:\:\:\:\: -\frac{1}{2}A_3^0(i_g;1_Q,2_{\bar{Q}}) \big( |\cmf((\wt{12})_Q,3_g,\hat{j}_g,\hat{\bar{i}}_Q)|^2+|\cmf(\hat{\bar{i}}_{\bar{Q}},3_g,\hat{j}_g,(\wt{12})_{\bar{Q}})|^2\nonumber\\
&&\:\:\:\:\:\:\:\:\:\:\:\:\:\:\:\:\:\:\:\:\:\:\:\:\:\:\:\:\:\:\: +|\cmf((\wt{12})_Q,\hat{j}_g,3_g\hat{\bar{i}}_Q)|^2+|\cmf(\hat{\bar{i}}_{\bar{Q}},\hat{j}_g,3_g(\wt{12})_{\bar{Q}})|^2\nonumber\\
&&\:\:\:\:\:\:\:\:\:\:\:\:\:\:\:\:\:\:\:\:\:\:\:\:\:\:\:\:\:\:\: +|\cmf((\wt{12})_Q,3_{\gamma},\hat{j}_{\gamma},\hat{\bar{i}}_Q)|^2+|\cmf(\hat{\bar{i}}_{\bar{Q}},3_{\gamma},\hat{j}_{\gamma},(\wt{12})_{\bar{Q}})|^2 \big) J_2^{(2)}(k_{\wt{12}},k_3)\nonumber\\
&&\:\:\:\:\:\:\:\:\:\:\:\:\:\:\:\:\:\:\:\: -D_3^0(i_g;3_g,1_Q) |\cmf((\wt{13})_Q,\hat{\bar{i}}_{\gamma},\hat{j}_{\gamma},2_{\bar{Q}})|^2 J_2^{(2)}(k_{\wt{13}},k_2)\nonumber\\
&&\:\:\:\:\:\:\:\:\:\:\:\:\:\:\:\:\:\:\:\: -D_3^0(i_g;3_g,2_{\bar{Q}}) |\cmf(1_Q,\hat{j}_{\gamma},\hat{\bar{i}}_{\gamma},(\wt{23})_{\bar{Q}})|^2 J_2^{(2)}(k_1,k_{\wt{23}})\nonumber\\
&&\:\:\:\:\:\:\:\:\:\:\:\:\:\:\:\:\:\:\:\: +\frac{1}{2N_c^2}A_3^0(i_g;1_Q,2_{\bar{Q}})\big( |\cmf((\wt{12})_Q,3_{\gamma},\hat{j}_{\gamma},\hat{\bar{i}}_{\bar{Q}})|^2\nonumber\\
&&\:\:\:\:\:\:\:\:\:\:\:\:\:\:\:\:\:\:\:\:\:\:\:\:\:\:\:\:\:\:\: +|\cmf(\hat{\bar{i}}_{\bar{Q}},3_{\gamma},\hat{j}_{\gamma},(\wt{12})_{\bar{Q}})|^2 \big) J_2^{(2)}(k_{\wt{12}},k_3)\Bigg]\nonumber\\
&&\:\:\:\:\:\:\:\:\:\:\:\:+\left(\frac{N_c^2+1}{N_c^2}\right)A_3^0(1_Q,3_g,2_{\bar{Q}})|\cmf((\wt{13})_Q,\hat{4}_{\gamma},\hat{5}_{\gamma},(\wt{23})_{\bar{Q}})|^2 J_2^{(2)}(k_{\wt{13}},k_{\wt{23}})\Bigg\}.\nonumber
\end{eqnarray}

\subsubsection{Consistency check}
We have performed a powerful check on all subtraction terms required for $t \bar{t}$ production given above. For each of those subtraction terms, which is a sum of terms multiplied by colour factors proportional to $N_{c}$, we have verified that 
it gives the correct non-colour ordered collinear or/and quasi-collinear
 behaviour. We have checked this feature for all collinear and quasi-collinear 
limits present in all subtraction terms presented above. 
We have verified that each subtraction term obey
\begin{equation}
{\rm d}\hat{\sigma}^S \stackrel{a||b}{\longrightarrow}g^2\:C\:\frac{P_{ab}(z)}{s_{ab}}\times|M_m^0|^2\times {\rm d}\Phi_m  \hspace{1mm} J_{m}^{(m)},
\end{equation}
where $C=C_A,C_F,T_R$ is the corresponding Casimir, $M_m^0$ 
is the {\it non-colour ordered}  reduced matrix element and $P_{ab}(z)$ stands for a massless or massive spitting function governing the particular collinear or quasi-collinear limit as defined in Section 4. 

Let us now check all collinear and quasi-collinear limits of the subtraction term 
${\rm d}\hat{\sigma}^{S}_{gg\rightarrow Q\bar{Q}g}$
given in eq.(\ref{eq:subtermttbarggg}) in this way.
This subtraction term has the following collinear and quasi-collinear limits: $1_Q||3_g$,$3_{g}||\hat{4}_{g}$ and $1_{Q}||\hat{4}_{g}$. 

The following relation between the non-colour ordered 
and colour ordered amplitudes squared for the process 
$gg\rightarrow Q\bar{Q}$ is needed. It reads,
\begin{eqnarray}\label{eq:identity}
&&|M_4^0(1_Q,\two,\hat{i}_g,\hat{j}_g)|^2=g^4\left(\frac{N_c^2-1}{N_c}\right)\bigg[ N_c^2 \bigg(|\cmf(1_Q,\hat{i}_g,\hat{j}_g,\two)|^2\nonumber\\
&&\:\:\:\:\:\:\:\:\:\:\:+|\cmf(1_Q,\hat{j}_g,\hat{i}_g,\two)|^2\bigg)-|\cmf(1_Q,\hat{i}_{\gamma},\hat{j}_{\gamma},\two)|^2\bigg].
\end{eqnarray}

When the final-state quasi-collinear limit  $1_Q||3_g$ is taken in the subtraction term given in eq.(\ref{eq:subtermttbarggg}),  we obtain
\begin{eqnarray}
&&{\rm d}\hat{\sigma}^{S}_{gg\rightarrow Q\bar{Q}g}\stackrel{1_Q||3_g}{\longrightarrow}g^6\left(\frac{N_c^2-1}{N_c}\right)^2\frac{P_{qg\rightarrow Q}(z,\mu_{qg}^2)}{s_{13}}\nonumber\\
&&\:\:\:\times\bigg[ N_c^2 \bigg(|\cmf((1+3)_Q,\hat{4}_g,\hat{5}_g,\two)|^2+|\cmf((1+3)_Q,\hat{5}_g,\hat{4}_g,\two)|^2\bigg)\nonumber\\
&&\:\:\:\:\:\:\:-|\cmf((1+3)_Q,\hat{4}_{\gamma},\hat{5}_{\gamma},\two)|^2\bigg]{\rm d}\Phi_2 (k_{(1+3)Q},k_{2\bar{Q}};p_{4g},p_{5g})J_2^{(2)}(k_{1+3},k_2) \\
&&\:\:\:=g^2\:C_F\:\frac{P_{qg\rightarrow Q}(z,\mu_{qg}^2)}{s_{13}}|M_4^0((1+3)_Q,\two,\hat{4}_g,\hat{5}_g)|^2 \nonumber\\
&&\:\:\:\:\:\:\:\:\:\:\:\:\:\:\:\times {\rm d}\Phi_2 (k_{(1+3)Q},k_{2\bar{Q}};p_{4g},p_{5g})J_2^{(2)}(k_{1+3},k_2),\nonumber
\label{eq:checkcol1}
\end{eqnarray}
where we have used eq.(\ref{eq:identity}) with the momentum of 
the massive final state quark $1_{Q}$ being given by $(k_1+k_3)$. 
In the limit $1_Q||3_g$, the contributing terms 
in eq.(\ref{eq:subtermttbarggg}) are proportional 
to the antenna $A^0_{3}(1_{Q},3_g,2_{\bar{Q}})$ multiplied by a 
reduced colour-ordered matrix elements squared 
involving the remapped momenta $k_{(\tilde{13})}$ and $k_{(\tilde{23})}$
In the $1_Q||3_g$ limit, those remapped momenta  
$k_{(\tilde{13})}$ and $k_{(\tilde{23})}$ are respectively 
given by $(k_1+k_3)$ and $k_2$.

Similarly, in the same subtraction term given in eq.(\ref{eq:subtermttbarggg})  
we can see that in the $3_g||\hat{4}_g$ limit we obtain
\begin{eqnarray}
&&{\rm d}\hat{\sigma}^{S}_{gg\rightarrow Q\bar{Q}g}\stackrel{3_g||\hat{4}_g}{\longrightarrow}g^6\frac{N_c^2-1}{N_c^2}\frac{P_{gg\leftarrow G}(z)}{s_{34}}\nonumber\\
&&\:\:\:\times\bigg[ N_c^2 \bigg(|\cmf(1_Q,(\widehat{4-3})_g,\hat{5}_g,\two)|^2+|\cmf(1_Q,\hat{5}_g,(\widehat{4-3})_g,\two)|^2\bigg)\nonumber\\
&&\:\:\:\:\:\:\:-|\cmf(1_Q,(\widehat{4-3})_{\gamma},\hat{5}_{\gamma},\two)|^2\bigg]{\rm d}\Phi_2 (k_{1_Q},k_{2\bar{Q}};p_{(4-3)g},p_{5g})J_2^{(2)}(k_1,k_2) \\
&&\:\:\:=g^2\:C_A\:\frac{P_{gg\rightarrow G}(z,\mu_{qg}^2)}{s_{34}}|M_4^0(1_Q,\two,(\widehat{4-3})_g,\hat{5}_g)|^2 \nonumber\\
&&\:\:\:\:\:\:\:\:\:\:\:\:\:\:\:\times {\rm d}\Phi_2 (k_{1Q},k_{2\bar{Q}};p_{(4-3)g},p_{5g})J_2^{(2)}(k_1,k_2).\nonumber
\end{eqnarray}
We have used eq.(\ref{eq:identity}) 
with incoming momenta $(p_4-k_3)$ and $p_5$. In the $3_g||\hat{4}_g$ limit, 
the contributing terms in eq.(\ref{eq:subtermttbarggg}) 
are proportional either to 
 $D_3^0(4_g;3_g,1_Q)$ or to $F_3^0(4_g,5_g;3_g)$. In the first case, 
the remapped momenta $(\tilde{13})$ and $\hat{\bar{4}}$ present 
in the reduced matrix elements multiplying by $D_3^0(4_g;3_g,1_Q)$ 
are given by $k_1$ and $(p_4-k_3)$ respectively. 
In the second case, the remapped momenta $\hat{\bar{4}}$ 
and $\hat{\bar{5}}$ present in the reduced matrix-element squared multiplying  $F_3^0(4_g,5_g;3_g)$, are given by $(p_4-k_3)$ and $p_5$ respectively.

Finally, for the $1_Q||\hat{4}_g$ limit, we see that
\begin{eqnarray}
&&{\rm d}\hat{\sigma}_{gg\rightarrow Q\bar{Q}g}\stackrel{1_Q||\hat{4}_g}{\longrightarrow}g^6\left(\frac{N_c^2-1}{N_c}\right)^2\frac{P_{q\bar{q}\leftarrow G}(z,\mu_{qg}^2)}{s_{14}}\nonumber\\
&&\:\:\:\times\bigg[ N_c^2 \bigg(|\cmf((\widehat{4-1})_{\bar{Q}},\hat{5}_g,3_g,\two)|^2+|\cmf((\widehat{4-1})_{\bar{Q}},3_g,\hat{5}_g,\two)|^2\bigg)\nonumber\\
&&\:\:\:\:\:\:\:-|\cmf((\widehat{4-1})_{\bar{Q}},3_{\gamma},\hat{5}_{\gamma},\two)|^2\bigg]{\rm d}\Phi_2 (k_{2\bar{Q}},k_{3g};p_{(4-1)\bar{Q}},p_{5g})J_2^{(2)}(k_2,k_3) \\
&&\:\:\:=g^2\:C_F\:\frac{P_{q\bar{q}\leftarrow G}(z,\mu_{qg}^2)}{s_{14}}  |M_4^0((\widehat{4-1})_{\bar{Q}},\two,3_g,\hat{5}_g)|^2 \nonumber\\
&&\:\:\:\:\:\:\:\:\:\:\:\:\:\:\:\times {\rm d}\Phi_2 (k_{2\bar{Q}},k_{3g};p_{(4-1)\bar{Q}},p_{5g})J_2^{(2)}(k_2,k_3).\nonumber
\label{eq:checkcol2}
\end{eqnarray}
Again, the relation between the full matrix element squared and 
the partial amplitudes given by eq.(\ref{eq:identity}) 
has been used. in this case the final state quark 
denoted by $1_Q$ and the gluon denoted by $4_g$ have been 
crossed to the initial state.

Since colour decomposition does not distinguish between initial 
and final state coloured particles, this crossings can be safely done 
with the relation still being true. The terms in eq.(\ref{eq:subtermttbarggg}) 
that contribute to this limit are those involving $A_3^0(4_g;1_Q,\two)$ 
and the corresponding remapped momenta appearing in the reduced squared
 amplitude multiplying this antenna 
become $k_{(\wt{12})}\rightarrow k_2$ and $p_{\hat{\bar{4}}}\rightarrow (p_4-k_1)$. 

These same verifications have been performed on all collinear 
and quasi-collinear limits of all subtraction terms listed 
in this section providing us with strong consistency check on our results
for the subtraction terms required to compute the cross section for 
$pp \rightarrow t\bar{t}$ at NLO.

\subsection{$t\bar{t}+jet$ production at LHC}
The real emission correction to the production of 
a massive quark-antiquark pair in association with a jet 
in a hadronic collision is given by,
\begin{eqnarray}
\lefteqn{{\rm d}\sigma^R=\int \frac{{\rm d}\xi_1}{\xi_1}\frac{{\rm d}\xi_2}{\xi_2}\bigg\{ \sum_q \left[ f_q(\xi_1)f_{\bar{q}}(\xi_2)\ds_{q\bar{q}\rightarrow Q\bar{Q}gg}+f_q(\xi_1)f_g(\xi_2)\ds_{qg\rightarrow Q\bar{Q}qg}\right.}\nonumber\\
&&\:\:\:\:\:\:\:\:\:\:\:\:\:\:\:\:\:\:\:\:\:\:\:\:\:\:\:\:\:\:\:\:\:\:\:\:\:\:\:\:\left.+f_{\bar{q}}(\xi_1)f_g(\xi_2)\ds_{\bar{q}g\rightarrow Q\bar{Q}\bar{q}g}+f_q(\xi_1)f_{\bar{q}}(\xi_2)\ds_{q\bar{q}\rightarrow Q\bar{Q}q\bar{q}}\right.\nonumber\\
&&\:\:\:\:\:\:\:\:\:\:\:\:\:\:\:\:\:\:\:\:\:\:\:\:\:\:\:\:\:\:\:\:\:\:\:\:\:\:\:\:\left.+f_q(\xi_1)f_q(\xi_2)\ds_{qq\rightarrow Q\bar{Q}qq}+f_{\bar{q}}(\xi_1)f_{\bar{q}}(\xi_2)\ds_{\bar{q}\bar{q}\rightarrow Q\bar{Q}\bar{q}\bar{q}}\right]\\
&&\:\:\:\:\:\:\:\:\:\:\:\:\:\:\:\:\:\:\:\:\:\:\:\:\:\:\:\:\:\:+\sum_{q\neq q'}\left[ f_q(\xi_1)f_q(\xi_2)\ds_{q\bar{q}\rightarrow Q\bar{Q}q'\bar{q}'}+f_q(\xi_1)f_{q'}(\xi_2)\ds_{qq'\rightarrow Q\bar{Q}qq'}\right.\nonumber\\
&&\:\:\:\:\:\:\:\:\:\:\:\:\:\:\:\:\:\:\:\:\:\:\:\:\:\:\:\:\:\:\:\:\:\:\:\:\:\:\:\:\left. +f_{\bar{q}}(\xi_q)f_{\bar{q}'}(\xi_2)\ds_{\bar{q}\bar{q}'\rightarrow Q\bar{Q}\bar{q}\bar{q}'}+f_q(\xi_1)f_{\bar{q}'}(\xi_2)\ds_{q\bar{q}'\rightarrow Q\bar{Q}q\bar{q}'}\right]\nonumber\\
&&\:\:\:\:\:\:\:\:\:\:\:\:\:\:\:\:\:\:\:\:\:\:\:\:\:\:\:\:\:\:\:\:+f_g(\xi_1)f_g(\xi_2)\left( \ds_{gg\rightarrow Q\bar{Q}gg}+ \ds_{gg\rightarrow Q\bar{Q}q\bar{q}}\right)\bigg\},\nonumber
\end{eqnarray}
with the partonic cross sections given by
\begin{eqnarray}
&&\ds_{q\bar{q}\rightarrow Q\bar{Q}gg}={\rm d}\Phi_4(k_Q,k_{\bar{Q}},k_{g_1},k_{g_2};p_q,p_{\bar{q}})|M^0_{q\bar{q}\rightarrow Q\bar{Q}gg}|^2 J_3^{(4)}(k_Q,k_{\bar{Q}},k_{g_1},k_{g_2})\\
&&\ds_{qg\rightarrow Q\bar{Q}qg}={\rm d}\Phi_4(k_Q,k_{\bar{Q}},k_g,k_q;p_q,p_g)|M^0_{qg\rightarrow Q\bar{Q}qg}|^2 J_3^{(4)}(k_Q,k_{\bar{Q}},k_q,k_g)\\
&&\ds_{q\bar{q}\rightarrow Q\bar{Q}q'\bar{q}'}={\rm d}\Phi_4(k_Q,k_{\bar{Q}},k_{q'},k_{\bar{q}'};p_q,p_{\bar{q}})|M^0_{q\bar{q}\rightarrow Q\bar{Q}q'\bar{q}'}|^2 J_3^{(4)}(k_Q,k_{\bar{Q}},k_{q'},k_{\bar{q}'})\\
&&\ds_{qq'\rightarrow Q\bar{Q}qq'}={\rm d}\Phi_4(k_Q,k_{\bar{Q}},k_{q},k_{q'};p_q,p_{q'})|M^0_{qq'\rightarrow Q\bar{Q}qq'}|^2 J_3^{(4)}(k_Q,k_{\bar{Q}},k_q,k_{q'})\\
&&\ds_{gg\rightarrow Q\bar{Q}gg}={\rm d}\Phi_4(k_Q,k_{\bar{Q}},k_{g_1},k_{g_2};p_{g_3},p_{g_4})|M^0_{gg\rightarrow Q\bar{Q}gg}|^2 J_3^{(4)}(k_Q,k_{\bar{Q}},k_{g_1},k_{g_2})\\
&&\ds_{gg\rightarrow Q\bar{Q}q\bar{q}}={\rm d}\Phi_4(k_Q,k_{\bar{Q}},k_q,k_{\bar{q}};p_{g_1},p_{g_2}) |M^0_{gg\rightarrow Q\bar{Q}q\bar{q}}|^2 J_3^{(4)}(k_Q,k_{\bar{Q}},k_q,k_{\bar{q}}).
\label{eq:partttjet}
\end{eqnarray}

The jet functions appearing in the partonic cross section 
are all of the type $J_{3}^{(4)}$ and correspond to the selection 
criteria of 3-jet events.
Out of four partons, from which 
two are a $Q\bar{Q}$ pair, an event with three jets is build. 
From these 3-jets, one jet is made of massless partons and 
two other jets have each  a heavy quark $Q$ or a heavy antiquark
 $\bar{Q}$ in them. In the corresponding subtraction terms,
the jet functions are of the type $J_{3}^{(3)}$ and those depend 
only on the hard final-state momenta appearing in the reduced 
matrix-element squared. 

In order to obtain the colour decomposition of the matrix element squared for the partonic processes defined above and given in eq.(\ref{eq:partttjet}),
 we will follow the same strategy as for $t \bar{t}$ production. 
We will use the colour decomposition of 
unphysical processes and then consider appropriate crossings.
For $t \bar{t}$ +1 jet production 
the unphysical processes to be considered are 
$0 \rightarrow Q\bar{Q}q \bar{q}q'\bar{q'}$, $0\rightarrow Q\bar{Q}q\bar{q}gg$ and $0\rightarrow Q\bar{Q}gggg$.

However, the presence of one additional parton in the final state 
introduces a few difficulties. In the first place, the number of partial amplitudes as well as the number of unresolved limits to subtract increases. 
Also, identical quark flavour contributions must be considered. 
But most importantly, in the contributions related 
to the partonic processes derived from 
$0 \rightarrow Q\bar{Q}q\bar{q}gg$ and $0\rightarrow Q\bar{Q}gggg$,  
the colour decomposition of the partonic amplitudes squared leads 
to interferences between partial amplitudes with different 
colour orderings that cannot be removed using any decoupling identities. 
Those require subtraction in an uncommon way in the antenna formalism, 
which we shall explain below.

\subsubsection{Interference terms}
\newcommand{\cmg}{{\cal M}}
The subtraction of infrared singularities in a colour ordered squared amplitude is easy within the antenna formalism: a suitable antenna function multiplied by a reduced squared matrix element with its momenta properly remapped accomplishes the task. However, this is not the case for interferences between
partial amplitudes with different coulour orderings.
Those interference terms which are most generally of the form 
$$\cmg_{n+1}^0(...,a,s,b,...)\cmg_{n+1}^0(...,c,s,d,...) ^{\dagger}$$
 with gluon $s$ colour connected to partons $a$ and $b$ in one amplitude and colour-connected to partons 
$c$ and $d$ in the other amplitude, lead to soft singularities 
when integrated over the phase space. 
Those singular behaviours cannot be straightforwardly 
subtracted with just one suitable antenna function multiplied 
by a reduced squared matrix element with remapped momenta. 

\parindent 1.5em
In order to understand how to subtract the soft singularities 
in these interferences terms, we appeal to the factorisation properties of colour ordered matrix elements at the amplitude level. Quite generally, when a gluon $s$ with helicity $\lambda$ becomes soft between partons $a$ and $b$, the colour ordered amplitude factorises \cite{Mangano,berendsgiele} as
\begin{equation}\label{eq:factamph}
\cmg_{n+1}^0(...,a,s_{\lambda},b,...)\stackrel{k_s\rightarrow 0}{\longrightarrow}\epsilon_{\lambda}(p_s)\cdot[J_a(p_s)-J_b(p_s)]\:\cmg_n^0(...,a,b,...),
\end{equation} 
where $\epsilon_{\lambda}(p_s)$ is the polarisation vector of the soft gluon, and $J^{\mu}_a(p_s)$ is the soft gluon current, given by
\begin{equation}
J_a^{\mu}(p_s)=\frac{p_a^{\mu}}{\sqrt{2}\:p_a\cdot p_s}.
\end{equation}
Eq. (\ref{eq:factamph}) holds in $d$ dimensions and is absolutely general: it does not depend on the identity of partons $a$ and $b$ (they can be either gluons, massive quarks or massless quarks) nor on their helicities. Therefore, summing over the helicities of the soft gluon, we find that the interference of two partial amplitudes with different colour orderings factorises in the soft limit, $k_s\rightarrow 0$, as
\begin{eqnarray}
\lefteqn{\sum_{\lambda=\pm}\cmg_{n+1}^0(...,a,s_{\lambda},b,...)\cmg_{n+1}^0(...,c,s_{\lambda},d,...)^{\dagger}}\label{eq:factamp}\nonumber\\
&&\stackrel{k_s\rightarrow 0}{\longrightarrow}\left( \sum_{\lambda=\pm}\epsilon^{\mu}_{\lambda}(p_s)\epsilon^{\nu}_{-\lambda}(p_s) \right)[J_a(p_s)-J_b(p_s)]_{\mu}[J_c(p_s)-J_d(p_s)]_{\nu}\label{eq:factamp}\\
&&\hspace{12mm}\times \cmg_{n}(...,a,b,...)\cmg_{n}(...,c,d,...)^{\dagger}.\nonumber
\end{eqnarray}
Since colour ordered amplitudes are gauge invariant, we can replace
\begin{equation}
 \sum_{\lambda=\pm}\epsilon^{\mu}_{\lambda}(p_s)\epsilon^{\nu}_{-\lambda}(p_s)\rightarrow -g^{\mu\nu}
\end{equation}
and thus obtain the following limit for the interference term:
\begin{eqnarray}\label{eikint}
\lefteqn{\cmg_{n+1}^0(...,a,s,b,...)\cmg_{n+1}^0(...,c,s,d,...)^{\dagger}}\\
&&\stackrel{k_s\rightarrow 0}{\longrightarrow}\left( \frac{s_{ad}}{s_{as}s_{ds}}+\frac{s_{bc}}{s_{bs}s_{cs}}-\frac{s_{ac}}{s_{as}s_{cs}}-\frac{s_{bd}}{s_{bs}s_{ds}}\right) \cmg_n^0(...,a,b,...)\cmg_n^0(...,c,d,...)^{\dagger}.\nonumber
\end{eqnarray}
In this soft limit, 
we see that the interference term given by 
$$\cmg_{n+1}^0(...,a,s,b,...)\cmg_{n+1}^0(...,c,s,d,...)^{\dagger}$$
factorises into a difference of four (half) eikonal factors times the
interference of the reduced (i.e where the soft gluon is absent)
interference term given by $$\cmg_n^0(...,a,b,...)\cmg_n^0(...,c,d,...)^{\dagger}.$$

Each of these half eikonal factors can be reproduced by a single
antenna function. The soft singularities of an interference term 
like
$\cmg_{n+1}^0(...,a,s,b,...)\cmg_{n+1}^0(...,c,s,d,...)^{\dagger}$ for
the case where $a$, $b$, $c$ and $d$ are all different can therefore be subtracted with
\begin{eqnarray}
&&\:\:\:\frac{1}{2}X_3^0(a,s,d) \cmg_{n,1}^0(...,\wt{as},\wt{bs},...)\cmg_{n,1}^0(...,\wt{cs},\wt{ds},...) ^{\dagger}\nonumber\\
&&+\frac{1}{2}X_3^0(b,s,c) \cmg_{n,2}^0(...,\wt{as},\wt{bs},...)\cmg_{n,2}^0(...,\wt{cs},\wt{ds},...) ^{\dagger}\label{eq:subtint}\\
&&-\frac{1}{2}X_3^0(a,s,c) \cmg_{n,3}^0(...,\wt{as},\wt{bs},...)\cmg_{n,3}^0(...,\wt{cs},\wt{ds},...) ^{\dagger}\nonumber\\
&&-\frac{1}{2}X_3^0(b,s,d) \cmg_{n,4}^0(...,\wt{as},\wt{bs},...)\cmg_{n,4}^0(...,\wt{cs},\wt{ds},...) ^{\dagger}.\nonumber
\end{eqnarray}
In this equation,
each of the antenna functions denoted as $X_{3}^0$ can be either
final-final, initial-final or initial-initial, depending on whether
the partons $a$, $b$, $c$ and $d$ are in the initial or final state,
and they can be either massive or massless depending on whether the
hard radiators are massive or massless. The absence of mass terms in
eq.(\ref{eikint}) is respected by eq.(\ref{eq:subtint}) even when one
of the hard radiators (and therefore all the antenna functions
involving that parton) is massive: 
In the soft limit, although a massive antenna yields a massive soft
eikonal factor including explicitly mass terms, as seen in Section 4,
in the combination of massive antennae required in
eq.(\ref{eq:subtint}) those mass terms cancel amongst each other. 

In eq.(\ref{eq:subtint}), the different labels on the reduced matrix elements  denoted by ${\cal M}^{(0)}_{n,i}$ do not mean that the matrix elements themselves are different, but that the momentum mapping is, in principle, different for each term. The remapping in each case is done in accordance with the type of antenna function involved.

\parindent 1.5em

A very important feature about this way of treating the interference terms is that not only soft singularities are subtracted with eq.(\ref{eq:subtint}), as it was originally intended, but also
no ``extra'' collinear singularities are introduced by this subtraction term itself.
The collinear limits of gluon $s$ collinear with partons $a$, $b$, $c$ and $d$ are correctly dealt with. Indeed, if $a$ and $b$ are different from $c$ and $d$, the interference term does not contain any collinear singularities.
It has square-root singularities that do not give rise to $\epsilon$ poles upon integration. In this case, the subtraction terms in eq.(\ref{eq:subtint}) do not contain any collinear singularities either: all the collinear limits introduced by the first two terms, are subtracted by the last two. And, in the case where two partons are equal in eq.(\ref{eq:factamp}),  for example, when $a$ and $c$ are the same, the interference term develops a collinear singularity when $a||s$. In the limit where $s$ becomes soft, setting $c=a$ in eq.(\ref{eq:factamp}) gives
\begin{eqnarray}
\lefteqn{\cmg_{n+1}^0(...,a,s,b,...)\cmg_{n+1}^0(...,a,s,d,...)^{\dagger}}\\
&&\stackrel{k_s\rightarrow 0}{\longrightarrow}\left( \frac{s_{ad}}{s_{as}s_{ds}}+\frac{s_{ab}}{s_{as}s_{bs}}-\frac{s_{bd}}{s_{bs}s_{ds}}-\frac{m_a^2}{s_{as}^2}\right)\cmg_n^0(...,a,b,...)\cmg_n^0(...,c,d,...) ^{\dagger}.\nonumber
\end{eqnarray}
The corresponding subtraction terms are
\begin{eqnarray}
&&\:\:\:\frac{1}{2}X_3^0(a,s,d) \cmg_{n,1}^0(...,\wt{as},\wt{bs},...)\cmg_{n,1}^0(...,\wt{as},\wt{ds},...) ^{\dagger}\nonumber\\
&&+\frac{1}{2}X_3^0(a,s,b) \cmg_{n,2}^0(...,\wt{as},\wt{bs},...)\cmg_{n,2}^0(...,\wt{as},\wt{ds},...) ^{\dagger}\\
&&-\frac{1}{2}X_3^0(b,s,d) \cmg_{n,3}^0(...,\wt{as},\wt{bs},...)\cmg_{n,3}^0(...,\wt{as},\wt{ds},...) ^{\dagger},\nonumber
\end{eqnarray}
from where it can be seen that these combination of subtraction terms also subtracts the $a||s$ (quasi)-collinear limit, in addition to the soft $s$ limit.

\parindent 1.5em

Taking this construction of the subtraction terms required for the 
interference terms of massive amplitudes into account, let us now present our results for the real contributions and their corresponding subtraction terms for all partonic processes involved in $t\bar{t}$+jet  production at NLO.  
We start with processes involving only quarks and derived from $0\rightarrow Q\bar{Q}q\bar{q}q'\bar{q}'$, which is the easiest case since only colour-ordered matrix-element squared are involved.  

\subsubsection{Processes derived from $0\rightarrow Q\bar{Q}q\bar{q}q'\bar{q}'$}\label{ttqqqq}
We choose to separate the colour decompositions of the real 
contributions related to processes derived from 
$0\rightarrow Q\bar{Q}q\bar{q}q'\bar{q}$
with and without identical quarks explicitly.

The colour decomposition for the unphysical partonic process 
$0\rightarrow Q\bar{Q}q\bar{q}q'\bar{q}'$
in the non-indentical flavour case ($q\neq q'$) is
given by,
\begin{eqnarray}
M_6^0(1_Q,2_{\bar{Q}},3_q,4_{\bar{q}},5_{q'},6_{\bar{q}'})&=&g^4\left[  \delta_{i_1,i_4}\delta_{i_3,i_6}\delta_{i_5,i_2}\cms(1_Q,4_{\bar{q}};;3_q,6_{\bar{q}'};5_{q'},2_{\bar{Q}})\right.\nonumber\\
&&\:\:\:\left. +\delta_{i_1,i_6}\delta_{i_3,i_2}\delta_{i_5,i_4}\cms(1_Q,6_{\bar{q}'};;3_q,2_{\bar{Q}};;5_{q'},4_{\bar{q}})\right.\nonumber\\
&&\left. -\frac{1}{N_c}\delta_{i_1,i_4}\delta_{i_3,i_2}\delta_{i_5,i_6}\cms(1_Q,4_{\bar{q}};;3_q,2_{\bar{Q}};;5_{q'},6_{\bar{q}'})\right.\nonumber\\
&&\left. -\frac{1}{N_c}\delta_{i_1,i_6}\delta_{i_3,i_4}\delta_{i_5,i_2}\cms(1_Q,6_{\bar{q}'};;3_q,4_{\bar{q}};;5_{q'},2_{\bar{Q}})\right.\\
&&\left. -\frac{1}{N_c}\delta_{i_1,i_2}\delta_{i_3,i_6}\delta_{i_5,i_4}\cms(1_Q,2_{\bar{Q}};;3_q,6_{\bar{q}'};;5_{q'},4_{\bar{q}})\right.\nonumber\\
&&\left. +\frac{1}{N_c^2}\delta_{i_1,i_2}\delta_{i_3,i_4}\delta_{i_5,i_6}\cms(1_Q,2_{\bar{Q}};;3_q,4_{\bar{q}};;5_{q'},6_{\bar{q}'})\right].\nonumber
\end{eqnarray} 
Squaring, and using the fact that the partial amplitudes satisfy
\begin{eqnarray}
\lefteqn{\cms(1_Q,2_{\bar{Q}};3_q,4_{\bar{q}};5_{q'},6_{\bar{q}'})}\nonumber\\
&&=\cms(1_Q,4_{\bar{q}};;3_q,6_{\bar{q}'};;5_{q'},2_{\bar{Q}})+\cms(1_Q,6_{\bar{q}'};;3_q,2_{\bar{Q}};;5_{q'},4_{\bar{q}})\\
&&=\frac{1}{2}\left(\cms(1_Q,4_{\bar{q}};;3_q,2_{\bar{Q}};5_{q'},6_{\bar{q}'})+\cms(1_Q,6_{\bar{q}'};;3_q,4_{\bar{q}};5_{q'},2_{\bar{Q}})+\cms(1_Q,2_{\bar{Q}};;3_q,6_{\bar{q}'};5_{q'},4_{\bar{q}})\right)\nonumber
\end{eqnarray}
gives
\begin{eqnarray}
\lefteqn{|M_6^0(1_Q,2_{\bar{Q}},3_q,4_{\bar{q}},5_{q'},6_{\bar{q}'})|^2=g^8(N_c^2-1)}\nonumber\\
&&\bigg\{ N_c \bigg( |\cms(1_Q,4_{\bar{q}};;3_q,6_{\bar{q}'};;5_{q'},2_{\bar{Q}})|^2+|\cms(1_Q,6_{\bar{q}'};;3_q,2_{\bar{Q}};;5_{q'},4_{\bar{q}})|^2\bigg) \nonumber\\
&&+\frac{1}{N_c}\bigg( |\cms(1_Q,4_{\bar{q}};;3_q,2_{\bar{Q}};;5_{q'},6_{\bar{q}'})|^2+|\cms(1_Q,6_{\bar{q}'};;3_q,4_{\bar{q}};;5_{q'},2_{\bar{Q}})|^2\label{eq:colourordqqqq}\\
&&\:\:\:\:\:\:\:\:\: +|\cms(1_Q,2_{\bar{Q}};;3_q,6_{\bar{q}'};;5_{q'},4_{\bar{q}})|^2-3|\cms(1_Q,2_{\bar{Q}};;3_q,4_{\bar{q}};;5_{q'},6_{\bar{q}'})|^2\bigg)\bigg\}\nonumber
\end{eqnarray}
which represents the colour decomposition for squared amplitude of the unphysical process $0\rightarrow Q\bar{Q}q \bar{q}q' \bar{q'}$ (with $q$ and $q'$ of different flavour). It has no interference terms. 
Depending on which partons are crossed in eq.(\ref{eq:colourordqqqq}) a sum over final state quark flavours might need to be performed. When a quark-antiquark pair of the same flavour is  crossed to the initial state resulting in the the colour ordered squared amplitude for the process $q\bar{q}\rightarrow Q\bar{Q}q' \bar{q'}$, the result must be multiplied \footnote{Note that the result needs to be multiplied by $(N_{F}-1)$ and not $N_{F}$  since we are explicitly separating the identical flavour contributions from the non-identical ones.} 
by $(N_{F}-1)$.

The identical flavour case is obtained taking the difference of two colour-ordered non-identical cases.  At the amplitude level,  we use the following relation
\begin{equation}
M_6^0(1_Q,2_{\bar{Q}},3_q,4_{\bar{q}},5_{q},6_{\bar{q}})=M_6^0(1_Q,2_{\bar{Q}},3_q,4_{\bar{q}},5_{q'},6_{\bar{q}'})-M_6^0(1_Q,2_{\bar{Q}},3_q,6_{\bar{q}},5_{q'},4_{\bar{q}'}),
\end{equation}
which, upon squaring, gives the colour decomposition of the matrix-element squared for the identical flavour case,
\newcommand{\re}{{\rm Re}}
\begin{eqnarray}
\lefteqn{|M_6^0(1_Q,2_{\bar{Q}},3_q,4_{\bar{q}},5_{q},6_{\bar{q}})|^2=g^8(N_c^2-1)}\nonumber\\
&&\times\bigg\{ N_c \bigg( |\cms(1_Q,4_{\bar{q}};;3_q,6_{\bar{q}'};;5_{q'},2_{\bar{Q}})|^2+|\cms(1_Q,6_{\bar{q}};;3_q,4_{\bar{q}'};;5_{q'},2_{\bar{Q}})|^2\nonumber\\
&&\:\:\:\:\:\:\:\:\:\:\:  +|\cms(1_Q,6_{\bar{q}'};;3_q,2_{\bar{Q}};;5_{q'},4_{\bar{q}})|^2+|\cms(1_Q,4_{\bar{q}'};;3_q,2_{\bar{Q}};;5_{q'},6_{\bar{q}})|^2  \bigg) \nonumber\\
&&\:\:\:\:+ 2\re(\cms(1_Q,4_{\bar{q}'};;3_q,2_{\bar{Q}};;5_{q'},6_{\bar{q}})\cms(1_Q,4_{\bar{q}};;3_q,2_{\bar{Q}};;5_{q'},6_{\bar{q}'})^{\dagger})\nonumber\\
&&\:\:\:\:+ 2\re(\cms(1_Q,4_{\bar{q}'};;3_q,6_{\bar{q}};;5_{q'},2_{\bar{Q}})\cms(1_Q,4_{\bar{q}};;3_q,6_{\bar{q}'};;5_{q'},2_{\bar{Q}})^{\dagger})\nonumber\\
&&\:\:\:\:+ 2\re(\cms(1_Q,6_{\bar{q}};;3_q,2_{\bar{Q}};;5_{q'},4_{\bar{q}'})\cms(1_Q,6_{\bar{q}'};;3_q,2_{\bar{Q}};;5_{q'},4_{\bar{q}})^{\dagger})\nonumber\\
&&\:\:\:\:+ 2\re(\cms(1_Q,6_{\bar{q}};;3_q,4_{\bar{q}'};;5_{q'},2_{\bar{Q}})\cms(1_Q,6_{\bar{q}'};;3_q,4_{\bar{q}};;5_{q'},2_{\bar{Q}})^{\dagger})\nonumber\\
&&\:\:\:\:- 2\re(\cms(1_Q,2_{\bar{Q}};;3_q,6_{\bar{q}};;5_{q'},4_{\bar{q}'})\cms(1_Q,2_{\bar{Q}};;3_q,4_{\bar{q}};;5_{q'},6_{\bar{q}'})^{\dagger})\label{eq:qqqqid}\\
&&\:\:+\frac{1}{N_c}\bigg( |\cms(1_Q,4_{\bar{q}};;3_q,2_{\bar{Q}};;5_{q'},6_{\bar{q}'})|^2+ |\cms(1_Q,6_{\bar{q}};;3_q,2_{\bar{Q}};;5_{q'},4_{\bar{q}'})|^2\nonumber\\
&&\:\:\:\:\:\:\:\:\:\:\: +|\cms(1_Q,6_{\bar{q}'};;3_q,4_{\bar{q}};;5_{q'},2_{\bar{Q}})|^2+|\cms(1_Q,4_{\bar{q}'};;3_q,6_{\bar{q}};;5_{q'},2_{\bar{Q}})|^2\nonumber\\
&&\:\:\:\:\:\:\:\:\:\:\: +|\cms(1_Q,2_{\bar{Q}};;3_q,6_{\bar{q}'};;5_{q'},4_{\bar{q}})|^2+|\cms(1_Q,2_{\bar{Q}};;3_q,4_{\bar{q}'};;5_{q'},6_{\bar{q}})|^2\nonumber\\
&&\:\:\:\:\:\:\:\:\:\:\: -3|\cms(1_Q,2_{\bar{Q}};;3_q,4_{\bar{q}};;5_{q'},6_{\bar{q}'})|^2-3|\cms(1_Q,2_{\bar{Q}};;3_q,6_{\bar{q}};;5_{q'},4_{\bar{q}'})|^2\bigg)\nonumber\\
&&\:\:-\frac{1}{N_c^2}\bigg(6\re(\cms(1_Q,2_{\bar{Q}};;3_q,6_{\bar{q}};;5_{q'},4_{\bar{q}'})\cms(1_Q,2_{\bar{Q}};;3_q,4_{\bar{q}};;5_{q'},6_{\bar{q}'})^{\dagger})\nonumber\\
&&\:\:\:\:\:\:\:\:\:\:\: -2\re(\cms(1_Q,2_{\bar{Q}};;3_q,6_{\bar{q}};;5_{q'},4_{\bar{q}'})\cms(1_Q,2_{\bar{Q}};;3_q,6_{\bar{q}'};;5_{q'},4_{\bar{q}})^{\dagger})\nonumber\\
&&\:\:\:\:\:\:\:\:\:\:\: -2\re(\cms(1_Q,2_{\bar{Q}};;3_q,4_{\bar{q}'};;5_{q'},6_{\bar{q}})\cms(1_Q,2_{\bar{Q}};;3_q,4_{\bar{q}};;5_{q'},6_{\bar{q}'})^{\dagger})\bigg)\bigg\}.\nonumber
\end{eqnarray}

The interference terms present in eq.(\ref{eq:qqqqid}) lead only 
to finite contributions when integrated over the phase space. Indeed, they lead 
to square root singularities which do not need subtraction. 

Furthermore, apart from those interference terms, the identical-quark flavour contributions involve all six colour-ordered matrix element squared appearing in the non-identical flavour case plus additional ones. 
In all cases, identical or non-identical quark and in all crossings required, 
the only singular behaviours which have to be captured by the antennae 
in the subtraction terms are final-final and initial-final collinear 
singularities between identical-flavour massless quark-(anti)quarks.
Due to the presence of a massive radiator in the final state, 
only one antenna, the quark-gluon massive antenna $E_{3}^{0}$ is required 
in final-final and initial-final configurations.

After the crossing of $3_q$ and $4_{\bar{q}}$ to the initial state, 
in the non-identical matrix element squared for 
the process $0 \rightarrow Q\bar{Q}q \bar{q}q'\bar{q}'$ 
given in eq.(\ref{eq:colourordqqqq})
we obtain the colour ordered matrix element squared for 
$q\bar{q}\rightarrow Q\bar{Q}q'\bar{q}'$.
Its subtraction term is,
\newcommand{\hatl}{\hat{l}}
\newcommand{\hatk}{\hat{k}_g}
\newcommand{\hatkph}{\hat{k}_{\gamma}}
\newcommand{\hatbl}{\hat{\bar{l}}}
\newcommand{\hatbk}{\hat{\bar{k}}}
\begin{eqnarray}
\lefteqn{{\rm d}\hat{\sigma}^S_{q\bar{q}\rightarrow Q\bar{Q}q'\bar{q}'}=\frac{g^8(N_c^2-1)(N_F-1)}{2}{\rm d}\Phi_4(k_{1Q},k_{2\bar{Q}},k_{5q'},k_{6\bar{q}'};p_{3\bar{q}},p_{4q})}\nonumber\\
&&\:\:\times\bigg\{ N_c\left[ E_3^0(1_Q,5_{q'},6_{\bar{q}'})\left( |\cm((\wt{15})_Q,\hat{4}_q;;\hat{3}_{\bar{q}},(\wt{56})_g,\two)|^2\right.\right.\nonumber\\
&&\:\:\:\:\:\:\:\:\:\:\:\:\:\:\:\:\:\:\:\:\left.\left. +|\cm((\wt{15})_Q,(\wt{56})_g,\hat{4}_q;;\hat{3}_{\bar{q}},\two)|^2\right)J_3^{(3)}(k_{\wt{15}},k_2,k_{\wt{56}})\right.\nonumber\\
&&\:\:\:\:\:\:\:\:\:\:\:\:\:\left.+E_3^0(\two,5_{q'},6_{\bar{q}'})\left( |\cm(1_Q,\hat{4}_q;;\hat{3}_{\bar{q}},(\wt{56})_g,(\wt{25})_{\bar{Q}})|^2\right.\right.\nonumber\\
&&\:\:\:\:\:\:\:\:\:\:\:\:\:\:\:\:\:\:\:\:\left.\left. +|\cm(1_Q,(\wt{56})_g,\hat{4}_q;;\hat{3}_{\bar{q}},(\wt{25})_{\bar{Q}})|^2\right)J_3^{(3)}(k_1,k_{\wt{25}},k_{\wt{56}})\right]\nonumber\\
&&\:\:\:\:+\frac{1}{N_c}\left[ E_3^0(1_Q,5_{q'},6_{\bar{q}'})\left( |\cm((\wt{15})_Q,\two;;\hat{3}_{\bar{q}},(\wt{56})_g,\hat{4}_q)|^2\right.\right.\nonumber\\
&&\:\:\:\:\:\:\:\:\:\:\:\:\:\:\:\:\:\:\:\:\left.\left. +|\cm((\wt{15})_Q,(\wt{56})_g,\two;;\hat{3}_{\bar{q}},\hat{4}_q)|^2\right.\right.\\
&&\:\:\:\:\:\:\:\:\:\:\:\:\:\:\:\:\:\:\:\:\left.\left. -2|\cm((\wt{15})_Q,\two,\hat{3}_{\bar{q}},\hat{4}_q,(\wt{56})_{\gamma})|^2\right)   J_3^{(3)}(k_{\wt{15}},k_2,k_{\wt{56}})\right.\nonumber\\
&&\:\:\:\:\:\:\:\:\:\:\:\:\:\left. +E_3^0(\two,5_{q'},6_{\bar{q}'})\left( |\cm(1_Q,(\wt{25})_{\bar{Q}};;\hat{3}_{\bar{q}},(\wt{56})_g,\hat{4}_q)|^2\right.\right.\nonumber\\
&&\:\:\:\:\:\:\:\:\:\:\:\:\:\:\:\:\:\:\:\:\left.\left. +|\cm(1_Q,(\wt{56})_g,(\wt{25})_{\bar{Q}};;\hat{3}_{\bar{q}},\hat{4}_q)|^2\right.\right.\nonumber\\
&&\:\:\:\:\:\:\:\:\:\:\:\:\:\:\:\:\:\:\:\:\left.\left. -2|\cm(1_Q,(\wt{25})_{\bar{Q}},\hat{3}_{\bar{q}},\hat{4}_q,(\wt{56})_{\gamma})|^2\right)J_3^{(3)}(k_1,k_{\wt{25}},k_{\wt{56}})\right]\bigg\}.\nonumber
\end{eqnarray}

If, instead, $4_{\bar{q}}$ and $6_{\bar{q}'}$ are crossed in eq.(\ref{eq:colourordqqqq}), we obtain the squared matrix element for the process $qq'\rightarrow Q\bar{Q}qq'$. The corresponding subtraction term reads
\begin{eqnarray}
\lefteqn{{\rm d}\hat{\sigma}^S_{qq'\rightarrow Q\bar{Q}qq'}=\frac{g^8(N_c^2-1)}{2}{\rm d}\Phi_4(k_{1Q},k_{2\bar{Q}},k_{3q},k_{5q'};p_{4q},p_{6q'})}\nonumber\\
&&\:\:\times\bigg\{ N_c \left[ E_3^0(4_q;3_q,1_Q)\left( |\cm((\wt{13})_Q,\hat{6}_{q'};;5_{q'},\hat{\bar{4}}_g,\two)|^2\right.\right.\nonumber\\
&&\:\:\:\:\:\:\:\:\:\:\:\:\:\:\:\:\:\:\:\:\left.\left.+|\cm((\wt{13})_Q,\hat{\bar{4}}_g,\hat{6}_{q'};;5_{q'},\two)|^2\right)J_3^{(3)}(k_{\wt{13}},k_2,k_5)\right.\nonumber\\
&&\:\:\:\:\:\:\:\:\:\:\:\:\:\left. +E_3^0(4_q;3_q,\two)\left( |\cm(1_Q,\hat{6}_{q'};;5_{q'},\hat{\bar{4}}_g,(\wt{23})_{\bar{Q}})|^2\right.\right.\nonumber\\
&&\:\:\:\:\:\:\:\:\:\:\:\:\:\:\:\:\:\:\:\:\left.\left.+|\cm(1_Q,\hat{\bar{4}}_g,\hat{6}_{q'};;5_{q'},(\wt{23})_{\bar{Q}})|^2\right)J_3^{(3)}(k_1,k_{\wt{23}},k_5)\right.\nonumber\\
&&\:\:\:\:\:\:\:\:\:\:\:\:\:\left. +E_3^0(6_{q'};5_{q'},1_Q)\left( |\cm((\wt{15})_Q,\hat{4}_q;;3_q,\hat{\bar{6}}_g,\two)|^2\right.\right.\nonumber\\
&&\:\:\:\:\:\:\:\:\:\:\:\:\:\:\:\:\:\:\:\:\left.\left. +|\cm((\wt{15})_Q,\hat{\bar{6}}_g,\hat{4}_q;;3_q,\two)|^2\right)J_3^{(3)}(k_{\wt{15}},k_2,k_3)\right.\nonumber\\
&&\:\:\:\:\:\:\:\:\:\:\:\:\:\left. +E_3^0(6_{q'};5_{q'},\two)\left( |\cm(1_Q,\hat{4}_q;;3_q,\hat{\bar{6}}_g,(\wt{25})_{\bar{Q}})|^2\right.\right.\nonumber\\
&&\:\:\:\:\:\:\:\:\:\:\:\:\:\:\:\:\:\:\:\:\left.\left. +|\cm(1_Q,\hat{\bar{6}}_g,\hat{4}_q;;3_q,(\wt{25})_{\bar{Q}})|^2\right)J_3^{(3)}(k_1,k_{\wt{25}},k_3)\right]\nonumber\\
&&\:\:\:\: +\frac{1}{N_c}\left[ E_3^0(4_q;3_q,1_Q)\left( |\cm((\wt{13})_Q,\two;;5_{q'},\hat{\bar{4}}_g,\hat{6}_{q'})|^2\right.\right.\\
&&\:\:\:\:\:\:\:\:\:\:\:\:\:\:\:\:\:\:\:\:\left.\left.+|\cm((\wt{13})_Q,\hat{\bar{4}}_g,\two;;5_{q'},\hat{6}_{q'})|^2\right.\right.\nonumber\\
&&\:\:\:\:\:\:\:\:\:\:\:\:\:\:\:\:\:\:\:\:\left.\left.-2|\cm((\wt{13})_Q,\two,5_{q'},\hat{6}_{q'},\hat{\bar{4}}_{\gamma})|^2 \right)J_3^{(3)}(k_{\wt{13}},k_2,k_5)\right.\nonumber\\
&&\:\:\:\:\:\:\:\:\:\:\:\:\:\left. +E_3^0(4_q;3_q,\two)\left( |\cm(1_Q,(\wt{23})_{\bar{Q}};;5_{q'},\hat{\bar{4}}_g,\hat{6}_{q'})|^2\right.\right.\nonumber\\
&&\:\:\:\:\:\:\:\:\:\:\:\:\:\:\:\:\:\:\:\:\left.\left.+|\cm(1_Q,\hat{\bar{4}}_g,(\wt{23})_{\bar{Q}};;5_{q'},\hat{6}_{q'})|^2\right.\right.\nonumber\\
&&\:\:\:\:\:\:\:\:\:\:\:\:\:\:\:\:\:\:\:\:\left.\left.-2|\cm(1_Q,(\wt{23})_{\bar{Q}},5_{q'},\hat{6}_{q'},\hat{\bar{4}}_{\gamma})|^2\right)J_3^{(3)}(k_1,k_{\wt{23}},k_5)\right.\nonumber\\
&&\:\:\:\:\:\:\:\:\:\:\:\:\:\left. +E_3^0(6_{q'};5_{q'},1_Q)\left( |\cm((\wt{15})_Q,\two;;3_q,\hat{\bar{6}}_g,\hat{4}_q)|^2\right.\right.\nonumber\\
&&\:\:\:\:\:\:\:\:\:\:\:\:\:\:\:\:\:\:\:\:\left.\left. +|\cm((\wt{15})_Q,\hat{\bar{6}}_g,\two;;3_q,\hat{4}_q)|^2\right.\right.\nonumber\\
&&\:\:\:\:\:\:\:\:\:\:\:\:\:\:\:\:\:\:\:\:\left.\left. -2 |\cm((\wt{15})_Q,\two,3_q,\hat{4}_q,\hat{\bar{6}}_{\gamma})|^2\right)J_3^{(3)}(k_{\wt{15}},k_2,k_3)\right.\nonumber\\
&&\:\:\:\:\:\:\:\:\:\:\:\:\:\left. +E_3^0(6_{q'};5_{q'},\two)\left( |\cm(1_Q,(\wt{25})_{\bar{Q}};;3_q,\hat{\bar{6}}_g,\hat{4}_q)|^2\right.\right.\nonumber\\
&&\:\:\:\:\:\:\:\:\:\:\:\:\:\:\:\:\:\:\:\:\left.\left. +|\cm(1_Q,\hat{\bar{6}}_g,(\wt{25})_{\bar{Q}};;3_q,\hat{4}_q)|^2\right.\right.\nonumber\\
&&\:\:\:\:\:\:\:\:\:\:\:\:\:\:\:\:\:\:\:\:\left.\left. +|\cm(1_Q,(\wt{25})_{\bar{Q}},3_q,\hat{4}_q,\hat{\bar{6}}_{\gamma})|^2 \right)J_3^{(3)}(k_1,k_{\wt{25}},k_3)\right]\bigg\}.\nonumber
\end{eqnarray}

For the identical flavour-case, the same two crossings, 
$3_q$ and $4_{\bar{q}}$ on one hand  and the crossing of $4_{\bar{q}}$ and $6_{\bar{q}}$ on the other hand  need to be considered.
As the real matrix-element in the identical quark case are given by those 
in the non-identical quark case plus additional terms, similarly the subtraction terms for the identical-quark case are obtained by adding to the subtraction terms valid for the non-identical case additional subtraction terms related to the new colour ordered matrix-element squared appearing only in the identical-flavour case only and given in eq.(\ref{eq:qqqqid}). 

If $3_q$ and $4_{\bar{q}}$ are crossed in the identical flavour squared matrix element given in eq.(\ref{eq:qqqqid}), the squared matrix element for $q\bar{q}\rightarrow Q\bar{Q}q\bar{q}$ is obtained in a colour ordered way. The subtraction term for this process is,
\begin{eqnarray}
\lefteqn{{\rm d}\hat{\sigma}^S_{q\bar{q}\rightarrow Q\bar{Q}q\bar{q}}=\frac{g^8(N_c^2-1)}{2}{\rm d}\Phi_4(k_{1Q},k_{2\bar{Q}},k_{5q},k_{6\bar{q}};p_{3\bar{q}},p_{4q})}\nonumber\\
&&\:\:\times\bigg\{ N_c\left[ E_3^0(1_Q,5_q,6_{\bar{q}})\left( |\cm((\wt{15})_Q,\hat{4}_q;;\hat{3}_{\bar{q}},(\wt{56})_g,\two)|^2\right.\right.\nonumber\\
&&\:\:\:\:\:\:\:\:\:\:\:\:\:\:\:\:\:\:\:\:\left.\left. +|\cm((\wt{15})_Q,(\wt{56})_g,\hat{4}_q;;\hat{3}_{\bar{q}},\two)|^2\right)J_3^{(3)}(k_{\wt{15}},k_2,k_{\wt{56}})\right.\nonumber\\
&&\:\:\:\:\:\:\:\:\:\:\:\:\:\left.+E_3^0(\two,5_q,6_{\bar{q}})\left( |\cm(1_Q,\hat{4}_q;;\hat{3}_{\bar{q}},(\wt{56})_g,(\wt{25})_{\bar{Q}})|^2\right.\right.\nonumber\\
&&\:\:\:\:\:\:\:\:\:\:\:\:\:\:\:\:\:\:\:\:\left.\left. +|\cm(1_Q,(\wt{56})_g,\hat{4}_q;;\hat{3}_{\bar{q}},(\wt{25})_{\bar{Q}})|^2\right)J_3^{(3)}(k_1,k_{\wt{25}},k_{\wt{56}})\right.\nonumber\\
&&\:\:\:\:\:\:\:\:\:\:\:\:\:\left.+E_3^0(4_q;5_q,1_Q)\left( |\cm((\wt{15})_Q,6_{\bar{q}};;\hat{3}_{\bar{q}},\hat{\bar{4}}_g,\two)|^2\right.\right.\nonumber\\
&&\:\:\:\:\:\:\:\:\:\:\:\:\:\:\:\:\:\:\:\:\left.\left. +|\cm((\wt{15})_Q,\hat{\bar{4}}_g,6_{\bar{q}};;\hat{3}_{\bar{q}},\two)|^2\right)J_3^{(3)}(k_{\wt{15}},k_2,k_6)\right.\nonumber\\
&&\:\:\:\:\:\:\:\:\:\:\:\:\:\left.+E_3^0(4_q;5_q,2_{\bar{Q}})\left( |\cm(1_Q,6_{\bar{q}};;\hat{3}_{\bar{q}},\hat{\bar{4}}_g,(\wt{25})_{\bar{Q}})|^2\right.\right.\nonumber\\
&&\:\:\:\:\:\:\:\:\:\:\:\:\:\:\:\:\:\:\:\:\left.\left. +|\cm(1_Q,\hat{\bar{4}}_g,6_{\bar{q}};;\hat{3}_{\bar{q}},(\wt{25})_{\bar{Q}})|^2\right)J_3^{(3)}(k_1,k_{\wt{25}},k_6)\right.\nonumber\\
&&\:\:\:\:\:\:\:\:\:\:\:\:\:\left.+E_3^0(3_{\bar{q}};6_{\bar{q}},1_Q)\left( |\cm((\wt{16})_Q,\hat{4}_q;;5_q,\hat{\bar{3}}_g,\two)|^2\right.\right.\nonumber\\
&&\:\:\:\:\:\:\:\:\:\:\:\:\:\:\:\:\:\:\:\:\left.\left. +|\cm((\wt{16})_Q,\hat{\bar{3}}_g,\hat{4}_q;;5_q,\two)|^2\right)J_3^{(3)}(k_{\wt{16}},k_2,k_5)\right.\nonumber\\
&&\:\:\:\:\:\:\:\:\:\:\:\:\:\left.+E_3^0(3_{\bar{q}};6_{\bar{q}},2_{\bar{Q}})\left( |\cm(1_Q,\hat{4}_q;;5_q,\hat{\bar{3}}_g,(\wt{26})_{\bar{Q}})|^2\right.\right.\nonumber\\
&&\:\:\:\:\:\:\:\:\:\:\:\:\:\:\:\:\:\:\:\:\left.\left. +|\cm(1_Q,\hat{\bar{3}}_g,\hat{4}_q;;5_q,(\wt{26})_{\bar{Q}})|^2\right)J_3^{(3)}(k_1,k_{\wt{26}},k_5)\right]\nonumber\\
&&\:\:\:\:+\frac{1}{N_c}\left[ E_3^0(1_Q,5_q,6_{\bar{q}})\left( |\cm((\wt{15})_Q,\two;;\hat{3}_{\bar{q}},(\wt{56})_g,\hat{4}_q)|^2\right.\right.\\
&&\:\:\:\:\:\:\:\:\:\:\:\:\:\:\:\:\:\:\:\:\left.\left. +|\cm((\wt{15})_Q,(\wt{56})_g,\two;;\hat{3}_{\bar{q}},\hat{4}_q)|^2\right.\right.\nonumber\\
&&\:\:\:\:\:\:\:\:\:\:\:\:\:\:\:\:\:\:\:\:\left.\left. -2|\cm((\wt{15})_Q,\two,\hat{3}_{\bar{q}},\hat{4}_q,(\wt{56})_{\gamma})|^2\right)   J_3^{(3)}(k_{\wt{15}},k_2,k_{\wt{56}})\right.\nonumber\\
&&\:\:\:\:\:\:\:\:\:\:\:\:\:\left. +E_3^0(\two,5_q,6_{\bar{q}})\left( |\cm(1_Q,(\wt{25})_{\bar{Q}};;\hat{3}_{\bar{q}},(\wt{56})_g,\hat{4}_q)|^2\right.\right.\nonumber\\
&&\:\:\:\:\:\:\:\:\:\:\:\:\:\:\:\:\:\:\:\:\left.\left. +|\cm(1_Q,(\wt{56})_g,(\wt{25})_{\bar{Q}};;\hat{3}_{\bar{q}},\hat{4}_q)|^2\right.\right.\nonumber\\
&&\:\:\:\:\:\:\:\:\:\:\:\:\:\:\:\:\:\:\:\:\left.\left. -2|\cm(1_Q,(\wt{25})_{\bar{Q}},\hat{3}_{\bar{q}},\hat{4}_q,(\wt{56})_{\gamma})|^2\right)J_3^{(3)}(k_1,k_{\wt{25}},k_{\wt{56}})\right.\nonumber\\
&&\:\:\:\:\:\:\:\:\:\:\:\:\:\left. +E_3^0(4_q;5_q,1_Q)\left( |\cm((\wt{15})_Q,\two;;\hat{3}_{\bar{q}},\hat{\bar{4}}_g,6_{\bar{q}})|^2\right.\right.\nonumber\\
&&\:\:\:\:\:\:\:\:\:\:\:\:\:\:\:\:\:\:\:\:\left.\left. +|\cm((\wt{15})_Q,\hat{\bar{4}}_g,\two;;\hat{3}_{\bar{q}},6_{\bar{q}})|^2\right.\right.\nonumber\\
&&\:\:\:\:\:\:\:\:\:\:\:\:\:\:\:\:\:\:\:\:\left.\left. -2|\cm((\wt{15})_Q,\two,\hat{3}_{\bar{q}},6_{\bar{q}},\hat{\bar{4}}_{\gamma})|^2\right)J_3^{(3)}(k_{\wt{15}},k_2,k_6)\right.\nonumber\\
&&\:\:\:\:\:\:\:\:\:\:\:\:\:\left. +E_3^0(4_q;5_q,\two)\left( |\cm(1_Q,(\wt{25})_{\bar{Q}};;\hat{3}_{\bar{q}},\hat{\bar{4}}_g,6_{\bar{q}})|^2\right.\right.\nonumber\\
&&\:\:\:\:\:\:\:\:\:\:\:\:\:\:\:\:\:\:\:\:\left.\left. +|\cm(1_Q,\hat{\bar{4}}_g,(\wt{25})_{\bar{Q}};;\hat{3}_{\bar{q}},6_{\bar{q}})|^2\right.\right.\nonumber\\
&&\:\:\:\:\:\:\:\:\:\:\:\:\:\:\:\:\:\:\:\:\left.\left. -2|\cm(1_Q,(\wt{25})_{\bar{Q}},\hat{3}_{\bar{q}},6_{\bar{q}},\hat{\bar{4}}_{\gamma})|^2\right)J_3^{(3)}(k_1,k_{\wt{25}},k_6)\right.\nonumber\\
&&\:\:\:\:\:\:\:\:\:\:\:\:\:\left. +E_3^0(3_{\bar{q}};6_{\bar{q}},1_Q)\left( |\cm((\wt{16})_Q,\two;;5_q,\hat{\bar{3}}_g,\hat{4}_q)|^2\right.\right.\nonumber\\
&&\:\:\:\:\:\:\:\:\:\:\:\:\:\:\:\:\:\:\:\:\left.\left. +|\cm((\wt{16})_Q,\hat{\bar{3}}_g,\two;;5_q,\hat{4}_{\bar{q}})|^2\right.\right.\nonumber\\
&&\:\:\:\:\:\:\:\:\:\:\:\:\:\:\:\:\:\:\:\:\left.\left. -2|\cm((\wt{16})_Q,\two,5_q,\hat{4}_q,\hat{\bar{3}}_{\gamma})|^2\right)J_3^{(3)}(k_{\wt{16}},k_2,k_5)\right.\nonumber\\
&&\:\:\:\:\:\:\:\:\:\:\:\:\:\left. +E_3^0(3_{\bar{q}};6_{\bar{q}},\two)\left( |\cm(1_Q,(\wt{26})_{\bar{Q}};;5_q,\hat{\bar{3}}_g,\hat{4}_q)|^2\right.\right.\nonumber\\
&&\:\:\:\:\:\:\:\:\:\:\:\:\:\:\:\:\:\:\:\:\left.\left. +|\cm(1_Q,\hat{\bar{3}}_g,(\wt{26})_{\bar{Q}};;5_q,\hat{4}_q)|^2\right.\right.\nonumber\\
&&\:\:\:\:\:\:\:\:\:\:\:\:\:\:\:\:\:\:\:\:\left.\left. -2|\cm(1_Q,(\wt{26})_{\bar{Q}},5_q,\hat{4}_q,\hat{\bar{3}}_{\gamma})|^2\right)J_3^{(3)}(k_1,k_{\wt{26}},k_5)\right]\bigg\}.\nonumber
\end{eqnarray}

The squared matrix element for $qq\rightarrow Q\bar{Q}qq$ in terms of colour ordered partial amplitudes is obtained by crossing $4_{\bar{q}}$ and $6_{\bar{q}}$ in  eq.(\ref{eq:qqqqid}). In this case, the subtraction term reads
\begin{eqnarray}
\lefteqn{{\rm d}\hat{\sigma}^S_{qq\rightarrow Q\bar{Q}qq}=\frac{g^8(N_c^2-1)}{2}{\rm d}\Phi_4(k_{1Q},k_{2\bar{Q}},k_{3q},k_{5q};p_{4q},p_{6q})}\nonumber\\
&&\:\:\times\bigg\{ N_c \left[ E_3^0(4_q;3_q,1_Q)\left( |\cm((\wt{13})_Q,\hat{6}_q;;5_q,\hat{\bar{4}}_g,\two)|^2\right.\right.\nonumber\\
&&\:\:\:\:\:\:\:\:\:\:\:\:\:\:\:\:\:\:\:\:\left.\left.+|\cm((\wt{13})_Q,\hat{\bar{4}}_g,\hat{6}_q;;5_q,\two)|^2\right)J_3^{(3)}(k_{\wt{13}},k_2,k_5)\right.\nonumber\\
&&\:\:\:\:\:\:\:\:\:\:\:\:\:\left. +E_3^0(4_q;3_q,\two)\left( |\cm(1_Q,\hat{6}_q;;5_q,\hat{\bar{4}}_g,(\wt{23})_{\bar{Q}})|^2\right.\right.\nonumber\\
&&\:\:\:\:\:\:\:\:\:\:\:\:\:\:\:\:\:\:\:\:\left.\left.+|\cm(1_Q,\hat{\bar{4}}_g,\hat{6}_q;;5_q,(\wt{23})_{\bar{Q}})|^2\right)J_3^{(3)}(k_1,k_{\wt{23}},k_5)\right.\nonumber\\
&&\:\:\:\:\:\:\:\:\:\:\:\:\:\left. +E_3^0(4_q;5_q,1_Q)\left( |\cm((\wt{15})_Q,\hat{6}_q;;3_q,\hat{\bar{4}}_g,\two)|^2\right.\right.\nonumber\\
&&\:\:\:\:\:\:\:\:\:\:\:\:\:\:\:\:\:\:\:\:\left.\left.+|\cm((\wt{15})_Q,\hat{\bar{4}}_g,\hat{6}_q;;3_q,\two)|^2\right)J_3^{(3)}(k_{\wt{15}},k_2,k_3)\right.\nonumber\\
&&\:\:\:\:\:\:\:\:\:\:\:\:\:\left. +E_3^0(4_q;5_q,\two)\left( |\cm(1_Q,\hat{6}_q;;3_q,\hat{\bar{4}}_g,(\wt{25})_{\bar{Q}})|^2\right.\right.\nonumber\\
&&\:\:\:\:\:\:\:\:\:\:\:\:\:\:\:\:\:\:\:\:\left.\left.+|\cm(1_Q,\hat{\bar{4}}_g,\hat{6}_q;;3_q,(\wt{25})_{\bar{Q}})|^2\right)J_3^{(3)}(k_1,k_{\wt{25}},k_3)\right.\nonumber\\
&&\:\:\:\:\:\:\:\:\:\:\:\:\:\left. +E_3^0(6_q;5_q,1_Q)\left( |\cm((\wt{15})_Q,\hat{4}_q;;3_q,\hat{\bar{6}}_g,\two)|^2\right.\right.\nonumber\\
&&\:\:\:\:\:\:\:\:\:\:\:\:\:\:\:\:\:\:\:\:\left.\left. +|\cm((\wt{15})_Q,\hat{\bar{6}}_g,\hat{4}_q;;3_q,\two)|^2\right)J_3^{(3)}(k_{\wt{15}},k_2,k_3)\right.\nonumber\\
&&\:\:\:\:\:\:\:\:\:\:\:\:\:\left. +E_3^0(6_q;3_q,1_Q)\left( |\cm((\wt{13})_Q,\hat{4}_q;;5_q,\hat{\bar{6}}_g,\two)|^2\right.\right.\nonumber\\
&&\:\:\:\:\:\:\:\:\:\:\:\:\:\:\:\:\:\:\:\:\left.\left. +|\cm((\wt{13})_Q,\hat{\bar{6}}_g,\hat{4}_q;;5_q,\two)|^2\right)J_3^{(3)}(k_{\wt{13}},k_2,k_5)\right.\nonumber\\
&&\:\:\:\:\:\:\:\:\:\:\:\:\:\left. +E_3^0(6_q;5_q,\two)\left( |\cm(1_Q,\hat{4}_q;;3_q,\hat{\bar{6}}_g,(\wt{25})_{\bar{Q}})|^2\right.\right.\nonumber\\
&&\:\:\:\:\:\:\:\:\:\:\:\:\:\:\:\:\:\:\:\:\left.\left. +|\cm(1_Q,\hat{\bar{6}}_g,\hat{4}_q;;3_q,(\wt{25})_{\bar{Q}})|^2\right)J_3^{(3)}(k_1,k_{\wt{25}},k_3)\right]\nonumber\\
&&\:\:\:\:\:\:\:\:\:\:\:\:\:\left. +E_3^0(6_q;3_q,\two)\left( |\cm(1_Q,\hat{4}_q;;5_q,\hat{\bar{6}}_g,(\wt{23})_{\bar{Q}})|^2\right.\right.\nonumber\\
&&\:\:\:\:\:\:\:\:\:\:\:\:\:\:\:\:\:\:\:\:\left.\left. +|\cm(1_Q,\hat{\bar{6}}_g,\hat{4}_q;;5_q,(\wt{23})_{\bar{Q}})|^2\right)J_3^{(3)}(k_1,k_{\wt{23}},k_5)\right]\nonumber\\
&&\:\:\:\: +\frac{1}{N_c}\left[ E_3^0(4_q;3_q,1_Q)\left( |\cm((\wt{13})_Q,\two;;5_q,\hat{\bar{4}}_g,\hat{6}_q)|^2\right.\right.\\
&&\:\:\:\:\:\:\:\:\:\:\:\:\:\:\:\:\:\:\:\:\left.\left.+|\cm((\wt{13})_Q,\hat{\bar{4}}_g,\two;;5_q,\hat{6}_q)|^2\right.\right.\nonumber\\
&&\:\:\:\:\:\:\:\:\:\:\:\:\:\:\:\:\:\:\:\:\left.\left.-2|\cm((\wt{13})_Q,\two,5_q,\hat{6}_q,\hat{\bar{4}}_{\gamma})|^2 \right)J_3^{(3)}(k_{\wt{13}},k_2,k_5)\right.\nonumber\\
&&\:\:\:\:\:\:\:\:\:\:\:\:\:\left. +E_3^0(4_q;3_q,\two)\left( |\cm(1_Q,(\wt{23})_{\bar{Q}};;5_q,\hat{\bar{4}}_g,\hat{6}_q)|^2\right.\right.\nonumber\\
&&\:\:\:\:\:\:\:\:\:\:\:\:\:\:\:\:\:\:\:\:\left.\left.+|\cm(1_Q,\hat{\bar{4}}_g,(\wt{23})_{\bar{Q}};;5_q,\hat{6}_q)|^2\right.\right.\nonumber\\
&&\:\:\:\:\:\:\:\:\:\:\:\:\:\:\:\:\:\:\:\:\left.\left.-2|\cm(1_Q,(\wt{23})_{\bar{Q}},5_q,\hat{6}_q,\hat{\bar{4}}_{\gamma})|^2\right)J_3^{(3)}(k_1,k_{\wt{23}},k_5)\right.\nonumber\\
&&\:\:\:\:\:\:\:\:\:\:\:\:\:\left. +E_3^0(4_q;5_q,1_Q)\left( |\cm((\wt{15})_Q,\two;;3_q,\hat{\bar{4}}_g,\hat{6}_q)|^2\right.\right.\nonumber\\
&&\:\:\:\:\:\:\:\:\:\:\:\:\:\:\:\:\:\:\:\:\left.\left.+|\cm((\wt{15})_Q,\hat{\bar{4}}_g,\two;;3_q,\hat{6}_q)|^2\right.\right.\nonumber\\
&&\:\:\:\:\:\:\:\:\:\:\:\:\:\:\:\:\:\:\:\:\left.\left.-2|\cm((\wt{15})_Q,\two,3_q,\hat{6}_q,\hat{\bar{4}}_{\gamma})|^2 \right)J_3^{(3)}(k_{\wt{15}},k_2,k_3)\right.\nonumber\\
&&\:\:\:\:\:\:\:\:\:\:\:\:\:\left. +E_3^0(4_q;5_q,\two)\left( |\cm(1_Q,(\wt{25})_{\bar{Q}};;3_q,\hat{\bar{4}}_g,\hat{6}_q)|^2\right.\right.\nonumber\\
&&\:\:\:\:\:\:\:\:\:\:\:\:\:\:\:\:\:\:\:\:\left.\left.+|\cm(1_Q,\hat{\bar{4}}_g,(\wt{25})_{\bar{Q}};;3_q,\hat{6}_q)|^2\right.\right.\nonumber\\
&&\:\:\:\:\:\:\:\:\:\:\:\:\:\:\:\:\:\:\:\:\left.\left.-2|\cm(1_Q,(\wt{25})_{\bar{Q}},3_q,\hat{6}_q,\hat{\bar{4}}_{\gamma})|^2\right)J_3^{(3)}(k_1,k_{\wt{25}},k_3)\right.\nonumber\\
&&\:\:\:\:\:\:\:\:\:\:\:\:\:\left. +E_3^0(6_q;5_q,1_Q)\left( |\cm((\wt{15})_Q,\two;;3_q,\hat{\bar{6}}_g,\hat{4}_q)|^2\right.\right.\nonumber\\
&&\:\:\:\:\:\:\:\:\:\:\:\:\:\:\:\:\:\:\:\:\left.\left. +|\cm((\wt{15})_Q,\hat{\bar{6}}_g,\two;;3_q,\hat{4}_q)|^2\right.\right.\nonumber\\
&&\:\:\:\:\:\:\:\:\:\:\:\:\:\:\:\:\:\:\:\:\left.\left. -2 |\cm((\wt{15})_Q,\two,3_q,\hat{4}_q,\hat{\bar{6}}_{\gamma})|^2\right)J_3^{(3)}(k_{\wt{15}},k_2,k_3)\right.\nonumber\\
&&\:\:\:\:\:\:\:\:\:\:\:\:\:\left. +E_3^0(6_q;5_q,\two)\left( |\cm(1_Q,(\wt{25})_{\bar{Q}};;3_q,\hat{\bar{6}}_g,\hat{4}_q)|^2\right.\right.\nonumber\\
&&\:\:\:\:\:\:\:\:\:\:\:\:\:\:\:\:\:\:\:\:\left.\left. +|\cm(1_Q,\hat{\bar{6}}_g,(\wt{25})_{\bar{Q}};;3_q,\hat{4}_q)|^2\right.\right.\nonumber\\
&&\:\:\:\:\:\:\:\:\:\:\:\:\:\:\:\:\:\:\:\:\left.\left. +|\cm(1_Q,(\wt{25})_{\bar{Q}},3_q,\hat{4}_q,\hat{\bar{6}}_{\gamma})|^2 \right)J_3^{(3)}(k_1,k_{\wt{25}},k_3)\right.\nonumber\\
&&\:\:\:\:\:\:\:\:\:\:\:\:\:\left. +E_3^0(6_q;3_q,1_Q)\left( |\cm((\wt{13})_Q,\two;;5_q,\hat{\bar{6}}_g,\hat{4}_q)|^2\right.\right.\nonumber\\
&&\:\:\:\:\:\:\:\:\:\:\:\:\:\:\:\:\:\:\:\:\left.\left. +|\cm((\wt{13})_Q,\hat{\bar{6}}_g,\two;;5_q,\hat{4}_q)|^2\right.\right.\nonumber\\
&&\:\:\:\:\:\:\:\:\:\:\:\:\:\:\:\:\:\:\:\:\left.\left. -2 |\cm((\wt{13})_Q,\two,5_q,\hat{4}_q,\hat{\bar{6}}_{\gamma})|^2\right)J_3^{(3)}(k_{\wt{13}},k_2,k_5)\right.\nonumber\\
&&\:\:\:\:\:\:\:\:\:\:\:\:\:\left. +E_3^0(6_q;3_q,\two)\left( |\cm(1_Q,(\wt{23})_{\bar{Q}};;5_q,\hat{\bar{6}}_g,\hat{4}_q)|^2\right.\right.\nonumber\\
&&\:\:\:\:\:\:\:\:\:\:\:\:\:\:\:\:\:\:\:\:\left.\left. +|\cm(1_Q,\hat{\bar{6}}_g,(\wt{23})_{\bar{Q}};;5_q,\hat{4}_q)|^2\right.\right.\nonumber\\
&&\:\:\:\:\:\:\:\:\:\:\:\:\:\:\:\:\:\:\:\:\left.\left. +|\cm(1_Q,(\wt{23})_{\bar{Q}},5_q,\hat{4}_q,\hat{\bar{6}}_{\gamma})|^2 \right)J_3^{(3)}(k_1,k_{\wt{23}},k_5)\right]\bigg\}.\nonumber
\end{eqnarray}

\subsubsection{Processes derived from $0\rightarrow Q\bar{Q}q\bar{q}gg$}
The colour decomposition for the unphysical process $0\rightarrow Q\bar{Q}q\bar{q}gg$ is 
\begin{eqnarray}
&&M_6^0(1_Q,2_{\bar{Q}},3_q,4_{\bar{q}},5_g,6_g)\nonumber\\
&&\:\:\:\: =2g^4\sum_{(i,j)\in P(5,6)}\bigg\{ \delta_{i_3,i_2}(T^{a_i}T^{a_j})_{i_1,i_4}\cms(1_Q,i_g,j_g,4_{\bar{q}};;3_q,\two)\nonumber\\
&&\:\:\:\:\:\:\:\:\:\:\:\:\:\:\:\:\:\:\:\:\:\:\:\:\:\:\:\:\:\:\:\: +(T^{a_i})_{i_1,i_4}(T^{a_j})_{i_3,i_2}\cms(1_Q,i_g,4_{\bar{q}};;3_q,j_g,\two)\nonumber\\
&&\:\:\:\:\:\:\:\:\:\:\:\:\:\:\:\:\:\:\:\:\:\:\:\:\:\:\:\:\:\:\:\:+ \delta_{i_1,i_4}(T^{a_i}T^{a_j})_{i_3,i_2}\cms(1_Q,4_{\bar{q}};;3_q,i_g,j_g,\two)\\
&&\:\:\:\:\:\:\:\:\:\:\:\:\:\:\:\:\:\:\:\:\:\:\:\:\:\:-\frac{1}{N_c}\delta_{i_3,i_4}(T^{a_i}T^{a_j})_{i_1,i_2}\cms(1_Q,i_g,j_g,\two;;3_q,4_{\bar{q}})\nonumber\\
&&\:\:\:\:\:\:\:\:\:\:\:\:\:\:\:\:\:\:\:\:\:\:\:\:\:\:-\frac{1}{N_c} (T^{a_i})_{i_1,i_2}(T^{a_j})_{i_3,i_4}\cms(1_Q,i_g,\two;;3_q,j_g,4_{\bar{q}})\nonumber\\
&&\:\:\:\:\:\:\:\:\:\:\:\:\:\:\:\:\:\:\:\:\:\:\:\:\:\:-\frac{1}{N_c} \delta_{i_1,i_2}(T^{a_i}T^{a_j})_{i_3,i_4}\cms(1_Q,\two;;3_q,i_g,j_g,4_{\bar{q}})\bigg\},\nonumber
\end{eqnarray}
and the partial amplitudes satisfy
\begin{eqnarray}
\lefteqn{\cms(1_Q,2_{\bar{Q}},3_q,4_{\bar{q}},5_{\gamma},6_{\gamma})}\nonumber\\
&&=\sum_{(i,j)\in P(5,6)} \left( \cms(1_Q,i_g,j_g,4_{\bar{q}};;3_q,\two)+\cms(1_Q,i_g,4_{\bar{q}};;3_q,j_g,\two)\right.\nonumber\\
&&\:\:\:\:\:\:\:\:\:\:\:\:\:\:\:\:\:\:\:\:\:\:\:\:\:\:\:\left.+\cms(1_Q,4_{\bar{q}};;3_q,i_g,j_g,\two)\right)\label{eq:idqqgg}\\
&&=\sum_{(k,l)\in P(5,6)} \left( \cms(1_Q,k_g,l_g,\two;;3_q,4_{\bar{q}})+\cms(1_Q,k_g,\two;;3_q,l_g,4_{\bar{q}})\right.\nonumber\\
&&\:\:\:\:\:\:\:\:\:\:\:\:\:\:\:\:\:\:\:\:\:\:\:\:\:\:\:\left.+\cms(1_Q,\two;;3_q,k_g,l_g,4_{\bar{q}})\right).\nonumber
\end{eqnarray}
Squaring, and using the identity eq.(\ref{eq:idqqgg}) to rearrange the result, we obtain
\begin{eqnarray}
\lefteqn{|M_6^0(1_Q,2_{\bar{Q}},3_q,4_{\bar{q}},5_g,6_g)|^2=\frac{g^8(N_c^2-1)}{N_c^2}}\nonumber\\
&&\times\sum_{(i,j)\in P(5,6)}\bigg\{ \nonumber\\
&&\:\:\:\:\:\:\:\:\:\:\:(N_c^2-1)\bigg[ N_c^2\bigg( |\cms(1_Q,i_g,j_g,4_{\bar{q}};;3_q,\two)|^2+ |\cms(1_Q,i_g,4_{\bar{q}};;3_q,j_g,\two)|^2\nonumber\\
&&\:\:\:\:\:\:\:\:\:\:\:\:\:\:\:\:\:\:\:\:\:\:\:\:\:\:\:\:\:\:\:\:\:\:\:\:\: +|\cms(1_Q,4_{\bar{q}};;3_q,i_g,j_g,2_{\bar{Q}})|^2\bigg)\nonumber\\
&&\:\:\:\:\:\:\:\:\:\:\:\:\:\:\:\:\:\:\:\:\:\:\:\:\:\:\:\: + |\cms(1_Q,i_g,j_g,\two;;3_q,4_{\bar{q}})|^2+ |\cms(1_Q,i_g,\two;;3_q,j_g,4_{\bar{q}})|^2\nonumber\\
&&\:\:\:\:\:\:\:\:\:\:\:\:\:\:\:\:\:\:\:\:\:\:\:\:\:\:\:\: + |\cms(1_Q,\two;;3_q,i_g,j_g,4_{\bar{q}})|^2\bigg]+|\cms(1_Q,\two,3_q,4_{\bar{q}},i_{\gamma},j_{\gamma})|^2 \nonumber\\
&&\:\:\:\:\:\:\:\:\:\:+N_c^2 \bigg[ 2\re(\cms(1_Q,i_g,j_g,4_{\bar{q}};;3_q,\two)\cms(1_Q,4_{\bar{q}};;3_q,i_g,j_g,2_{\bar{Q}})^{\dagger})\nonumber\\
&&\:\:\:\:\:\:\:\:\:\:\:\: +2\re(\cms(1_Q,i_g,j_g,4_{\bar{q}};;3_q,\two)\cms(1_Q,4_{\bar{q}};;3_q,j_g,i_g,2_{\bar{Q}})^{\dagger})\nonumber\\
&&\:\:\:\:\:\:\:\:\:\:\:\: +\re(\cms(1_Q,i_g,4_{\bar{q}};;3_q,j_g,\two) \cms(1_Q,j_g,4_{\bar{q}};;3_q,i_g,\two)^{\dagger})\nonumber\\
&&\:\:\:\:\:\:\:\:\:\:\:\: -\re(\cms(1_Q,i_g,j_g,4_{\bar{q}};;3_q,\two)\cms(1_Q,j_g,i_g,4_{\bar{q}};;3_q,\two)^{\dagger})\nonumber\\
&&\:\:\:\:\:\:\:\:\:\:\:\: -\re( \cms(1_Q,4_{\bar{q}};;3_q,i_g,j_g,2_{\bar{Q}})\cms(1_Q,4_{\bar{q}};;3_q,j_g,i_g,2_{\bar{Q}})^{\dagger})\nonumber\\
&&\:\:\:\:\:\:\:\:\:\:\:\: -2\re(\cms(1_Q,i_g,j_g,4_{\bar{q}};;3_q,\two)\cms(1_Q,i_g,j_g,\two;;3_q,4_{\bar{q}})^{\dagger})\nonumber\\
&&\:\:\:\:\:\:\:\:\:\:\:\: -2\re(\cms(1_Q,i_g,4_{\bar{q}};;3_q,j_g,\two)\cms(1_Q,i_g,j_g,\two;;3_q,4_{\bar{q}}))^{\dagger})\nonumber\\
&&\:\:\:\:\:\:\:\:\:\:\:\: -2\re( \cms(1_Q,4_{\bar{q}};;3_q,i_g,j_g,2_{\bar{Q}})\cms(1_Q,i_g,j_g,\two;;3_q,4_{\bar{q}})^{\dagger})\nonumber\\
&&\:\:\:\:\:\:\:\:\:\:\:\: -2\re( \cms(1_Q,i_g,j_g,4_{\bar{q}};;3_q,\two)\cms(1_Q,i_g,\two;;3_q,j_g,4_{\bar{q}})^{\dagger})\label{eq:qqgg}\\
&&\:\:\:\:\:\:\:\:\:\:\:\: -2\re(\cms(1_Q,i_g,4_{\bar{q}};;3_q,j_g,\two)\cms(1_Q,i_g,\two;;3_q,j_g,4_{\bar{q}})^{\dagger})\nonumber\\
&&\:\:\:\:\:\:\:\:\:\:\:\: -2\re(\cms(1_Q,i_g,4_{\bar{q}};;3_q,j_g,\two)\cms(1_Q,j_g,\two;;3_q,i_g,4_{\bar{q}})^{\dagger})\nonumber\\
&&\:\:\:\:\:\:\:\:\:\:\:\: -2\re(\cms(1_Q,4_{\bar{q}};;3_q,i_g,j_g,2_{\bar{Q}}) \cms(1_Q,j_g,\two;;3_q,i_g,4_{\bar{q}})^{\dagger})\nonumber\\
&&\:\:\:\:\:\:\:\:\:\:\:\: -2\re(\cms(1_Q,i_g,j_g,4_{\bar{q}};;3_q,\two) \cms(1_Q,\two;;3_q,i_g,j_g,4_{\bar{q}}))^{\dagger})\nonumber\\
&&\:\:\:\:\:\:\:\:\:\:\:\: -2\re(\cms(1_Q,4_{\bar{q}};;3_q,i_g,j_g,2_{\bar{Q}})\cms(1_Q,\two;;3_q,i_g,j_g,4_{\bar{q}})^{\dagger})\nonumber\\
&&\:\:\:\:\:\:\:\:\:\:\:\: -2\re(\cms(1_Q,i_g,4_{\bar{q}};;3_q,j_g,\two)\cms(1_Q,\two;;3_q,j_g,i_g,4_{\bar{q}})^{\dagger})\bigg]\nonumber\\
&&\:\:\:\:\:\:\:\:\:\:\:+ \re(\cms(1_Q,i_g,\two;;3_q,j_g,4_{\bar{q}})\cms(1_Q,j_g,\two;;3_q,i_g,4_{\bar{q}})^{\dagger})\nonumber\\
&&\:\:\:\:\:\:\:\:\:\:\:- \re(\cms(1_Q,i_g,j_g,\two;;3_q,4_{\bar{q}})\cms(1_Q,j_g,i_g,\two;;3_q,4_{\bar{q}})^{\dagger})\nonumber\\
&&\:\:\:\:\:\:\:\:\:\:\:- \re(\cms(1_Q,\two;;3_q,i_g,j_g,4_{\bar{q}})\cms(1_Q,\two;;3_q,j_g,i_g,4_{\bar{q}})^{\dagger})\nonumber\\
&&\:\:\:\:\:\:\:\:\:\:\: +2\re (\cms(1_Q,i_g,j_g,\two;;3_q,4_{\bar{q}}) \cms(1_Q,\two;;3_q,i_g,j_g,4_{\bar{q}})^{\dagger})\nonumber\\
&&\:\:\:\:\:\:\:\:\:\:\: +2\re (\cms(1_Q,i_g,j_g,\two;;3_q,4_{\bar{q}}) \cms(1_Q,\two;;3_q,j_g,i_g,4_{\bar{q}})^{\dagger})\bigg\}.\nonumber
\end{eqnarray}

Besides colour-ordered amplitudes squared this expression includes 
interference terms between amplitudes with different colour orderings which  
require subtraction. 

After appropriate crossings, for each of the colour-ordered matrix-element squared appearing on the right hand side of this equation, 
an antenna capturing the single unresolved radiation is determined.
For the soft radiation behaviour present in the interference terms, 
the subtraction terms are constructed with the difference 
of four antennae as explained in Section 7.2.1.
For the subtraction terms, we will need all types of massive 
final-final and initial final antennae defined in Section 3  
in addition to massless antennae  in all three (final-final, 
initial-final and initial-initial) configurations which were defined 
in \cite{ourant,Daleo,Joao}.

By crossing $3_q$ and $4_{\bar{q}}$ in eq.(\ref{eq:qqgg}) to the initial state, the matrix element for the process $q\bar{q}\rightarrow Q\bar{Q}gg$ is obtained. From the unresolved limits of this squared matrix element we obtain the following subtraction term,

\newcommand{\hatf}{\hat{4}_q}
\newcommand{\hatt}{\hat{3}_{\bar{q}}}
\newcommand{\hatbf}{\hat{\bar{4}}_q}
\newcommand{\hatbt}{\hat{\bar{3}}_{\bar{q}}}

\begin{eqnarray}
\lefteqn{{\rm d}\hat{\sigma}_{q\bar{q}\rightarrow Q\bar{Q}gg}^S=g^8(N_c^2-1){\rm d}\Phi_4(k_{1Q},k_{2\bar{Q}},k_{5g},k_{6g};p_{3\bar{q}},p_{4q})\sum_{(i,j)\in(5,6)}\bigg\{ }\nonumber\\
&&\:\:\:N_c^2\left[ A_3^0(4_q;1_Q,i_g)|\cm((\wt{1i})_Q,\hat{\bar{4}}_q;;\hat{3}_{\bar{q}},j_g,2_{\bar{Q}})|^2 J_3^{(3)}(k_{\wt{1i}},k_2,k_j)\right.\nonumber\\
&&\left. \:\:\:\:\:\:\:+A_3^0(3_{\bar{q}};2_{\bar{Q}},i_g)|\cm(1_Q,j_g,\hat{4}_q;;\hat{\bar{3}}_{\bar{q}},(\wt{2i})_{\bar{Q}})|^2 J_3^3(k_1,k_{\wt{2i}},k_j)\right.\nonumber\\
&&\left. \:\:\:\:\:\:\:+d_3^0(1_Q,i_g,j_g)|\cm((\wt{1i})_Q,(\wt{ij})_g,\hat{4}_q;;\hat{3}_{\bar{q}},2_{\bar{Q}})|^2 J_3^{(3)}(k_{\wt{1i}},k_{\wt{ij}},k_2)\right.\nonumber\\
&&\left. \:\:\:\:\:\:\:+d_3^0(2_{\bar{Q}},i_g,j_g)|\cm(1_Q,\hatf;;\hatt,(\wt{ij})_g,(\wt{2i})_{\bar{Q}})|^2 J_3^{(3)}(k_1,k_{\wt{2i}},k_{\wt{ij}})\right.\nonumber\\
&&\left. \:\:\:\:\:\:\:+d_3^0(4_q;j_g,i_g)|\cm(1_Q,(\wt{ij})_g,\hatbf;;\hatt,2_{\bar{Q}})|^2 J_3^{(3)}(k_1,k_2,k_{\wt{ij}})\right.\nonumber\\
&&\left. \:\:\:\:\:\:\:+d_3^0(3_{\bar{q}};i_g,j_g)|\cm(1_Q,\hatf;;\hatbt,(\wt{ij})_g,2_{\bar{Q}})|^2 J_3^{(3)}(k_1,k_2,k_{\wt{ij}})\right]\nonumber\\
&&\:\:\:+A_3^0(1_Q,i_g,2_{\bar{Q}})\left( |\cm((\wt{1i})_Q,(\wt{2i})_{\bar{Q}};;\hatt,j_g,\hatf)|^2 \right.\nonumber\\
&&\:\:\:\:\:\:\:\:\:\:\:\:\:\: \left. +2\re(\cm((\wt{1i})_Q,\hatf;;\hatt,j_g,(\wt{2i})_{\bar{Q}})\cm((\wt{1i})_Q,j_g,\hatf;;\hatt,(\wt{2i})_{\bar{Q}})^{\dagger})\right.\nonumber\\
&&\:\:\:\:\:\:\:\:\:\:\:\:\:\: \left. -2\re(\cm((\wt{1i})_Q,(\wt{2i})_{\bar{Q}};;\hatt,j_g,\hatf)\cm((\wt{1i})_Q,\hatf;;\hatt,j_g,(\wt{2i})_{\bar{Q}})^{\dagger})\right.\nonumber\\
&&\:\:\:\:\:\:\:\:\:\:\:\:\:\: \left. -2\re(\cm((\wt{1i})_Q,(\wt{2i})_{\bar{Q}};;\hatt,j_g,\hatf)\cm((\wt{1i})_Q,j_g,\hatf;;\hatt,(\wt{2i})_{\bar{Q}})^{\dagger})\right)J_3^{(3)}(k_{\wt{1i}},k_{\wt{2i}},k_j)\nonumber\\
&&\:\:\:-A_3^0(4_q;1_Q,i_g)\left( |\cm((\wt{1i})_Q,j_g,\hatbf;;\hatt,2_{\bar{Q}})|^2+|\cm((\wt{1i})_Q,\hatbf;;\hatt,j_g,2_{\bar{Q}})|^2 \right.\nonumber\\
&&\:\:\:\:\:\:\:\:\:\:\:\:\:\: \left. +2\re(\cm((\wt{1i})_Q,\hatbf;;\hatt,j_g,2_{\bar{Q}})\cm((\wt{1i})_Q,2_{\bar{Q}};;\hatt,j_g,\hatbf)^{\dagger})\right.\nonumber\\
&&\:\:\:\:\:\:\:\:\:\:\:\:\:\: \left. +2\re(\cm((\wt{1i})_Q,\hatbf;;\hatt,j_g,2_{\bar{Q}})\cm((\wt{1i})_Q,j_g,2_{\bar{Q}};;\hatt,\hatbf)^{\dagger})\right)J_3^{(3)}(k_{\wt{1i}},k_2,k_j)\nonumber\\
&&\:\:\:+A_3^0(3_{\bar{q}};1_Q,i_g)\left( 2\re( \cm((\wt{1i})_Q,\hatf;;\hatbt,j_g,2_{\bar{Q}}) \cm((\wt{1i})_Q,j_g,2_{\bar{Q}};;\hatbt,\hatf)^{\dagger})\right.\nonumber\\
&&\:\:\:\:\:\:\:\:\:\:\:\:\:\: \left. +2\re( \cm((\wt{1i})_Q,j_g,\hatf;;\hatbt,2_{\bar{Q}}) \cm((\wt{1i})_Q,2_{\bar{Q}};;\hatbt,j_g,\hatf)^{\dagger})\right.\nonumber\\
&&\:\:\:\:\:\:\:\:\:\:\:\:\:\: \left. -2\re( \cm((\wt{1i})_Q,j_g,\hatf;;\hatbt,2_{\bar{Q}}) \cm((\wt{1i})_Q,\hatf;;\hatbt,j_g,2_{\bar{Q}})^{\dagger})\right)J_3^{(3)}(k_{\wt{1i}},k_2,k_j)\nonumber\\
&&\:\:\:+A_3^0(4_q;2_{\bar{Q}},i_g)\left( 2\re(\cm(1_Q,\hatbf;;\hatt,j_g,(\wt{2i})_{\bar{Q}}) \cm(1_Q,(\wt{2i})_{\bar{Q}};;\hatt,j_g,\hatbf)^{\dagger})\right.\nonumber\\
&&\:\:\:\:\:\:\:\:\:\:\:\:\:\: \left. +2\re(\cm(1_Q,j_g,\hatbf;;\hatt,(\wt{2i})_{\bar{Q}}) \cm(1_Q,j_g,(\wt{2i})_{\bar{Q}};;\hatt,\hatbf)^{\dagger})\right.\nonumber\\
&&\:\:\:\:\:\:\:\:\:\:\:\:\:\: \left. -2\re(\cm(1_Q,\hatbf;;\hatt,j_g,(\wt{2i})_{\bar{Q}}) \cm(1_Q,j_g,\hatbf;;\hatt(\wt{2i})_{\bar{Q}})^{\dagger})\right)J_3^{(3)}(k_1,k_{\wt{2i}},k_j)\nonumber\\
&&\:\:\:-A_3^0(3_{\bar{q}};2_{\bar{Q}},i_g)\left( |\cm(1_Q,\hatf;;\hatbt,j_g,(\wt{2i})_{\bar{Q}})|^2+|\cm(1_Q,j_g,\hatf;;\hatbt,(\wt{2i})_{\bar{Q}})|^2\right.\nonumber\\
&&\:\:\:\:\:\:\:\:\:\:\:\:\:\: \left. +2\re(\cm(1_Q,j_g,\hatf;;\hatbt,(\wt{2i})_{\bar{Q}}) \cm(1_Q,(\wt{2i})_{\bar{Q}};;\hatbt,j_g,\hatf)^{\dagger})\right.\nonumber\\
&&\:\:\:\:\:\:\:\:\:\:\:\:\:\: \left. +2\re(\cm(1_Q,j_g,\hatf;;\hatbt,(\wt{2i})_{\bar{Q}}) \cm(1_Q,j_g,(\wt{2i})_{\bar{Q}};;\hatbt,\hatf)^{\dagger})\right)J_3^{(3)}(k_1,k_{\wt{2i}},k_j)\nonumber\\
&&\:\:\:+A_3^0(3_{\bar{q}},4_q;i_g)\left( |\cm(\tilde{1}_Q,\tilde{j}_g,\tilde{2}_{\bar{Q}};;\hatbt,\hatbf)|^2\right.\nonumber\\
&&\:\:\:\:\:\:\:\:\:\:\:\:\:\: \left. +2\re(\cm(\tilde{1}_Q,\hatbf;;\hatbt,\tilde{j}_g,\tilde{2}_{\bar{Q}})\cm(\tilde{1}_Q,j_g,\hatbf;;\hatbt,\tilde{2}_{\bar{Q}})^{\dagger})\right.\nonumber\\
&&\:\:\:\:\:\:\:\:\:\:\:\:\:\: \left. -2\re(\cm(\tilde{1}_Q,\hatbf;;\hatbt,\tilde{j}_g,\tilde{2}_{\bar{Q}})\cm(\tilde{1}_Q,\tilde{j}_g,\tilde{2}_{\bar{Q}};;\hatbt,\hatbf)^{\dagger})\right.\nonumber\\
&&\:\:\:\:\:\:\:\:\:\:\:\:\:\: \left. -2\re(\cm(\tilde{1}_Q,\tilde{j}_g,\hatbf;;\hatbt,\tilde{2}_{\bar{Q}})\cm(\tilde{1}_Q,\tilde{j}_g,\tilde{2}_{\bar{Q}};;\hatbt,\hatbf)^{\dagger})\right)J_3^{(3)}(\tilde{k}_1,\tilde{k}_2,\tilde{k}_j)\nonumber\\
&&\:\:\:+d_3^0(1_Q,i_g,j_g)\left( |\cm((\wt{1i})_Q,(\wt{ij})_g,2_{\bar{Q}};;\hatt,\hatf)|^2 \right.\nonumber\\
&&\:\:\:\:\:\:\:\:\:\:\:\:\:\: \left. +2\re(\cm((\wt{1i})_Q,\hatf;;\hatt,(\wt{ij})_g,2_{\bar{Q}}) \cm((\wt{1i})_Q,2_{\bar{Q}};;\hatt,(\wt{ij})_g,\hatf)^{\dagger})\right.\nonumber\\
&&\:\:\:\:\:\:\:\:\:\:\:\:\:\: \left. -2\re(\cm((\wt{1i})_Q,\hatf;;\hatt,(\wt{ij})_g,2_{\bar{Q}}) \cm((\wt{1i})_Q,(\wt{ij})_g,2_{\bar{Q}};;\hatt,\hatf)^{\dagger})\right.\nonumber\\
&&\:\:\:\:\:\:\:\:\:\:\:\:\:\: \left. -2\re(\cm((\wt{1i})_Q,(\wt{ij})_g,\hatf;;\hatt,2_{\bar{Q}}) \cm((\wt{1i})_Q,2_{\bar{Q}};;\hatt,(\wt{ij})_g,\hatf)^{\dagger})\right.\nonumber\\
&&\:\:\:\:\:\:\:\:\:\:\:\:\:\: \left. -2\re(\cm((\wt{1i})_Q,(\wt{ij})_g,\hatf;;\hatt,2_{\bar{Q}}) \cm((\wt{1i})_Q,(\wt{ij})_g,2_{\bar{Q}};;\hatt,\hatf)^{\dagger})\right)J_3^{(3)}(k_{\wt{1i}},k_2,k_{\wt{ij}})\nonumber\\
&&\:\:\:+d_3^0(2_{\bar{Q}},i_g,j_g) \left( |\cm(1_Q,(\wt{ij})_g,(\wt{2i})_{\bar{Q}};;\hatt,\hatf)|^2 \right.\nonumber\\
&&\:\:\:\:\:\:\:\:\:\:\:\:\:\: \left. +2\re( \cm(1_Q,(\wt{ij})_g,\hatf;;\hatt,(\wt{2i})_{\bar{Q}}) \cm(1_Q, (\wt{2i})_{\bar{Q}};;\hatt,(\wt{ij})_g,\hatf)^{\dagger})\right.\nonumber\\
&&\:\:\:\:\:\:\:\:\:\:\:\:\:\: \left. -2\re( \cm(1_Q,\hatf;;\hatt,(\wt{ij})_g,(\wt{2i})_{\bar{Q}}) \cm(1_Q, (\wt{2i})_{\bar{Q}};;\hatt,(\wt{ij})_g,\hatf)^{\dagger})\right.\nonumber\\
&&\:\:\:\:\:\:\:\:\:\:\:\:\:\: \left. -2\re( \cm(1_Q,\hatf;;\hatt,(\wt{ij})_g,(\wt{2i})_{\bar{Q}}) \cm(1_Q, (\wt{ij})_g,(\wt{2i})_{\bar{Q}};;\hatt,\hatf)^{\dagger})\right.\nonumber\\
&&\:\:\:\:\:\:\:\:\:\:\:\:\:\: \left. -2\re( \cm(1_Q,(\wt{ij})_g,\hatf;;\hatt,(\wt{2i})_{\bar{Q}}) \cm(1_Q, (\wt{ij})_g,(\wt{2i})_{\bar{Q}};;\hatt,\hatf)^{\dagger})\right)J_3^{(3)}(k_1,k_{\wt{2i}},k_{\wt{ij}})\nonumber\\
&&\:\:\:+d_3^0(4_q;i_g,j_g)\left( |\cm(1_Q,2_{\bar{Q}};;\hatt,(\wt{ij})_g,\hatbf)|^2 \right.\nonumber\\
&&\:\:\:\:\:\:\:\:\:\:\:\:\:\: \left. +2\re( \cm(1_Q,\hatbf;;\hatt,(\wt{ij})_g,2_{\bar{Q}})\cm(1_Q,(\wt{ij})_g,2_{\bar{Q}};;\hatt,\hatbf)^{\dagger})\right.\nonumber\\
&&\:\:\:\:\:\:\:\:\:\:\:\:\:\: \left. -2\re( \cm(1_Q,\hatbf;;\hatt,(\wt{ij})_g,2_{\bar{Q}})\cm(1_Q,2_{\bar{Q}};;\hatt,(\wt{ij})_g,\hatbf)^{\dagger})\right.\nonumber\\
&&\:\:\:\:\:\:\:\:\:\:\:\:\:\: \left. -2\re( \cm(1_Q,(\wt{ij})_g,\hatbf;;\hatt,2_{\bar{Q}})\cm(1_Q,2_{\bar{Q}};;\hatt,(\wt{ij})_g,\hatbf)^{\dagger})\right.\nonumber\\
&&\:\:\:\:\:\:\:\:\:\:\:\:\:\: \left. -2\re( \cm(1_Q,(\wt{ij})_g,\hatbf;;\hatt,2_{\bar{Q}})\cm(1_Q,(\wt{ij})_g,2_{\bar{Q}};;\hatt,\hatbf)^{\dagger})\right)J_3^{(3)}(k_1,k_2,k_{\wt{ij}})\nonumber\\
&&\:\:\:+d_3^0(3_{\bar{q}};i_g,j_g)\left( |\cm(1_Q,2_{\bar{Q}};;\hatbt,(\wt{ij})_g,\hatf)|^2 \right.\nonumber\\
&&\:\:\:\:\:\:\:\:\:\:\:\:\:\: \left. +2\re( \cm(1_Q,(\wt{ij})_g,\hatf;;\hatbt,2_{\bar{Q}}) \cm(1_Q,(\wt{ij})_g,2_{\bar{Q}};;\hatbt,\hatf)^{\dagger})\right.\nonumber\\
&&\:\:\:\:\:\:\:\:\:\:\:\:\:\: \left. -2\re( \cm(1_Q,\hatf;;\hatbt,(\wt{ij})_g,2_{\bar{Q}}) \cm(1_Q,(\wt{ij})_g,2_{\bar{Q}};;\hatbt,\hatf)^{\dagger})\right.\nonumber\\
&&\:\:\:\:\:\:\:\:\:\:\:\:\:\: \left. -2\re( \cm(1_Q,\hatf;;\hatbt,(\wt{ij})_g,2_{\bar{Q}}) \cm(1_Q,2_{\bar{Q}};;\hatbt,(\wt{ij})_g,\hatf)^{\dagger})\right.\nonumber\\
&&\:\:\:\:\:\:\:\:\:\:\:\:\:\: \left. -2\re( \cm(1_Q,(\wt{ij})_g,\hatf;;\hatbt,2_{\bar{Q}}) \cm(1_Q,2_{\bar{Q}};;\hatbt,(\wt{ij})_g,\hatf)^{\dagger})\right)J_3^{(3)}(k_1,k_2,k_{\wt{ij}})\nonumber\\
&&\:\:\:+\frac{1}{N_c^2}\left[ A_3^0(1_Q,i_g,2_{\bar{Q}})\left( 2 |\cm((\wt{1i})_Q,(\wt{2i})_{\bar{Q}},\hatt,\hatf,j_{\gamma})|^2-|\cm((\wt{1i})_Q,(\wt{2i})_{\bar{Q}};\hatt,j_g,\hatf)|^2 \right.\right.\nonumber\\
&&\:\:\:\:\:\:\:\:\:\:\:\:\:\:\:\:\:\:\:\:\:\:\:\:\: \left.\left. -|\cm((\wt{1i})_Q,j_g,(\wt{2i})_{\bar{Q}};;\hatt,\hatf)|^2 \right)J_3^{(3)}(k_{\wt{1i}},k_{\wt{2i}},k_j)\right.\nonumber\\
&&\:\:\:\:\:\:\:\:\:\:\:\left. +A_3^0(3_{\bar{q}},4_q;i_g)\left( 2|\cm(\tilde{1}_Q,\tilde{2}_{\bar{Q}},\hatbt,\hatbf,\tilde{j}_{\gamma})|^2-|\cm(\tilde{1}_Q,\tilde{2}_{\bar{Q}};;\hatbt,\tilde{j}_g,\hatbf)|^2\right.\right.\nonumber\\
&&\:\:\:\:\:\:\:\:\:\:\:\:\:\:\:\:\:\:\:\:\:\:\:\:\: \left.\left. -|\cm(\tilde{1}_Q,\tilde{j}_g,\tilde{2}_{\bar{Q}};;\hatbt,\hatbf)|^2\right)J_3^{(3)}(\tilde{k}_1,\tilde{k}_2,\tilde{k}_j)\right.\nonumber\\
&&\:\:\:\:\:\:\:\:\:\:\:\left. +A_3^0(4_q;1_Q,i_g)2 \re( \cm((\wt{1i})_Q,2_{\bar{Q}};;\hatt,j_g,\hatbf) \cm((\wt{1i})_Q,j_g,2_{\bar{Q}};;\hatt,\hatbf)^{\dagger})J_3^{(3)}(k_{\wt{1i}},k_2,k_j)\right.\nonumber\\
&&\:\:\:\:\:\:\:\:\:\:\:\left. +A_3^0(3_{\bar{q}};2_{\bar{Q}},i_g) 2\re ( \cm(1_Q,(\wt{2i})_{\bar{Q}};;\hatbt,j_g,\hatf) \cm(1_Q,j_g,(\wt{2i})_{\bar{Q}};;\hatbt,\hatf)^{\dagger})J_3^{(3)}(k_1,k_{\wt{2i}},k_j)\right.\nonumber\\
&&\:\:\:\:\:\:\:\:\:\:\:\left. -A_3^0(3_{\bar{q}};1_Q,i_g) 2\re ( \cm((\wt{1i})_Q,2_{\bar{Q}};;\hatbt,j_g,\hatf)\cm((\wt{1i})_Q,j_g,2_{\bar{Q}};;\hatbt,\hatf)^{\dagger})J_3^{(3)}(k_{\wt{1i}},k_2,k_j)\right.\nonumber\\
&&\:\:\:\:\:\:\:\:\:\:\:\left. -A_3^0(4_q;2_{\bar{Q}},i_g) 2\re( \cm(1_Q,(\wt{2i})_{\bar{Q}};;\hatt,j_g,\hatbf)\cm(1_Q,j_g,(\wt{2i})_{\bar{Q}};;\hatt,\hatbf)^{\dagger})J_3^{(3)}(k_1,k_{\wt{2i}},k_j)\right]\bigg\}\nonumber
\end{eqnarray}

\parindent 1.5em
When $4_q$ and $6_g$ are crossed to the initial state in eq.(\ref{eq:qqgg}), the squared matrix element for $qg\rightarrow Q\bar{Q}qg$ in terms of colour ordered subamplitudes is obtained. The subtraction term in this case reads,
\newcommand{\hats}{\hat{6}_g}
\newcommand{\hatbfg}{\hat{\bar{4}}_g}
\newcommand{\hatbs}{\hat{\bar{6}}}
\begin{eqnarray}
\lefteqn{{\rm d}\hat{\sigma}_{qg\rightarrow Q\bar{Q}qg}^S=g^8(N_c^2-1){\rm d}\Phi_4(k_{1Q},k_{2\bar{Q}},k_{3q},k_{5g};p_{4q},p_{6g}) }\nonumber\\
&&\bigg\{N_c^2\left[ A_3^0(3_q,5_g,2_{\bar{Q}})|\cm(1_Q,\hats,\hatf;;(\wt{35})_q,(\wt{25})_{\bar{Q}})|^2 J_3^{(3)}(k_1,k_{\wt{25}},k_{\wt{35}})\right.\nonumber\\ 
&&\left. \:\:\:\:\:\:\:+A_3^0(4_q;1_Q,5_g)|\cm((\wt{15})_Q,\hatbf;;3_q,\hats,\two)|^2 J_3^{(3)}(k_{\wt{15}},k_2,k_3)\right.\nonumber\\
&&\left. \:\:\:\:\:\:+\frac{1}{2}A_3^0(6_g;1_Q,\two)\left( |\cm((\wt{12})_Q,5_g,\hatf;;3_q,\hat{\bar{6}}_Q)|^2+|\cm((\wt{12})_Q,\hatf;;3_q,5_g,\hat{\bar{6}}_Q)|^2\right.\right.\nonumber\\
&&\:\:\:\:\:\:\:\:\:\:\:\:\:\:\: \left.\left. +|\cm(\hat{\bar{6}}_{\bar{Q}},5_g,\hatf;;3_q,(\wt{12})_{\bar{Q}})|^2+|\cm(\hat{\bar{6}}_{\bar{Q}},\hatf;;3_q,5_g,(\wt{12})_{\bar{Q}})|^2\right)J_3^{(3)}(k_{\wt{12}},k_3,k_5)\right.\nonumber\\
&&\:\:\:\:\:\:\left.+A_3^0(4_q,6_g;3_q)\left( |\cm(\tilde{1}_Q,\hatbf;;\hatbs_{\bar{q}},\tilde{5}_g,\tilde{2}_{\bar{Q}})|^2+|\cm(\tilde{1}_Q,\tilde{5}_g,\hatbf;;\hatbs_{\bar{q}},\tilde{2}_{\bar{Q}})|^2\right)J_3^{(3)}(\tilde{k}_1,\tilde{k}_2,\tilde{k}_5)\right.\nonumber\\
&&\:\:\:\:\:\:\left. +D_3^0(6_g;5_g,1_Q)|\cm((\wt{15})_Q,\hatbs_g,\hatf;;3_q,\two)|^2 J_3^{(3)}(k_{\wt{15}},k_2,k_3)\right.\nonumber\\
&&\:\:\:\:\:\:\left. +D_3^0(6_g;5_g,\two)|\cm(1_Q,\hatf;;3_q,\hatbs_g,(\wt{25})_{\bar{Q}})|^2 J_3^{(3)}(k_1,k_{\wt{25}},k_3)\right.\nonumber\\
&&\:\:\:\:\:\:\left. +D_3^0(6_g;5_g,3_q)|\cm(1_q,\hatf;;(\wt{35})_q,\hatbs_g,\two)|^2 J_3^{(3)}(k_1,k_2,k_{\wt{35}})\right.\nonumber\\
&&\:\:\:\:\:\:\left. +D_3^0(4_q,6_g;5_g)|\cm(\tilde{1}_Q,\hatbs_g,\hatbf;;\tilde{3}_q,\tilde{2}_{\bar{Q}})|^2 J_3^{(3)}(\tilde{k}_1,\tilde{k}_2,\tilde{k}_3)\right.\nonumber\\
&&\:\:\:\:\:\:\left. +\frac{1}{2}E_3^0(4_q;3_q,1_Q)\left( |\cm((\wt{13})_Q,5_g,\hats,\hatbfg,\two)|^2+|\cm((\wt{13})_Q,5_g,\hatbfg,\hats,\two)|^2\right.\right.\nonumber\\
&&\:\:\:\:\:\:\:\:\:\:\:\:\:\:\: \left.\left.+|\cm((\wt{13})_Q,\hatbfg,5_g,\hats,\two)|^2+|\cm((\wt{13})_Q,\hats,5_g,\hatbfg,\two)|^2\right.\right.\nonumber\\
&&\:\:\:\:\:\:\:\:\:\:\:\:\:\:\: \left.\left.+|\cm((\wt{13})_Q,\hats,\hatbfg,5_g,\two)|^2+|\cm((\wt{13})_Q,\hatbfg,\hats,5_g,\two)|^2\right)J_3^{(3)}(k_{\wt{13}},k_2,k_5)\right.\nonumber\\
&&\:\:\:\:\:\:\left. +\frac{1}{2}E_3^0(4_q;3_q,\two)\left( |\cm(1_Q,5_g,\hats,\hatbfg,(\wt{23})_{\bar{Q}})|^2+|\cm(1_Q,5_g,\hatbfg,\hats,(\wt{23})_{\bar{Q}})|^2\right.\right.\nonumber\\
&&\:\:\:\:\:\:\:\:\:\:\:\:\:\:\: \left.\left. +|\cm(1_Q,\hats,5_g,\hatbfg,(\wt{23})_{\bar{Q}})|^2+|\cm(1_Q,\hatbfg,5_g,\hats,(\wt{23})_{\bar{Q}})|^2\right.\right.\nonumber\\
&&\:\:\:\:\:\:\:\:\:\:\:\:\:\:\: \left.\left. +|\cm(1_Q,\hats,\hatbfg,5_g,(\wt{23})_{\bar{Q}})|^2+|\cm(1_Q,\hatbfg,\hats,5_g,(\wt{23})_{\bar{Q}})|^2\right)J_3^{(3)}(k_1,k_{\wt{23}},k_5)\right]\nonumber\\
&&\:\:\:+A_3^0(1_Q,5_g,\two)\left( |\cm((\wt{15})_Q,(\wt{25})_{\bar{Q}};;3_g,\hats,\hatf)|^2\right.\nonumber\\
&&\:\:\:\:\:\:\:\:\:\:\:\:\:\: \left. +2\re ( \cm((\wt{15})_Q,\hatf;;3_q,\hats,(\wt{25})_{\bar{Q}}) \cm((\wt{15})_Q,\hats,\hatf;;3_q,(\wt{25})_{\bar{Q}})^{\dagger})\right.\nonumber\\
&&\:\:\:\:\:\:\:\:\:\:\:\:\:\: \left. -2\re ( \cm((\wt{15})_Q,\hatf;;3_q,\hats,(\wt{25})_{\bar{Q}}) \cm((\wt{15})_Q,(\wt{25})_{\bar{Q}};;3_g,\hats,\hatf)^{\dagger})\right.\nonumber\\
&&\:\:\:\:\:\:\:\:\:\:\:\:\:\: \left. -2\re ( \cm((\wt{15})_Q,\hats,\hatf;;3_q,(\wt{25})_{\bar{Q}}) \cm((\wt{15})_Q,(\wt{25})_{\bar{Q}};;3_g,\hats,\hatf)^{\dagger})\right)J_3^{(3)}(k_{\wt{15}},k_{\wt{25}},k_3)\nonumber\\
&&\:\:\:+A_3^0(1_Q,5_g,3_q)\left( 2\re( \cm((\wt{15})_Q,\hatf;;(\wt{35})_q,\hats,\two) \cm((\wt{15})_Q,\hats,\two;;(\wt{35})_q,\hatf)^{\dagger})\right.\nonumber\\
&&\:\:\:\:\:\:\:\:\:\:\:\:\:\: \left. +2\re( \cm((\wt{15})_Q,\hats,\hatf;;(\wt{35})_q,\two) \cm((\wt{15})_Q,\two;;(\wt{35})_q,\hats,\hatf)^{\dagger})\right.\nonumber\\
&&\:\:\:\:\:\:\:\:\:\:\:\:\:\: \left. -2\re( \cm((\wt{15})_Q,\hatf;;(\wt{35})_q,\hats,\two) \cm((\wt{15})_Q,\hats,\hatf;;(\wt{35})_q,\two)^{\dagger})\right)J_3^{(0)}(k_{\wt{15}},k_2,k_{\wt{35}})\nonumber\\
&&\:\:\:-A_3^0(3_q,5_g,\two)\left( |\cm(1_Q,\hatf;;(\wt{35})_q,\hats,(\wt{25})_{\bar{Q}})|^2\right.\nonumber\\
&&\:\:\:\:\:\:\:\:\:\:\:\:\:\: \left. +|\cm(1_Q,\hats,\hatf;;(\wt{35})_q,(\wt{25})_{\bar{Q}})|^2\right.\nonumber\\
&&\:\:\:\:\:\:\:\:\:\:\:\:\:\: \left. +2\re ( \cm(1_Q,\hats,\hatf;;(\wt{35})_q,(\wt{25})_{\bar{Q}}) \cm(1_Q,(\wt{25})_{\bar{Q}};;(\wt{35})_q,\hats,\hatf)^{\dagger})\right.\nonumber\\
&&\:\:\:\:\:\:\:\:\:\:\:\:\:\: \left. +2\re ( \cm(1_Q,\hats,\hatf;;(\wt{35})_q,(\wt{25})_{\bar{Q}}) \cm(1_Q,\hats,(\wt{25})_{\bar{Q}};;(\wt{35})_q,\hatf)^{\dagger})\right)J_3^{(3)}(k_1,k_{\wt{25}},k_{\wt{35}})\nonumber\\
&&\:\:\: -A_3^0(4_q;1_Q,5_g)\left( |\cm((\wt{15})_Q,\hats,\hatbf;;3_q,\two)|^2\right.\nonumber\\
&&\:\:\:\:\:\:\:\:\:\:\:\:\:\: \left. +|\cm((\wt{15})_Q,\hatbf;;3_q,\hats,\two)|^2\right.\nonumber\\
&&\:\:\:\:\:\:\:\:\:\:\:\:\:\: \left. +2\re ( \cm((\wt{15})_Q,\hatbf;;3_q,\hats,\two) \cm((\wt{15})_Q,\two;;3_q,\hats,\hatbf)^{\dagger})\right.\nonumber\\
&&\:\:\:\:\:\:\:\:\:\:\:\:\:\: \left. +2\re ( \cm((\wt{15})_Q,\hatbf;;3_q,\hats,\two) \cm((\wt{15})_Q,\hats,\two;;3_q,\hatbf)^{\dagger})\right) J_3^{(3)}(k_{\wt{15}},k_2,k_3)\nonumber\\
&&\:\:\: +A_3^0(4_q;\two,5_g)\left( 2\re ( \cm(1_Q,\hatbf;;3_q,\hats,(\wt{25})_{\bar{Q}}) \cm(1_Q,(\wt{25})_{\bar{Q}};;3_q,\hats,\hatbf)^{\dagger})\right.\nonumber\\
&&\:\:\:\:\:\:\:\:\:\:\:\:\:\: \left. +2\re ( \cm(1_Q,\hats,\hatbf;;3_q,(\wt{25})_{\bar{Q}}) \cm(1_Q,\hats,(\wt{25})_{\bar{Q}};;3_q,\hatbf)^{\dagger})\right.\nonumber\\
&&\:\:\:\:\:\:\:\:\:\:\:\:\:\: \left. -2\re (  \cm(1_Q,\hatbf;;3_q,\hats,(\wt{25})_{\bar{Q}}) \cm(1_Q,\hats,\hatbf;;3_q,(\wt{25})_{\bar{Q}})^{\dagger})\right)J_3^{(3)}(k_1,k_{\wt{25}},k_3)\nonumber\\
&&\:\:\: +A_3^0(4_q;3_q,5_g)\left( |\cm(1_Q,\hats,\two;(\wt{35})_q,\hatbf)|^2\right.\nonumber\\
&&\:\:\:\:\:\:\:\:\:\:\:\:\:\: \left. +2\re ( \cm(1_Q,\hatbf;;(\wt{35})_q,\hats,\two) \cm(1_Q,\hats,\hatbf;;(\wt{35})_q,\two)^{\dagger})\right.\nonumber\\
&&\:\:\:\:\:\:\:\:\:\:\:\:\:\: \left. -2\re ( \cm(1_Q,\hatbf;;(\wt{35})_q,\hats,\two) \cm(1_Q,\hats,\two;;(\wt{35})_q,\hatbf)^{\dagger})\right.\nonumber\\
&&\:\:\:\:\:\:\:\:\:\:\:\:\:\: \left. -2\re ( \cm(1_Q,\hats,\hatbf;;(\wt{35})_q,\two) \cm(1_Q,\hats,\two;;(\wt{35})_q,\hatbf)^{\dagger})\right)J_3^{(3)}(k_1,k_2,k_{\wt{35}})\nonumber\\
&&\:\:\: +\frac{1}{2}A_3^0(6_g;1_Q,\two)\left( |\cm((\wt{12})_Q,\hatbs_Q;;3_q,5_g,\hatf)|^2+|\cm((\wt{12})_Q,5_g,\hatbs_Q;;3_q,\hatf)|^2 \right.\nonumber\\
&&\:\:\:\:\:\:\:\:\:\:\:\:\:\:\: \left. +|\cm(\hatbs_{\bar{Q}},(\wt{12})_{\bar{Q}};;3_q,5_g,\hatf)|^2+|\cm(\hatbs_{\bar{Q}},5_g,(\wt{12})_{\bar{Q}};;3_q,\hatf)|^2\right.\nonumber\\
&&\:\:\:\:\:\:\:\:\:\:\:\:\:\:\: \left. -|\cm((\wt{12})_Q,\hatf;;3_q,5_g,\hatbs_Q)|^2-|\cm((\wt{12})_Q,5_g,\hatf;;3_q,\hatbs_Q)|^2\right.\nonumber\\
&&\:\:\:\:\:\:\:\:\:\:\:\:\:\:\: \left. -|\cm(\hatbs_{\bar{Q}},\hatf;;3_q,5_g,(\wt{12})_{\bar{Q}})|^2-|\cm(\hatbs_{\bar{Q}},5_g,\hatf;;3_q,(\wt{12})_{\bar{Q}})|^2\right.\nonumber\\
&&\:\:\:\:\:\:\:\:\:\:\:\:\:\:\: \left. -2|\cm((\wt{12})_Q,\hatbs_Q,3_q,\hatf,5_{\gamma})|^2-2|\cm(\hatbs_{\bar{Q}},(\wt{12})_{\bar{Q}},3_q,\hatf,5_{\gamma})|^2 \right) J_3^{(3)}(k_{\wt{12}},k_3,k_5)\nonumber\\
&&\:\:\:+A_3^0(4_q,6_g;3_q)\left( |\cm(\tilde{1}_Q,\tilde{5}_g,\tilde{2}_{\bar{Q}};;\hatbs_{\bar{q}},\hatbf)|^2+|\cm(\tilde{1}_Q,\tilde{2}_{\bar{Q}};;\hatbs_{\bar{q}},\tilde{5}_g,\hatbf)|^2\right.\nonumber\\
&&\:\:\:\:\:\:\:\:\:\:\:\:\:\:\: \left. -|\cm(\tilde{1}_Q,\tilde{5}_g,\hatbf;;\hatbs_{\bar{q}},\tilde{2}_{\bar{Q}})|^2-|\cm(\tilde{1}_Q,\hatbf;;\hatbs_{\bar{q}},\tilde{5}_g,\tilde{2}_{\bar{Q}})|^2\right.\nonumber\\
&&\:\:\:\:\:\:\:\:\:\:\:\:\:\:\: \left. -2|\cm(\tilde{1}_Q,\tilde{2}_{\bar{Q}},\hatbs_{\bar{q}},\hatbf,\tilde{5}_{\gamma})|^2\right)J_3^0(\tilde{k}_1,\tilde{k}_2,\tilde{k}_5)\nonumber\\
&&\:\:\:+D_3^0(6_g;5_g,1_Q)\left( |\cm((\wt{15})_Q,\hatbs_g,\two;;3_q,\hatf)|^2\right.\nonumber\\
&&\:\:\:\:\:\:\:\:\:\:\:\:\:\:\: \left. +2\re ( \cm((\wt{15})_Q,\hatf;;3_q,\hatbs_g,\two) \cm((\wt{15})_Q,\two;;3_q,\hatbs_g,\hatf)^{\dagger})\right.\nonumber\\
&&\:\:\:\:\:\:\:\:\:\:\:\:\:\:\: \left. -2\re ( \cm((\wt{15})_Q,\hatf;;3_q,\hatbs_g,\two) \cm((\wt{15})_Q,\hatbs_g,\two;;3_q,\hatf)^{\dagger})\right.\nonumber\\
&&\:\:\:\:\:\:\:\:\:\:\:\:\:\:\: \left. -2\re ( \cm((\wt{15})_Q,\hatbs_g,\hatf;;3_q,\two) \cm((\wt{15})_Q,\hatbs_g,\two;;3_q,\hatf)^{\dagger})\right.\nonumber\\
&&\:\:\:\:\:\:\:\:\:\:\:\:\:\:\: \left. -2\re ( \cm((\wt{15})_Q,\hatbs_g,\hatf;;3_q,\two) \cm((\wt{15})_Q,\two;;3_q,\hatbs_g,\hatf)^{\dagger})\right)J_3^{(3)}(k_{\wt{15}},k_2,k_3)\nonumber\\
&&\:\:\:+D_3^0(6_g;5_g,2_{\bar{Q}})\left( |\cm(1_q,\hatbs_g,(\wt{25})_{\bar{Q}};;3_q,\hatf)|^2\right.\nonumber\\
&&\:\:\:\:\:\:\:\:\:\:\:\:\:\:\: \left. +2\re ( \cm(1_Q,\hatbs_g,\hatf;;3_q,(\wt{25})_{\bar{Q}}) \cm(1_Q,(\wt{25})_{\bar{Q}};;3_q,\hatbs_g,\hatf)^{\dagger})\right.\nonumber\\
&&\:\:\:\:\:\:\:\:\:\:\:\:\:\:\: \left. -2\re ( \cm(1_Q,\hatf;;3_q,\hatbs_g,(\wt{25})_{\bar{Q}}) \cm(1_Q,(\wt{25})_{\bar{Q}};;3_q,\hatbs_g,\hatf)^{\dagger})\right.\nonumber\\
&&\:\:\:\:\:\:\:\:\:\:\:\:\:\:\: \left. -2\re ( \cm(1_Q,\hatf;;3_q,\hatbs_g,(\wt{25})_{\bar{Q}}) \cm(1_Q,\hatbs_g,(\wt{25})_{\bar{Q}};;3_q,\hatf)^{\dagger})\right.\nonumber\\
&&\:\:\:\:\:\:\:\:\:\:\:\:\:\:\: \left. -2\re ( \cm(1_Q,\hatbs_g,\hatf;;3_q,(\wt{25})_{\bar{Q}}) \cm(1_Q,\hatbs_g,(\wt{25})_{\bar{Q}};;3_q,\hatf)^{\dagger})\right)J_3^{(3)}(k_1,k_{\wt{25}},k_3)\nonumber\\
&&\:\:\:+D_3^0(6_g;5_g,3_q)\left( |\cm(1_Q,\two;;(\wt{35})_q,\hatbs_g,\hatf)|^2\right.\nonumber\\
&&\:\:\:\:\:\:\:\:\:\:\:\:\:\:\: \left. +2\re ( \cm(1_q,\hatbs_g,\hatf;;(\wt{35})_q,\two) \cm(1_Q,\hatbs_g,\two;;(\wt{35})_q,\hatf)^{\dagger})\right.\nonumber\\
&&\:\:\:\:\:\:\:\:\:\:\:\:\:\:\: \left. -2\re ( \cm(1_q,\hatf;;(\wt{35})_q,\hatbs_g,\two) \cm(1_Q,\two;;(\wt{35})_q,\hatbs_g,\hatf)^{\dagger})\right.\nonumber\\
&&\:\:\:\:\:\:\:\:\:\:\:\:\:\:\: \left. -2\re ( \cm(1_q,\hatf;;(\wt{35})_q,\hatbs_g,\two) \cm(1_Q,\hatbs_g,\two;;(\wt{35})_q,\hatf)^{\dagger})\right.\nonumber\\
&&\:\:\:\:\:\:\:\:\:\:\:\:\:\:\: \left. -2\re ( \cm(1_q,\hatbs_g,\hatf;;(\wt{35})_q,\two) \cm(1_Q,\two;;(\wt{35})_q,\hatbs_g,\hatf)^{\dagger})\right)J_3^{(3)}(k_1,k_2,k_{\wt{35}})\nonumber\\
&&\:\:\:+D_3^0(4_q,6_g;5_g)\left( |\cm(\tilde{1}_Q,\tilde{2}_{\bar{Q}};;\tilde{3}_q,\hatbs_g,\hatbf)|^2\right.\nonumber\\
&&\:\:\:\:\:\:\:\:\:\:\:\:\:\:\: \left. +2\re ( \cm(\tilde{1}_Q,\hatbf;;\tilde{3}_q,\hatbs_g,\tilde{2}_{\bar{Q}}) \cm(\tilde{1}_Q,\hatbs_g,\tilde{2}_{\bar{Q}};;\tilde{3}_q,\hatbf)^{\dagger})\right.\nonumber\\
&&\:\:\:\:\:\:\:\:\:\:\:\:\:\:\: \left. -2\re ( \cm(\tilde{1}_Q,\hatbf;;\tilde{3}_q,\hatbs_g,\tilde{2}_{\bar{Q}}) \cm(\tilde{1}_Q,\tilde{2}_{\bar{Q}};;\tilde{3}_q,\hatbs_g,\hatbf)^{\dagger})\right.\nonumber\\
&&\:\:\:\:\:\:\:\:\:\:\:\:\:\:\: \left. -2\re ( \cm(\tilde{1}_Q,\hatbs_g,\hatbf;;\tilde{3}_q,\tilde{2}_{\bar{Q}}) \cm(\tilde{1}_Q,\tilde{2}_{\bar{Q}};;\tilde{3}_q,\hatbs_g,\hatbf)^{\dagger})\right.\nonumber\\
&&\:\:\:\:\:\:\:\:\:\:\:\:\:\:\: \left. -2\re ( \cm(\tilde{1}_Q,\hatbs_g,\hatbf;;\tilde{3}_q,\tilde{2}_{\bar{Q}}) \cm(\tilde{1}_Q,\hatbs_g,\tilde{2}_{\bar{Q}};;\tilde{3}_q,\hatbf)^{\dagger})\right)J_3^{(3)}(\tilde{k}_1,\tilde{k}_2,\tilde{k}_3)\nonumber\\
&&\:\:\:-\frac{1}{2}E_3^0(4_q;3_q,1_Q)\left( |\cm((\wt{13})_Q,5_g,\hats,\hat{\bar{4}}_{\gamma},\two)|^2+|\cm((\wt{13})_Q,\hats,5_g,\hat{\bar{4}}_{\gamma},\two)|^2\right.\nonumber\\
&&\:\:\:\:\:\:\:\:\:\:\:\:\:\:\: \left. +|\cm((\wt{13})_Q,5_g,\hat{\bar{4}}_g,\hat{6}_{\gamma},\two)|^2+|\cm((\wt{13})_Q,\hat{\bar{4}}_g,5_g,\hat{6}_{\gamma},\two)|^2\right.\nonumber\\
&&\:\:\:\:\:\:\:\:\:\:\:\:\:\:\: \left. +|\cm((\wt{13})_Q,\hats,\hat{\bar{4}}_g,5_{\gamma},\two|^2+|\cm((\wt{13})_Q,\hat{\bar{4}}_g,\hats,5_{\gamma},\two|^2\right)J_3^{(3)}(k_{\wt{13}},k_2,k_5)\nonumber\\
&&\:\:\:-\frac{1}{2}E_3^0(4_q;3_q,2_{\bar{Q}})\left( |\cm(1_Q,5_g,\hats,\hat{\bar{4}}_{\gamma},(\wt{23})_{\bar{Q}})|^2+|\cm(1_Q,\hats,5_g,\hat{\bar{4}}_{\gamma},(\wt{23})_{\bar{Q}})|^2\right.\nonumber\\
&&\:\:\:\:\:\:\:\:\:\:\:\:\:\:\left. +|\cm(1_Q,5_g,\hat{\bar{4}}_g,\hat{6}_{\gamma},(\wt{23})_{\bar{Q}})|^2 +|\cm(1_Q,\hat{\bar{4}}_g,5_g,\hat{6}_{\gamma},(\wt{23})_{\bar{Q}})|^2\right.\nonumber\\
&&\:\:\:\:\:\:\:\:\:\:\:\:\:\:\left. +|\cm(1_Q,\hats,\hat{\bar{4}}_g,5_{\gamma},(\wt{23})_{\bar{Q}})|^2+|\cm(1_Q,\hat{\bar{4}}_g,\hats,5_{\gamma},(\wt{23})_{\bar{Q}})|^2\right)J_3^{(3)}(k_1,k_{\wt{23}},k_5)\nonumber\\
&&\:\:\:+\frac{1}{N_c^2}\left[ A_3^0(1_Q,5_g,2_{\bar{Q}})\left( 2|\cm((\wt{15})_Q,(\wt{25})_{\bar{Q}},3_q,\hatf,\hat{6}_{\gamma})|^2\right.\right.\nonumber\\
&&\:\:\:\:\:\:\:\:\:\:\:\:\left.\left. -|\cm(\wt{15})_Q,(\wt{25})_{\bar{Q}};;3_q,\hats,\hatf)|^2-|\cm(\wt{15})_Q,\hats,(\wt{25})_{\bar{Q}};;3_q,\hatf)|^2\right)J_3^{(3)}(k_{\wt{15}},k_{\wt{25}},k_3)\right.\nonumber\\
&&\:\:\:\:\:\:\left. +A_3^0(3_q,5_g,\two)2\re ( \cm(1_Q,(\wt{25})_{\bar{Q}};;(\wt{35})_q,\hats,\hatf)\cm(1_Q,\hats,(\wt{25})_{\bar{Q}};;(\wt{35})_q,\hatf)^{\dagger})J_3^{(3)}(k_1,k_{\wt{25}},k_{\wt{35}})\right.\nonumber\\
&&\:\:\:\:\:\:\left. -A_3^0(1_Q,5_g,3_q)2\re ( \cm ((\wt{15})_Q,\two;;(\wt{35})_q,\hats,\hatf)\cm ((\wt{15})_Q,\hats,\two;;(\wt{35})_q,\hatf)^{\dagger})J_3^{(3)}(k_{\wt{15}},k_2,k_{\wt{35}})\right.\nonumber\\
&&\:\:\:\:\:\:\left. +A_3^0(4_q;1_Q,5_g)2\re (\cm((\wt{15})_Q,\two;;3_q,\hats,\hatbf)\cm((\wt{15})_Q,\hats,\two;;3_q,\hatbf)^{\dagger})J_3^{(3)}(k_{\wt{15}},k_2,k_3)\right.\nonumber\\
&&\:\:\:\:\:\:\left. -A_3^0(4_q;2_{\bar{Q}},5_g) 2\re( \cm(1_Q,(\wt{25})_{\bar{Q}};;3_q,\hats,\hatbf) \cm(1_Q,\hats,(\wt{25})_{\bar{Q}};;3_q,\hatbf)^{\dagger})J_3^{(3)}(k_1,k_{\wt{25}},k_3)\right.\nonumber\\
&&\:\:\:\:\:\:\left. +A_3^0(4_q;3_q,5_g)\left( 2|\cm(1_Q,\two,(\wt{35})_q,\hatf,\hat{6}_{\gamma})|^2\right.\right.\nonumber\\
&&\:\:\:\:\:\:\:\:\:\:\:\:\:\left.\left. -|\cm(1_Q,\hats,\two;(\wt{35})_q,\hatf)|^2-|\cm(1_Q,\two;;(\wt{35})_q,\hats,\hatf)|^2\right)J_3^{(3)}(k_1,k_2,k_{\wt{35}})\right.\nonumber\\
&&\:\:\:\:\:\:\left. +\frac{1}{2}A_3^0(6_g;1_Q,\two)\left( 2|\cm((\wt{12})_Q,\hat{\bar{6}}_Q,3_q,\hatf,5_{\gamma})|^2+2|\cm(\hat{\bar{6}}_{\bar{Q}}(\wt{12})_{\bar{Q}},3_q,\hatf,5_{\gamma})|^2\right.\right.\nonumber\\
&&\:\:\:\:\:\:\:\:\:\:\:\:\:\left.\left. -|\cm((\wt{12})_Q,5_g,\hat{\bar{6}}_Q;;3_q,\hatf)|^2-|\cm(\hat{\bar{6}}_{\bar{Q}},5_g,(\wt{12})_{\bar{Q}};;3_q,\hatf)|^2\right.\right.\nonumber\\
&&\:\:\:\:\:\:\:\:\:\:\:\:\:\left.\left. -|\cm((\wt{12})_Q,\hat{\bar{6}}_Q;;3_q,5_g,\hatf)|^2-|\cm(\hat{\bar{6}}_{\bar{Q}},(\wt{12})_{\bar{Q}};;3_q,5_g,\hatf)|^2\right)J_3^{(3)}(k_{\wt{12}},k_3,k_5)\right.\nonumber\\
&&\:\:\:\:\:\:\left. +A_3^0(4_q,6_g;3_q)\left( 2|\cm(\tilde{1}_Q,\tilde{2}_{\bar{Q}},\hat{\bar{6}}_{\bar{q}},\hatbf,\tilde{5}_{\gamma})|^2\right.\right.\nonumber\\
&&\:\:\:\:\:\:\:\:\:\:\:\:\:\left.\left. -|\cm(\tilde{1}_Q,\tilde{5}_{g},\tilde{2}_{\bar{Q}};;\hat{\bar{6}}_{\bar{q}},\hatbf)|^2-|\cm(\tilde{1}_Q,\tilde{2}_{\bar{Q}};;\hat{\bar{6}}_{\bar{q}},\tilde{5}_{g},\hatbf)|^2\right)J_3^{(3)}(\tilde{k}_1,\tilde{k}_2,\tilde{k}_5)\right.\nonumber\\
&&\:\:\:\:\:\:\left. +\frac{1}{2}E_3^0(4_q;3_q,1_Q)|\cm((\wt{13})_Q,\two,\hat{\bar{4}}_{\gamma},5_{\gamma},\hat{6}_{\gamma})|^2 J_3^{(3)}(k_{\wt{13}},k_2,k_5)\right.\nonumber\\
&&\:\:\:\:\:\:\left. +\frac{1}{2}E_3^0(4_q;3_q,2_{\bar{Q}})|\cm(1_Q,(\wt{25})_{\bar{Q}},\hat{\bar{4}}_{\gamma},5_{\gamma},\hat{6}_{\gamma})|^2 J_3^{(3)}(k_1,k_{\wt{23}},k_5)\right]\bigg\}\nonumber
\end{eqnarray}

Finally, when gluons $5_g$ and $6_g$ are crossed to the initial state in eq.(\ref{eq:qqgg}), the squared matrix element for $gg\rightarrow Q\bar{Q}q\bar{q}$ is obtained.
This partonic process only contains collinear and quasi-collinear limits but no soft limits as no gluons are present in the final state. 
Using the decoupling identities given 
in eqs.(\ref{eq:id1}, \ref{eq:id3}, \ref{eq:id2}), the reduced matrix elements multiplying each antenna function can be rewritten in a fairly compact form with no interference terms left. After doing this simplification the subtraction term for this process reads,
\newcommand{\hati}{\hat{i}}
\newcommand{\hatj}{\hat{j}}
\newcommand{\hatbi}{\hat{\bar{i}}}
\newcommand{\four}{4_{\bar{q}}}
\begin{eqnarray}
\lefteqn{{\rm d}\hat{\sigma}_{gg\rightarrow Q\bar{Q}q\bar{q}}^S=g^8\:N_F(N_c^2-1){\rm d}\Phi_4(k_{1Q},k_{2\bar{Q}},k_{3q},k_{4\bar{q}};p_{5g},p_{6g}) }\nonumber\\
&&\times\bigg\{\sum_{(i,j)\in P(5,6)}\bigg[ \nonumber\\
&&\:\:\:\:\:\:\:\:\:\:\:\:N_c^2\left[ \frac{1}{2}A_3^0(i_g;1_Q,2_{\bar{Q}})\left( |\cm((\wt{12})_Q,\four;;3_q,\hatj_g,\hatbi_Q)|^2+ |\cm(\hatbi_{\bar{Q}},\four;;3_q,\hatj_g,(\wt{12})_{\bar{Q}})|^2\right.\right.\nonumber\\
&&\:\:\:\:\:\:\:\:\:\:\:\:\:\:\:\:\:\:\:\:\:\:\:\:\:\:\:\left.\left.+|\cm((\wt{12})_Q,\hatj_g,\four;;3_q,\hatbi_Q)|^2+|\cm(\hatbi_{\bar{Q}},\hatj_g\four;;3_q,(\wt{12})_{\bar{Q}})|^2\right)J_3^{(3)}(k_{\wt{12}},k_3,k_4)\right.\nonumber\\
&&\:\:\:\:\:\:\:\:\:\:\:\:\:\:\:\:\:\left. +\frac{1}{2}A_3^0(i_g;3_q,\four)\left( |\cm(1_Q,\hatbi_q;;(\wt{34})_q,\hatj_g,\two )|^2+|\cm(1_Q,(\wt{34})_{\bar{q}};;\hatbi_{\bar{q}},\hatj_g,\two)|^2\right.\right.\nonumber\\
&&\:\:\:\:\:\:\:\:\:\:\:\:\:\:\:\:\:\:\:\:\:\:\:\:\:\:\:\left.\left. +|\cm(1_Q,\hatj_g,\hatbi_q;;(\wt{34})_q,\two )|^2+|\cm(1_Q,\hatj_g,(\wt{34})_{\bar{q}};;\hatbi_{\bar{q}},\two)|^2\right)J_3^{(3)}(k_1,k_2,k_{\wt{34}})\right.\nonumber\\
&&\:\:\:\:\:\:\:\:\:\:\:\:\:\:\:\:\:\left. +\frac{1}{2}E_3^0(1_Q,3_q,\four)\left( |\cm((\wt{13})_Q,\hati_g,\hatj_g,(\wt{34})_g,\two|^2+|\cm((\wt{13})_Q,\hati_g,(\wt{34})_g,\hatj_g,\two|^2\right.\right.\nonumber\\
&&\:\:\:\:\:\:\:\:\:\:\:\:\:\:\:\:\:\:\:\:\:\:\:\:\:\:\:\left.\left. +|\cm((\wt{13})_Q,(\wt{34})_g,\hati_g,\hatj_g,\two|^2\right)J_3^{(3)}(k_{\wt{13}},k_2,k_{\wt{34}})\right.\nonumber\\
&&\:\:\:\:\:\:\:\:\:\:\:\:\:\:\:\:\:\left. +\frac{1}{2}E_3^0(\two,3_q,\four)\left( |\cm(1_Q,\hati_g,\hatj_g,(\wt{34})_g,(\wt{23})_{\bar{Q}})|^2 + |\cm(1_Q,\hati_g,(\wt{34})_g,\hatj_g,(\wt{23})_{\bar{Q}})|^2\right.\right.\nonumber\\
&&\:\:\:\:\:\:\:\:\:\:\:\:\:\:\:\:\:\:\:\:\:\:\:\:\:\:\:\left.\left. + |\cm(1_Q,(\wt{34})_g,\hati_g,\hatj_g,(\wt{23})_{\bar{Q}})|^2\right)J_3^{(3)}(k_1,k_{\wt{23}},k_{\wt{34}})\right]\nonumber\\
&&\:\:\:\:\:\:\:\:\:\:\:\:+\frac{1}{2}A_3^0(i_g;1_Q,\two)\left( |\cm((\wt{12})_Q,\hatbi_Q;;3_q,\hatj_g,\four)|^2+|\cm(\hatbi_{\bar{Q}},(\wt{12})_{\bar{Q}};;3_q,\hatj_g,\four)|^2\right.\nonumber\\
&&\:\:\:\:\:\:\:\:\:\:\:\:\:\:\:\:\:\:\:\:\:\:\:\:\left.  +|\cm((\wt{12})_Q,\hatj_g,\hatbi_Q;;3_q,\four)|^2+|\cm(\hatbi_{\bar{Q}},\hatj_g,(\wt{12})_{\bar{Q}};;3_q,\four)|^2\right.\nonumber\\
&&\:\:\:\:\:\:\:\:\:\:\:\:\:\:\:\:\:\:\:\:\:\:\:\:\left. - |\cm((\wt{12})_Q,\four;;3_q,\hatj_g,\hatbi_Q)|^2- |\cm(\hatbi_{\bar{Q}},\four;;3_q,\hatj_g,(\wt{12})_{\bar{Q}})|^2\right.\\
&&\:\:\:\:\:\:\:\:\:\:\:\:\:\:\:\:\:\:\:\:\:\:\:\:\left.-|\cm((\wt{12})_Q,\hatj_g,\four;;3_q,\hatbi_Q)|^2.-|\cm(\hatbi_{\bar{Q}},\hatj_g\four;;3_q,(\wt{12})_{\bar{Q}})|^2\right.\nonumber\\
&&\:\:\:\:\:\:\:\:\:\:\:\:\:\:\:\:\:\:\:\:\:\:\:\:\left. -2|\cm((\wt{12})_Q,\hatbi_Q,3_q,\four,\hatj_{\gamma})|^2-2|\cm(\hatbi_{\bar{Q}},(\wt{12})_{\bar{Q}},3_q,\four,\hatj_{\gamma})|^2\right)J_3^{(3)}(k_{\wt{12}},k_3,k_4)\nonumber\\
&&\:\:\:\:\:\:\:\:\:\:\:\:+\frac{1}{2}A_3^0(i_g;3_q,\four)\left( |\cm(1_Q,\two;;(\wt{34})_q,\hatj_g,\hatbi_q)|^2+|\cm(1_Q,\two;;\hatbi_{\bar{q}},\hatj_g,(\wt{34})_{\bar{q}})|^2\right.\nonumber\\
&&\:\:\:\:\:\:\:\:\:\:\:\:\:\:\:\:\:\:\:\:\:\:\:\:\left. +|\cm(1_Q,\hatj_g,\two;;(\wt{34})_q,\hatbi_q)|^2+|\cm(1_Q,\hatj_g,\two;;\hatbi_{\bar{q}},(\wt{34})_{\bar{q}})|^2\right.\nonumber\\
&&\:\:\:\:\:\:\:\:\:\:\:\:\:\:\:\:\:\:\:\:\:\:\:\:\left. -|\cm(1_Q,\hatbi_q;;(\wt{34})_q,\hatj_g,\two )|^2-|\cm(1_Q,(\wt{34})_{\bar{q}};;\hatbi_{\bar{q}},\hatj_g,\two)|^2\right.\nonumber\\
&&\:\:\:\:\:\:\:\:\:\:\:\:\:\:\:\:\:\:\:\:\:\:\:\:\left. -|\cm(1_Q,\hatj_g,\hatbi_q;;(\wt{34})_q,\two )|^2-|\cm(1_Q,\hatj_g,(\wt{34})_{\bar{q}};;\hatbi_{\bar{q}},\two)|^2\right.\nonumber\\
&&\:\:\:\:\:\:\:\:\:\:\:\:\:\:\:\:\:\:\:\:\:\:\:\:\left. -2|\cm(1_Q,\two,(\wt{34})_q,\hatbi_q,\hatj_{\gamma})|^2-2|\cm(1_Q,\two,\hatbi_{\bar{q}},(\wt{34})_{\bar{q}},\hatj_{\gamma})|^2\right)J_3^{(3)}(k_1,k_2,k_{\wt{34}})\nonumber\\
&&\:\:\:\:\:\:\:\:\:\:\:\:-\frac{1}{2}E_3^0(1_Q,3_q,\four)\left( |\cm((\wt{13})_Q,\hati_g,\hatj_g,(\wt{34})_{\gamma},\two)|^2+|\cm((\wt{13})_Q,\hati_g,(\wt{34})_g,\hatj_{\gamma},\two)|^2\right.\nonumber\\
&&\:\:\:\:\:\:\:\:\:\:\:\:\:\:\:\:\:\:\:\:\:\:\:\:\left. +|\cm((\wt{13})_Q,(\wt{34})_g,\hati_g,\hatj_{\gamma},\two)|^2\right)J_3^{(3)}(k_{\wt{13}},k_2,k_{\wt{34}})\nonumber\\
&&\:\:\:\:\:\:\:\:\:\:\:\:-\frac{1}{2}E_3^0(\two,3_q,\four)\left( |\cm(1_Q,\hati_g,\hatj_g,(\wt{34})_{\gamma},(\wt{23})_{\bar{Q}})|^2 +|\cm(1_Q,\hati_g,(\wt{34})_g,\hatj_{\gamma},(\wt{23})_{\bar{Q}})|^2\right.\nonumber\\
&&\:\:\:\:\:\:\:\:\:\:\:\:\:\:\:\:\:\:\:\:\:\:\:\:\left. +|\cm(1_Q,(\wt{34})_g,\hati_g,\hatj_{\gamma},(\wt{23})_{\bar{Q}})|^2\right)J_3^{(3)}(k_1,k_{\wt{23}},k_{\wt{34}})\nonumber\\
&&\:\:\:\:\:\:\:\:\:\:\:\:-\frac{1}{2N_c^2}\left[A_3^0(i_g;1_Q,\two)\left( |\cm((\wt{12})_Q,\hatbi_Q;;3_q,\hatj_g,\four)|^2+|\cm(\hatbi_{\bar{Q}},(\wt{12})_{\bar{Q}};;3_q,\hatj_g,\four)|^2\right.\right.\nonumber\\
&&\:\:\:\:\:\:\:\:\:\:\:\:\:\:\:\:\:\:\:\:\:\:\:\:\:\:\:\left.\left.  +|\cm((\wt{12})_Q,\hatj_g,\hatbi_Q;;3_q,\four)|^2+|\cm(\hatbi_{\bar{Q}},\hatj_g,(\wt{12})_{\bar{Q}};;3_q,\four)|^2\right.\right.\nonumber\\
&&\:\:\:\:\:\:\:\:\:\:\:\:\:\:\:\:\:\:\:\:\:\:\:\:\:\:\:\left.\left. -2|\cm((\wt{12})_Q,\hatbi_Q,3_q,\four,\hatj_{\gamma})|^2-2|\cm(\hatbi_{\bar{Q}},(\wt{12})_{\bar{Q}},3_q,\four,\hatj_{\gamma})|^2\right)J_3^{(3)}(k_{\wt{12}},k_3,k_4)\right.\nonumber\\
&&\:\:\:\:\:\:\:\:\:\:\:\:\:\:\:\:\left. +A_3^0(i_g;3_q,\four)\left( |\cm(1_Q,\two;;(\wt{34})_q,\hatj_g,\hatbi_q)|^2+|\cm(1_Q,\two;;\hatbi_{\bar{q}},\hatj_g,(\wt{34})_{\bar{q}})|^2\right.\right.\nonumber\\
&&\:\:\:\:\:\:\:\:\:\:\:\:\:\:\:\:\:\:\:\:\:\:\:\:\:\:\:\left.\left. +|\cm(1_Q,\hatj_g,\two;;(\wt{34})_q,\hatbi_q)|^2+|\cm(1_Q,\hatj_g,\two;;\hatbi_{\bar{q}},(\wt{34})_{\bar{q}})|^2\right.\right.\nonumber\\
&&\:\:\:\:\:\:\:\:\:\:\:\:\:\:\:\:\:\:\:\:\:\:\:\:\:\:\:\left.\left. -2|\cm(1_Q,\two,(\wt{34})_q,\hatbi_q,\hatj_{\gamma})|^2-2|\cm(1_Q,\two,\hatbi_{\bar{q}},(\wt{34})_{\bar{q}},\hatj_{\gamma})|^2\right)J_3^{(3)}(k_1,k_2,k_{\wt{34}})\right]\nonumber\\
&&\:\:\:\:\:\:\:\:\:\:\:\:-\frac{N_c^2-1}{4N_c^2}\left( E_3^0(1_Q,3_q,\four)|\cm((\wt{13})_Q,(\wt{34})_{\gamma},\hat{i}_{\gamma},\hat{j}_{\gamma},\two)|^2 J_3^{(3)}(k_{\wt{13}},k_2,k_{\wt{34}})\right.\nonumber\\
&&\:\:\:\:\:\:\:\:\:\:\:\:\:\:\:\:\:\:\:\:\:\:\:\:\:\:\:\:\:\left. +E_3^0(\two,3_q,\four)|\cm(1_Q,(\wt{34})_{\gamma},\hat{i}_{\gamma},\hat{j}_{\gamma},(\wt{23})_{\bar{Q}})|^2 J_3^{(3)}(k_1,k_{\wt{23}},k_{\wt{34}})\right)\bigg]\bigg\}.\nonumber
\end{eqnarray}
For all three crossings considered, we have performed the consistency check explained in Section 7.1.3. We have checked that in all collinear 
or quasi-collinear limits contained in these subtraction terms, the sum of the terms contributing to a given limit collapses 
to the appropriate colour factor multiplying a given massless or massive  
splitting function, corresponding to the given collinear splitting,  
times the appropriate non-colour ordered matrix-element squared.

\subsubsection{Partonic process $gg\rightarrow Q\bar{Q}gg$}
The colour decomposition for the unphysical process $0\rightarrow Q\bar{Q}gggg$ is
\begin{equation}
M_6^0(1_Q,2_{\bar{Q}},3_g,4_g,5_g,6_g)=4g^4\sum_{(i,j,k,l)\in P(3,4,5,6)} (T^{a_i}T^{a_j}T^{a_k}T^{a_l})_{i_1,i_2}{\cal M}(i,j,k,l),
\end{equation}
where we used ${\cal M}(i,j,k,l)=\cms(1_Q,i_g,j_g,k_g,l_g,\two)$ for simplification of the formulae. Squaring gives
\begin{eqnarray}
\lefteqn{|M_6^0(1_Q,2_{\bar{Q}},3_g,4_g,5_g,6_g)|^2=\frac{g^8(N_c^2-1)}{N_c^3}}\nonumber\\
&&\times\bigg\{ \sum_{(i,j,k,l)\in P(3,4,5,6)} \bigg[ N_c^6 |{\cal M}(i,j,k,l)|^2 -N_c^4 |{\cal M}(i,j,k;;l)|^2+\frac{N_c^2}{2!}|{\cal M}(i,j;;k,l)|^2\nonumber\\
&&\:\:\:\:\:\:\:\:\:\:\:\:\:\:\:\:\:\:\:\:\:\:\:\:\:\:\:\:\:\:\:\:\:\:\:\:\:-N_c^4 \re\bigg(\big( {\cal M}(j,i,l,k)+ {\cal M}(j,l,i,k)+ {\cal M}(j,l,k,i)\nonumber\\
&&\:\:\:\:\:\:\:\:\:\:\:\:\:\:\:\:\:\:\:\:\:\:\:\:\:\:\:\:\:\:\:\:\:\:\:\:\:\:\:\:\:\:\:\:\:\:\:\:\:\:\:\:\:+{\cal M}(k,i,l,j)+{\cal M}(k,j,l,i)+{\cal M}(l,i,k,j)\label{eq:gggg}\\
&&\:\:\:\:\:\:\:\:\:\:\:\:\:\:\:\:\:\:\:\:\:\:\:\:\:\:\:\:\:\:\:\:\:\:\:\:\:\:\:\:\:\:\:\:\:\:\:\:\:\:\:\:\:+{\cal M}(l,j,i,k)+{\cal M}(l,k,j,i)\big)\times{\cal M}(i,j,k,l)^{\dagger} \bigg)\nonumber\\
&&\:\:\:\:\:\:\:\:\:\:\:\:\:\:\:\:\:\:\:\:\:\:\:\:\:\:\:\:\:\:\:\:\:\:\:\:\:+\frac{(N_c^4-3N_c^2-1)}{4!}|\bar{{\cal M}}(i,j,k,l)|^2\bigg]\bigg\},\nonumber
\end{eqnarray}
where when the gluon labelled with $l$, when the gluons labelled with $l$ and with $k$ or when all four gluons decouple we have respectively, 
\begin{eqnarray}
{\cal M}(i,j,k;;l)&=&\cms(1_Q,i_g,j_g,k_g,l_{\gamma},2_{\bar{Q}})\\
&=&{\cal M}(i,j,k,l)+{\cal M}(i,j,l,k)+{\cal M}(i,l,j,k)+{\cal M}(l,i,j,k)\nonumber\\ \nonumber\\
{\cal M}(i,j;;k,l)&=&\cms(1_Q,i_g,j_g,k_{\gamma},l_{\gamma},2_{\bar{Q}})\nonumber\\
&=&{\cal M}(i,j,k,l)+{\cal M}(i,j,l,k)+{\cal M}(i,k,j,l)+{\cal M}(i,l,j,k)\\
&&+{\cal M}(i,k,l,j)+{\cal M}(i,l,k,j)+{\cal M}(k,i,j,l)+{\cal M}(l,i,j,k)\nonumber\\
&&+{\cal M}(k,i,l,j)+{\cal M}(l,i,k,j)+{\cal M}(k,l,i,j)+{\cal M}(l,k,i,j)\nonumber\\ \nonumber\\
\bar{{\cal M}}(3,4,5,6)&=&\cms(1_Q,3_{\gamma},4_{\gamma},5_{\gamma},6_{\gamma},\two)\\
&=&\sum_{(i,j,k,l)\in P(3,4,5,6)}\cms(1_Q,i_g,j_g,k_g,l_g,\two)\nonumber
\end{eqnarray}
This colour-ordered decomposition for the squared amplitude for 
$gg\rightarrow Q\bar{Q}gg$ has also been derived in \cite{Mangano}.
Our result is in complete agreement with it. 

Crossing gluons $5_g$ and $6_g$ in eq.(\ref{eq:gggg}) gives the squared matrix element for $gg\rightarrow Q\bar{Q}gg$ reading,
\begin{eqnarray}
\lefteqn{|M_6^0(1_Q,2_{\bar{Q}},3_g,4_g,\hat{5}_g,\hat{6}_g)|^2=\frac{g^8(N_c^2-1)}{N_c^3}}\nonumber\\
&&\times\bigg\{ \sum_{(i,j)\in P(3,4),(k,l)\in P(5,6)}\bigg[ N_c^6 \bigg( |{\cal M}(i,j,\hat{k},\hat{l})|^2+|{\cal M}(i,\hat{k},j,\hat{l})|^2+|{\cal M}(i,\hat{k},\hat{l},j)|^2\nonumber\\
&&\hspace{15mm}+|{\cal M}(\hat{k},i,j\hat{l})|^2+|{\cal M}(\hat{k},i,\hat{l},j)|^2+|{\cal M}(\hat{k},\hat{l},i,j)|^2\bigg)\nonumber\\
&&\hspace{7mm}-N_c^4 \bigg( |{\cal M}(i,j,\hat{k};;\hat{l})|^2+|{\cal M}(i,\hat{k},j;;\hat{l})|^2+|{\cal M}(i,\hat{k},\hat{l};;j)|^2\nonumber\\
&&\hspace{15mm}+|{\cal M}(\hat{k},i,j;;\hat{l})|^2+|{\cal M}(\hat{k},i,\hat{l};;j)|^2+|{\cal M}(\hat{k},\hat{l},i;;j)|^2\bigg)\nonumber\\
&&\hspace{7mm}-N_c^4 \bigg[ \re\bigg(\big( {\cal M}(j,i,\hat{l},\hat{k})+ {\cal M}(j,\hat{l},i,\hat{k})+ {\cal M}(j,\hat{l},\hat{k},i)+{\cal M}(\hat{k},i,\hat{l},j)\nonumber\\
&&\hspace{23mm} +{\cal M}(\hat{k},j,\hat{l},i)+{\cal M}(\hat{l},i,\hat{k},j)+{\cal M}(\hat{l},j,i,\hat{k})+{\cal M}(\hat{l},\hat{k},j,i)\big)\times{\cal M}(i,j,\hat{k},\hat{l})^{\dagger} \bigg)\nonumber\\
&&\hspace{14mm}+\re\bigg(\big( {\cal M}(\hat{k},i,\hat{l},j)+ {\cal M}(\hat{k},\hat{l},i,j)+ {\cal M}(\hat{k},\hat{l},j,i)+{\cal M}(j,i,\hat{l},\hat{k})\nonumber\\
&&\hspace{23mm} +{\cal M}(j,\hat{k},\hat{l},i)+{\cal M}(\hat{l},i,j,\hat{k})+{\cal M}(\hat{l},\hat{k},i,j)+{\cal M}(\hat{l},j,\hat{k},i)\big)\times{\cal M}(i,\hat{k},j,\hat{l})^{\dagger} \bigg)\nonumber\\
&&\hspace{14mm}+\re\bigg(\big( {\cal M}(\hat{k},i,j,\hat{l})+ {\cal M}(\hat{k},j,i,\hat{l})+ {\cal M}(\hat{k},j,\hat{l},i)+{\cal M}(\hat{l},i,j,\hat{k})\nonumber\\
&&\hspace{23mm} +{\cal M}(\hat{l},\hat{k},j,i)+{\cal M}(j,i,\hat{l},\hat{k})+{\cal M}(j,\hat{k},i,\hat{l})+{\cal M}(j,\hat{l},\hat{k},i)\big)\times{\cal M}(i,\hat{k},\hat{l},j)^{\dagger} \bigg)\nonumber\\
&&\hspace{14mm}+\re\bigg(\big( {\cal M}(i,\hat{k},\hat{l},j)+ {\cal M}(i,\hat{l},\hat{k},j)+ {\cal M}(i,\hat{l},j,\hat{k})+{\cal M}(j,\hat{k},\hat{l},i)\nonumber\\
&&\hspace{23mm} +{\cal M}(j,i,\hat{l},\hat{k})+{\cal M}(\hat{l},\hat{k},j,i)+{\cal M}(\hat{l},i,\hat{k},j)+{\cal M}(\hat{l},j,i,\hat{k})\big)\times{\cal M}(\hat{k},i,j,\hat{l})^{\dagger} \bigg)\nonumber\\
&&\hspace{14mm}+\re\bigg(\big( {\cal M}(i,\hat{k},j,\hat{l})+ {\cal M}(i,j,\hat{k},\hat{l})+ {\cal M}(i,j,\hat{l},\hat{k})+{\cal M}(\hat{l},\hat{k},j,i)\nonumber\\
&&\hspace{23mm} +{\cal M}(\hat{l},i,j,\hat{k})+{\cal M}(j,\hat{k},\hat{l},i)+{\cal M}(j,i,\hat{k},\hat{l})+{\cal M}(j,\hat{l},i,\hat{k})\big)\times{\cal M}(\hat{k},i,\hat{l},j)^{\dagger} \bigg)\nonumber\\
&&\hspace{14mm}+\re\bigg(\big( {\cal M}(\hat{l},\hat{k},j,i)+ {\cal M}(\hat{l},j,\hat{k},i)+ {\cal M}(\hat{l},j,i,\hat{k})+{\cal M}(i,\hat{k},j,\hat{l})\nonumber\\
&&\hspace{23mm} +{\cal M}(i,\hat{l},j,\hat{k})+{\cal M}(j,\hat{k},i,\hat{l})+{\cal M}(j,\hat{l},\hat{k},i)+{\cal M}(j,i,\hat{l},\hat{k})\big)\times{\cal M}(\hat{k},\hat{l},i,j)^{\dagger} \bigg)\bigg]\nonumber\\
&&\hspace{7mm}+N_c^2\bigg[ \frac{1}{2}|{\cal M}(i,j;;\hat{k},\hat{l})|^2+|{\cal M}(i,\hat{k};;j,\hat{l})|^2+|{\cal M}(\hat{k},i;;j,\hat{l})|^2+\frac{1}{2}|{\cal M}(\hat{k},\hat{l};;i,k)|^2\bigg]\nonumber\\
&&\hspace{7mm}+\frac{(N_c^4-3N_c^2-1)}{4}|\bar{{\cal M}}(i,j,\hat{k},\hat{l})|^2\bigg]\bigg\},\nonumber\\
\end{eqnarray}
The expression above is obtained as follows. 
We start by expanding the sum over permutations for all four gluons 
present in the final state  in the colour decomposition 
for the unphysical process $0\rightarrow Q\bar{Q}gggg$ , given in eq.(\ref{eq:gggg}). Then, gluons $5_g$ and $6_g$ are crossed to the initial state. 
Finally, the terms are recombined in a double sum over permutations
for the two initial state and the two final state gluons.
The corresponding subtraction term reads,
\begin{eqnarray}
\lefteqn{{\rm d}\hat{\sigma}_{gg\rightarrow Q\bar{Q}gg}^S=g^8(N_c^2-1){\rm d}\Phi_4(k_{1Q},k_{2\bar{Q}},k_{3g},k_{4g};p_{5g},p_{6g})}\nonumber\\
&&\times\sum_{(i,j)\in P(3,4),(k,l)\in P(5,6)}\bigg\{ \nonumber\\
&&\:\:N_c^3\bigg[ \frac{1}{2}A_3^0(l_g;1_Q,\two)\left( |\cm((\wt{12})_Q,i_g,j_g,\hatk,\hatbl_Q)|^2+|\cm(\hatbl_{\bar{Q}},i_g,j_g,\hatk,(\wt{12})_{\bar{Q}})|^2\right.\nonumber\\
&&\:\:\:\:\:\:\:\:\:\:\:\:\:\:\:\:\:\:\:\:\left.+|\cm((\wt{12})_Q,i_g,\hatk,j_g,\hatbl_Q)|^2+|\cm(\hatbl_{\bar{Q}},i_g,\hatk,j_g,(\wt{12})_{\bar{Q}})|^2\right.\nonumber\\
&&\:\:\:\:\:\:\:\:\:\:\:\:\:\:\:\:\:\:\:\:\left.+|\cm((\wt{12})_Q,\hatk,i_g,j_g,\hatbl_Q)|^2+|\cm(\hatbl_{\bar{Q}},\hatk,i_g,j_g,(\wt{12})_{\bar{Q}})|^2\right)J_3^{(3)}(k_{\wt{12}},k_3,k_4)\nonumber\\
&&\:\:\:\:\:\:\:\:\:\:\: +d_3^0(1_Q,i_g,j_g)|\cm((\wt{1i})_Q,(\wt{ij})_g,\hatk,\hatl_g,\two)|^2 J_3^{(3)}(k_{\wt{1i}},k_2,k_{\wt{ij}})\nonumber\\
&&\:\:\:\:\:\:\:\:\:\:\: +d_3^0(\two,i_g,j_g)|\cm(1_Q,\hatk,\hatl_g,(\wt{ij})_g,(\wt{2i})_{\bar{Q}})|^2 J_3^{(3)}(k_1,k_{\wt{2i}},k_{\wt{ij}})\nonumber\\
&&\:\:\:\:\:\:\:\:\:\:\: +D_3^0(l_g;i_g,1_Q)\left( |\cm((\wt{1i})_Q,\hatbl_g,j_g,\hatk,\two)|^2\right.\nonumber\\
&&\:\:\:\:\:\:\:\:\:\:\:\:\:\:\:\:\:\:\:\:\left.+|\cm((\wt{1i})_Q,\hatbl_g,\hatk,j_g,\two)\right)|^2J_3^{(3)}(k_{\wt{1i}},k_2,k_j)\nonumber\\
&&\:\:\:\:\:\:\:\:\:\:\: +D_3^0(l_g;i_g,\two)\left( |\cm(1_Q,j_g,\hatk,\hatbl_g,(\wt{2i})_{\bar{Q}})|^2\right.\nonumber\\
&&\:\:\:\:\:\:\:\:\:\:\:\:\:\:\:\:\:\:\:\:\left.+ |\cm(1_Q,\hatk,j_g,\hatbl_g,(\wt{2i})_{\bar{Q}})|^2\right)J_3^{(3)}(k_1,k_{\wt{2i}},k_j)\nonumber\\
&&\:\:\:\:\:\:\:\:\:\:\: +f_3^0(l_g;i_g,j_g)\left(|\cm(1_Q,(\wt{ij})_g,\hatbl_g,\hatk,\two)|^2+|\cm(1_Q,\hatbl_g,(\wt{ij})_g,\hatk,\two)|^2\right.\nonumber\\
&&\:\:\:\:\:\:\:\:\:\:\:\:\:\:\:\:\:\:\:\:\left.+|\cm(1_Q,\hatk,(\wt{ij})_g,\hatbl_g,\two)|^2+|\cm(1_Q,\hatk,\hatbl_g,(\wt{ij})_g,\two)|^2\right)J_3^{(3)}(k_1,k_2,k_{\wt{ij}})\nonumber\\
&&\:\:\:\:\:\:\:\:\:\:\: +F_3^0(k_g,l_g;i_g)\left( |\cm(\tilde{1}_Q,\hatbk_g,\hatbl_g,\tilde{j}_g,\tilde{2}_{\bar{Q}})|^2+|\cm(\tilde{1}_Q,\tilde{j}_g,\hatbk_g,\hatbl_g,\tilde{2}_{\bar{Q}})|^2\right)J_3^{(3)}(\tilde{k}_1,\tilde{k}_2,\tilde{k}_j)\bigg]\nonumber\\
&&-N_c\bigg[ A_3^0(1_Q,i_g,\two)\left( |\cm((\wt{1i})_Q,j_g,\hatk,\hatl_g,(\wt{2i})_{\bar{Q}})|^2-(1/2)|\cm((\wt{1i})_Q,j_{\gamma},\hat{k}_{\gamma},\hatl_{\gamma},(\wt{2i})_{\bar{Q}})|^2\right.\nonumber\\
&&\:\:\:\:\:\:\:\:\:\:\:\:\:\:\:\:\:\:\:\:\left. +|\cm((\wt{1i})_Q,\hatk,j_g,\hatl_g,(\wt{2i})_{\bar{Q}})|^2+|\cm((\wt{1i})_Q,\hatk,\hatl_g,j_g,(\wt{2i})_{\bar{Q}})|^2\right.\nonumber\\
&&\:\:\:\:\:\:\:\:\:\:\:\:\:\:\:\:\:\:\:\:\left. +\re( \cm((\wt{1i})_Q,j_g,\hatk,\hatl_g,(\wt{2i})_{\bar{Q}})\cm((\wt{1i})_Q,j_g,\hatl_g,\hatk,(\wt{2i})_{\bar{Q}})^{\dagger})\right.\nonumber\\
&&\:\:\:\:\:\:\:\:\:\:\:\:\:\:\:\:\:\:\:\:\left. +\re( \cm((\wt{1i})_Q,\hatk,j_g,\hatl_g,(\wt{2i})_{\bar{Q}})\cm((\wt{1i})_Q,\hatl_g,j_g,\hatk,(\wt{2i})_{\bar{Q}})^{\dagger})\right.\nonumber\\
&&\:\:\:\:\:\:\:\:\:\:\:\:\:\:\:\:\:\:\:\:\left. +\re( \cm((\wt{1i})_Q,\hatk,\hatl_g,j_g,(\wt{2i})_{\bar{Q}})\cm((\wt{1i})_Q,\hatl_g,\hatk,j_g,(\wt{2i})_{\bar{Q}})^{\dagger})\right.\nonumber\\
&&\:\:\:\:\:\:\:\:\:\:\:\:\:\:\:\:\:\:\:\:\left. +2\re( \cm((\wt{1i})_Q,j_g,\hatk,\hatl_g,(\wt{2i})_{\bar{Q}})\cm((\wt{1i})_Q,\hatk,j_g,\hatl_g,(\wt{2i})_{\bar{Q}})^{\dagger})\right.\nonumber\\
&&\:\:\:\:\:\:\:\:\:\:\:\:\:\:\:\:\:\:\:\:\left. +2\re( \cm((\wt{1i})_Q,j_g,\hatk,\hatl_g,(\wt{2i})_{\bar{Q}})\cm((\wt{1i})_Q,\hatl_g,\hatk,j_g,(\wt{2i})_{\bar{Q}})^{\dagger})\right.\nonumber\\
&&\:\:\:\:\:\:\:\:\:\:\:\:\:\:\:\:\:\:\:\:\left.+2\re( \cm((\wt{1i})_Q,\hatk,j_g,\hatl_g,(\wt{2i})_{\bar{Q}})\cm((\wt{1i})_Q,\hatk,\hatl_g,j_g,(\wt{2i})_{\bar{Q}})^{\dagger})\right)J_3^{(3)}(k_{\wt{1i}},k_{\wt{2i}},k_j)\nonumber\\
&&\:\:\:\:\:\:\:\:\:\:\:+\frac{1}{2}A_3^0(l_g;1_Q,2_{\bar{Q}})\left( |\cm((\wt{12})_Q,i_g,j_g,\hatk,\hatbl_Q)|^2+|\cm(\hatbl_{\bar{Q}},i_g,j_g,\hatk,(\wt{12})_{\bar{Q}})|^2\right.\nonumber\\
&&\:\:\:\:\:\:\:\:\:\:\:\:\:\:\:\:\:\:\:\:\left.+|\cm((\wt{12})_Q,i_g,\hatk,j_g,\hatbl_Q)|^2+|\cm(\hatbl_{\bar{Q}},i_g,\hatk,j_g,(\wt{12})_{\bar{Q}})|^2\right.\nonumber\\
&&\:\:\:\:\:\:\:\:\:\:\:\:\:\:\:\:\:\:\:\:\left.+|\cm((\wt{12})_Q,\hatk,i_g,j_g,\hatbl_Q)|^2+|\cm(\hatbl_{\bar{Q}},\hatk,i_g,j_g,(\wt{12})_{\bar{Q}})|^2\right.\nonumber\\
&&\:\:\:\:\:\:\:\:\:\:\:\:\:\:\:\:\:\:\:\:\left. +|\cm((\wt{12})_Q,i_g,j_g,\hatkph,\hatbl_Q)|^2 +|\cm(\hatbl_{\bar{Q}},i_g,j_g,\hatkph,(\wt{12})_{\bar{Q}})|^2\right.\nonumber\\
&&\:\:\:\:\:\:\:\:\:\:\:\:\:\:\:\:\:\:\:\:\left. +|\cm((\wt{12})_Q,i_g,\hatk,j_{\gamma},\hatbl_Q)|^2 +|\cm(\hatbl_{\bar{Q}},i_g,\hatk,j_{\gamma},(\wt{12})_{\bar{Q}})|^2\right.\nonumber\\
&&\:\:\:\:\:\:\:\:\:\:\:\:\:\:\:\:\:\:\:\:\left. +|\cm((\wt{12})_Q,\hatk,i_g,j_{\gamma},\hatbl_Q)|^2+|\cm(\hatbl_{\bar{Q}},\hatk,i_g,j_{\gamma},(\wt{12})_{\bar{Q}})|^2 \right.\nonumber\\
&&\:\:\:\:\:\:\:\:\:\:\:\:\:\:\:\:\:\:\:\:\left. -(1/2)|\cm((\wt{12})_Q,i_{\gamma},j_{\gamma},\hat{k}_{\gamma},\hatbl_Q)|^2\right.\nonumber\\
&&\:\:\:\:\:\:\:\:\:\:\:\:\:\:\:\:\:\:\:\:\left.-(1/2)|\cm(\hatbl_{\bar{Q}}),i_{\gamma},j_{\gamma},\hat{k}_{\gamma},(\wt{12})_{\bar{Q}})|^2\right)J_3^{(3)}(k_{\wt{12}},k_i,k_j)\nonumber\\
&&\:\:\:\:\:\:\:\:\:\:\: +d_3^0(1_Q,i_g,j_g)\left( |\cm((\wt{1i})_Q,(\wt{ij})_g,\hatk,\hatl_{\gamma},\two)|^2 \right.\nonumber\\
&&\:\:\:\:\:\:\:\:\:\:\:\:\:\:\:\:\:\:\:\:\left.+\re(\cm((\wt{1i})_Q,\hatk,(\wt{ij})_g,\hatl_g,\two) \cm((\wt{1i})_Q,\hatl_g,(\wt{ij})_g,\hatk,\two)^{\dagger})\right.\nonumber\\
&&\:\:\:\:\:\:\:\:\:\:\:\:\:\:\:\:\:\:\:\:\left.-\re(\cm((\wt{1i})_Q,(\wt{ij})_g,\hatk,\hatl_g,\two) \cm((\wt{1i})_Q,(\wt{ij})_g,\hatl_g,\hatk,\two)^{\dagger})\right.\nonumber\\
&&\:\:\:\:\:\:\:\:\:\:\:\:\:\:\:\:\:\:\:\:\left.-\re(\cm((\wt{1i})_Q,\hatk,\hatl_g,(\wt{ij})_g,\two) \cm((\wt{1i})_Q,\hatl_g,\hatk,(\wt{ij})_g,\two)^{\dagger})\right.\nonumber\\
&&\:\:\:\:\:\:\:\:\:\:\:\:\:\:\:\:\:\:\:\:\left.+2\re(\cm((\wt{1i})_Q,\hatk,\hatl_g,(\wt{ij})_g,\two) \cm((\wt{1i})_Q,\hatl_g,(\wt{ij})_g,\hatk,\two)^{\dagger})\right.\nonumber\\
&&\:\:\:\:\:\:\:\:\:\:\:\:\:\:\:\:\:\:\:\:\left.-2\re(\cm((\wt{1i})_Q,(\wt{ij})_g,\hatk,\hatl_g,\two) \cm((\wt{1i})_Q,\hatl_g,(\wt{ij})_g,\hatk,\two)^{\dagger})\right.\nonumber\\
&&\:\:\:\:\:\:\:\:\:\:\:\:\:\:\:\:\:\:\:\:\left.-2\re(\cm((\wt{1i})_Q,(\wt{ij})_g,\hatk,\hatl_g,\two) \cm((\wt{1i})_Q,\hatl_g,\hatk,(\wt{ij})_g,\two)^{\dagger})\right)J_3^{(3)}(k_{\wt{1i}},k_2,k_{\wt{ij}})\nonumber\\
&&\:\:\:\:\:\:\:\:\:\:\:+d_3^0(\two,i_g,j_g)\left( |\cm(1_Q,\hatk,(\wt{ij})_g,\hatl_{\gamma},(\wt{2i})_{\bar{Q}})|^2 \right.\nonumber\\
&&\:\:\:\:\:\:\:\:\:\:\:\:\:\:\:\:\:\:\:\:\left.+\re( \cm(1_Q,\hatk,(\wt{ij})_g,\hatl_g,(\wt{2i})_{\bar{Q}}) \cm(1_Q,\hatl_g,(\wt{ij})_g,\hatk,(\wt{2i})_{\bar{Q}})^{\dagger})\right.\nonumber\\
&&\:\:\:\:\:\:\:\:\:\:\:\:\:\:\:\:\:\:\:\:\left.-\re( \cm(1_Q,(\wt{ij})_g,\hatk,\hatl_g,(\wt{2i})_{\bar{Q}}) \cm(1_Q,(\wt{ij})_g,\hatl_g,\hatk,(\wt{2i})_{\bar{Q}})^{\dagger})\right.\nonumber\\
&&\:\:\:\:\:\:\:\:\:\:\:\:\:\:\:\:\:\:\:\:\left. -\re( \cm(1_Q,\hatk,\hatl_g,(\wt{ij})_g,(\wt{2i})_{\bar{Q}}) \cm(1_Q,\hatl_g,\hatk,(\wt{ij})_g,(\wt{2i})_{\bar{Q}})^{\dagger})\right.\nonumber\\
&&\:\:\:\:\:\:\:\:\:\:\:\:\:\:\:\:\:\:\:\:\left. +2\re( \cm(1_Q,(\wt{ij})_g,\hatk,\hatl_g,(\wt{2i})_{\bar{Q}}) \cm(1_Q,\hatl_g,(\wt{ij})_g,\hatk,(\wt{2i})_{\bar{Q}})^{\dagger})\right.\nonumber\\
&&\:\:\:\:\:\:\:\:\:\:\:\:\:\:\:\:\:\:\:\:\left.-2\re( \cm(1_Q,\hatk,\hatl_g,(\wt{ij})_g,(\wt{2i})_{\bar{Q}}) \cm(1_Q,\hatl_g,(\wt{ij})_g,\hatk,(\wt{2i})_{\bar{Q}})^{\dagger})\right.\nonumber\\
&&\:\:\:\:\:\:\:\:\:\:\:\:\:\:\:\:\:\:\:\:\left.-2\re( \cm(1_Q,(\wt{ij})_g,\hatk,\hatl_g,(\wt{2i})_{\bar{Q}}) \cm(1_Q,\hatl_g,\hatk,(\wt{ij})_g,(\wt{2i})_{\bar{Q}})^{\dagger})\right) J_3^{(3)}(k_1,k_{\wt{2i}},k_{\wt{ij}})\nonumber\\
&&\:\:\:\:\:\:\:\:\:\:\: +D_3^0(l_g;i_g,1_Q)\left( |\cm((\wt{1i})_Q,\hatbl_g,j_g,\hatkph,\two)|^2+|\cm((\wt{1i})_Q,\hatbl_g,\hatk,j_{\gamma},\two)|^2\right.\nonumber\\
&&\:\:\:\:\:\:\:\:\:\:\:\:\:\:\:\:\:\:\:\:\left.+ 2\re (\cm((\wt{1i})_Q,j_g,\hatbl_g,\hatk,\two) \cm((\wt{1i})_Q,\hatk,j_g,\hatbl_g,\two)^{\dagger}) \right.\nonumber\\
&&\:\:\:\:\:\:\:\:\:\:\:\:\:\:\:\:\:\:\:\:\left. + 2\re (\cm((\wt{1i})_Q,j_g,\hatbl_g,\hatk,\two) \cm((\wt{1i})_Q,\hatk,\hatbl_g,j_g,\two)^{\dagger})\right.\nonumber\\
&&\:\:\:\:\:\:\:\:\:\:\:\:\:\:\:\:\:\:\:\:\left.+ 2\re (\cm((\wt{1i})_Q,j_g,\hatk,\hatbl_g,\two) \cm((\wt{1i})_Q,\hatk,\hatbl_g,j_g,\two)^{\dagger})\right.\nonumber\\
&&\:\:\:\:\:\:\:\:\:\:\:\:\:\:\:\:\:\:\:\:\left.- 2\re (\cm((\wt{1i})_Q,j_g,\hatbl_g,\hatk,\two) \cm((\wt{1i})_Q,\hatbl_g,\hatk,j_g,\two)^{\dagger}) \right.\nonumber\\
&&\:\:\:\:\:\:\:\:\:\:\:\:\:\:\:\:\:\:\:\:\left.- 2\re (\cm((\wt{1i})_Q,j_g,\hatk,\hatbl_g,\two) \cm((\wt{1i})_Q,\hatbl_g,\hatk,j_g,\two)^{\dagger}) \right.\nonumber\\
&&\:\:\:\:\:\:\:\:\:\:\:\:\:\:\:\:\:\:\:\:\left.- 2\re (\cm((\wt{1i})_Q,\hatbl_g,j_g,\hatk,\two) \cm((\wt{1i})_Q,\hatbl_g,\hatk,j_g,\two)^{\dagger}) \right.\nonumber\\
&&\:\:\:\:\:\:\:\:\:\:\:\:\:\:\:\:\:\:\:\:\left. - 2\re (\cm((\wt{1i})_Q,j_g,\hatk,\hatbl_g,\two) \cm((\wt{1i})_Q,\hatk,j_g,\hatbl_g,\two)^{\dagger})\right.\nonumber\\
&&\:\:\:\:\:\:\:\:\:\:\:\:\:\:\:\:\:\:\:\:\left. - 2\re (\cm((\wt{1i})_Q,\hatbl_g,j_g,\hatk,\two) \cm((\wt{1i})_Q,\hatk,j_g,\hatbl_g,\two)^{\dagger})\right.\nonumber\\
&&\:\:\:\:\:\:\:\:\:\:\:\:\:\:\:\:\:\:\:\:\left.- 2\re (\cm((\wt{1i})_Q,\hatbl_g,j_g,\hatk,\two) \cm((\wt{1i})_Q,\hatk,\hatbl_g,j_g,\two)^{\dagger}) \right)|^2J_3^{(3)}(k_{\wt{1i}},k_2,k_j)\nonumber\\
&&\:\:\:\:\:\:\:\:\:\:\: +D_3^0(l_g;i_g,\two)\left( |\cm(1_Q,j_g,\hatbl_g,\hatkph,(\wt{2i})_{\bar{Q}})|^2+ |\cm(1_Q,\hatk,\hatbl_g,j_{\gamma},(\wt{2i})_{\bar{Q}})|^2\right.\nonumber\\
&&\:\:\:\:\:\:\:\:\:\:\:\:\:\:\:\:\:\:\:\:\left.+2\re (\cm(1_Q,j_g,\hatbl_g,\hatk,(\wt{2i})_{\bar{Q}}) \cm(1_Q,\hatbl_g,\hatk,j_g,(\wt{2i})_{\bar{Q}})^{\dagger})\right.\nonumber\\
&&\:\:\:\:\:\:\:\:\:\:\:\:\:\:\:\:\:\:\:\:\left. +2\re (\cm(1_Q,j_g,\hatbl_g,\hatk,(\wt{2i})_{\bar{Q}}) \cm(1_Q,\hatk,\hatbl_g,j_g,(\wt{2i})_{\bar{Q}})^{\dagger})\right.\nonumber\\
&&\:\:\:\:\:\:\:\:\:\:\:\:\:\:\:\:\:\:\:\:\left. +2\re (\cm(1_Q,\hatbl_g,j_g,\hatk,(\wt{2i})_{\bar{Q}}) \cm(1_Q,\hatk,\hatbl_g,j_g,(\wt{2i})_{\bar{Q}})^{\dagger})\right.\nonumber\\
&&\:\:\:\:\:\:\:\:\:\:\:\:\:\:\:\:\:\:\:\:\left. - 2\re (\cm(1_Q,j_g,\hatk,\hatbl_g,(\wt{2i})_{\bar{Q}}) \cm(1_Q,\hatbl_g,\hatk,j_g,(\wt{2i})_{\bar{Q}})^{\dagger})\right.\nonumber\\
&&\:\:\:\:\:\:\:\:\:\:\:\:\:\:\:\:\:\:\:\:\left.- 2\re (\cm(1_Q,\hatbl_g,j_g,\hatk,(\wt{2i})_{\bar{Q}}) \cm(1_Q,\hatbl_g,\hatk,j_g,(\wt{2i})_{\bar{Q}})^{\dagger})\right.\nonumber\\
&&\:\:\:\:\:\:\:\:\:\:\:\:\:\:\:\:\:\:\:\:\left. - 2\re (\cm(1_Q,j_g,\hatbl_g,\hatk,(\wt{2i})_{\bar{Q}}) \cm(1_Q,\hatk,j_g,\hatbl_g,(\wt{2i})_{\bar{Q}})^{\dagger})\right.\nonumber\\
&&\:\:\:\:\:\:\:\:\:\:\:\:\:\:\:\:\:\:\:\:\left. - 2\re (\cm(1_Q,j_g,\hatk,\hatbl_g,(\wt{2i})_{\bar{Q}}) \cm(1_Q,\hatk,j_g,\hatbl_g,(\wt{2i})_{\bar{Q}})^{\dagger})\right.\nonumber\\
&&\:\:\:\:\:\:\:\:\:\:\:\:\:\:\:\:\:\:\:\:\left. - 2\re (\cm(1_Q,\hatbl_g,j_g,\hatk,(\wt{2i})_{\bar{Q}}) \cm(1_Q,\hatk,j_g,\hatbl_g,(\wt{2i})_{\bar{Q}})^{\dagger})\right.\nonumber\\
&&\:\:\:\:\:\:\:\:\:\:\:\:\:\:\:\:\:\:\:\:\left. - 2\re (\cm(1_Q,j_g,\hatk,\hatbl_g,(\wt{2i})_{\bar{Q}}) \cm(1_Q,\hatk,\hatbl_g,j_g,(\wt{2i})_{\bar{Q}})^{\dagger})\right)J_3^{(3)}(k_1,k_{\wt{2i}},k_j)\nonumber\\
&&\:\:\:\:\:\:\:\:\:\:\: +f_3^0(l_g;i_g,j_g)\left(|\cm(1_Q,(\wt{ij})_g,\hatbl_g,\hatkph,\two)|^2+|\cm(1_Q,\hatbl_g,(\wt{ij})_g,\hatkph,\two)|^2\right.\nonumber\\
&&\:\:\:\:\:\:\:\:\:\:\:\:\:\:\:\:\:\:\:\:\left. +4\re(\cm(1_Q,(\wt{ij})_g,\hatk,\hatbl_g,\two)\cm(1_Q,\hatbl_g,\hatk,(\wt{ij})_g,\two)^{\dagger})\right.\nonumber\\
&&\:\:\:\:\:\:\:\:\:\:\:\:\:\:\:\:\:\:\:\:\left. - 4\re(\cm(1_Q,(\wt{ij})_g,\hatbl_g,\hatk,\two)\cm(1_Q,\hatk,(\wt{ij})_g,\hatbl_g,\two)^{\dagger})\right.\nonumber\\
&&\:\:\:\:\:\:\:\:\:\:\:\:\:\:\:\:\:\:\:\:\left. - 4\re(\cm(1_Q,\hatbl_g,(\wt{ij})_g,\hatk,\two)\cm(1_Q,\hatk,(\wt{ij})_g,\hatbl_g,\two)^{\dagger})\right.\nonumber\\
&&\:\:\:\:\:\:\:\:\:\:\:\:\:\:\:\:\:\:\:\:\left. - 4\re(\cm(1_Q,(\wt{ij})_g,\hatbl_g,\hatk,\two)\cm(1_Q,\hatk,\hatbl_g,(\wt{ij})_g,\two)^{\dagger})\right.\nonumber\\
&&\:\:\:\:\:\:\:\:\:\:\:\:\:\:\:\:\:\:\:\:\left. - 4\re(\cm(1_Q,\hatbl_g,(\wt{ij})_g,\hatk,\two)\cm(1_Q,\hatk,\hatbl_g,(\wt{ij})_g,\two)^{\dagger})\right)J_3^{(3)}(k_1,k_2,k_{\wt{ij}})\nonumber\\
&&\:\:\:\:\:\:\:\:\:\:\:  +F_3^0(k_g,l_g;i_g)\left( |\cm(\tilde{1}_Q,\hatbk_g,\hatbl_g,\tilde{j}_{\gamma},\tilde{2}_{\bar{Q}})|^2\right.\nonumber\\
&&\:\:\:\:\:\:\:\:\:\:\:\:\:\:\:\:\:\:\:\:\left.+2\re(\cm(\tilde{1}_Q,\hatbk_g,\tilde{j}_g,\hatbl_g,\tilde{2}_{\bar{Q}}) \cm(\tilde{1}_Q,\hatbl_g,\tilde{j}_g,\hatbk_g,\tilde{2}_{\bar{Q}})^{\dagger})\right.\nonumber\\
&&\:\:\:\:\:\:\:\:\:\:\:\:\:\:\:\:\:\:\:\:\left. -4\re(\cm(\tilde{1}_Q,\tilde{j}_g,\hatbk_g,\hatbl_g,\tilde{2}_{\bar{Q}}) \cm(\tilde{1}_Q,\hatbk_g,\hatbl_g,\tilde{j}_g,\tilde{2}_{\bar{Q}})^{\dagger})\right.\nonumber\\
&&\:\:\:\:\:\:\:\:\:\:\:\:\:\:\:\:\:\:\:\:\left. -4\re(\cm(\tilde{1}_Q,\tilde{j}_g,\hatbk_g,\hatbl_g,\tilde{2}_{\bar{Q}}) \cm(\tilde{1}_Q,\hatbl_g,\hatbk_g,\tilde{j}_g,\tilde{2}_{\bar{Q}})^{\dagger})\right)J_3^{(3)}(\tilde{k}_1,\tilde{k}_2,\tilde{k}_j)\bigg]\nonumber\\
&&+\frac{1}{N_c}\bigg[A_3^0(1_Q,i_g,\two)\left( |\cm((\wt{1i})_Q,j_g,\hatk,\hatl_{\gamma},(\wt{2i})_{\bar{Q}})|^2+ |\cm((\wt{1i})_Q,\hatk,j_g,\hatl_{\gamma},(\wt{2i})_{\bar{Q}})|^2\right.\nonumber\\
&&\:\:\:\:\:\:\:\:\:\:\:\:\:\:\:\:\:\:\:\:\left.+ |\cm((\wt{1i})_Q,\hatk,\hatl_g,j_{\gamma},(\wt{2i})_{\bar{Q}})|^2\right.\nonumber\\
&&\:\:\:\:\:\:\:\:\:\:\:\:\:\:\:\:\:\:\:\:\left. -(3/2)|\cm((\wt{1i})_Q,j_{\gamma},\hat{k}_{\gamma},\hatl_{\gamma},(\wt{2i})_{\bar{Q}})|^2 \right)J_3^{(3)}(k_{\wt{1i}},k_{\wt{2i}},k_j)\nonumber\\
&&\:\:\:\:\:\:\:\:\:\:\:+\frac{1}{2}A_3^0(l_g;1_Q,\two)\left( |\cm((\wt{12})_Q,i_g,j_g,\hat{k}_{\gamma},\hatbl_Q)|^2+|\cm(\hatbl_{\bar{Q}},i_g,j_g,\hat{k}_{\gamma},(\wt{12})_{\bar{Q}})|^2\right.\nonumber\\
&&\:\:\:\:\:\:\:\:\:\:\:\:\:\:\:\:\:\:\:\:\left. +|\cm((\wt{12})_Q,i_g,\hat{k}_g,j_{\gamma},\hatbl_Q)|^2+|\cm(\hatbl_{\bar{Q}},i_g,\hat{k}_g,j_{\gamma},(\wt{12})_{\bar{Q}})|^2\right.\nonumber\\
&&\:\:\:\:\:\:\:\:\:\:\:\:\:\:\:\:\:\:\:\:\left. +|\cm((\wt{12})_Q,\hat{k}_g,i_g,j_{\gamma},\hatbl_Q)|^2+|\cm(\hatbl_{\bar{Q}},\hat{k}_g,i_g,j_{\gamma},(\wt{12})_{\bar{Q}})|^2 \right)J_3^{(3)}(k_{\wt{12}},k_i,k_j)\nonumber\\
&&\:\:\:\:\:\:\:\:\:\:\:+\frac{1}{2} d_3^0(1_Q,i_g,j_g) |\cm((\wt{1i})_Q,(\wt{ij})_{\gamma},\hat{k}_{\gamma},\hatl_{\gamma},\two)|^2 J_3^{(3)}(k_{\wt{1i}},k_2,k_{\wt{ij}})\nonumber\\
&&\:\:\:\:\:\:\:\:\:\:\:+\frac{1}{2} d_3^0(\two,i_g,j_g) |\cm(1_Q,(\wt{ij})_{\gamma},\hat{k}_{\gamma},\hatl_{\gamma},(\wt{2i})_{\bar{Q}})|^2 J_3^{(3)}(k_1,k_{\wt{2i}},k_{\wt{ij}}) \nonumber\\
&&\:\:\:\:\:\:\:\:\:\:\:+D_3^0(l_g;i_g,1_Q)|\cm((\wt{1i})_Q,j_{\gamma},\hat{k}_{\gamma},\hatbl_{\gamma},2_{\bar{Q}})|^2 J_3^{(3)}(k_{\wt{1i}},k_2,k_j)\nonumber\\
&&\:\:\:\:\:\:\:\:\:\:\:+D_3^0(l_g;i_g,\two)|\cm(1_Q,j_{\gamma},\hat{k}_{\gamma},\hatbl_{\gamma},(\wt{2i})_{\bar{Q}})|^2 J_3^{(3)}(k_1,k_{\wt{2i}},k_j)\bigg]\nonumber\\
&&-\frac{1}{2N_c^3}\bigg[ A_3^0(1_Q,i_g,\two)|\cm((\wt{1i})_Q,j_{\gamma},\hat{k}_{\gamma},\hatl_{\gamma},(\wt{2i})_{\bar{Q}})|^2 J_3^{(3)}(k_{\wt{1i}},k_{\wt{2i}},k_j)\nonumber\\
&&\:\:\:\:\:\:\:\:\:\:\:+\frac{1}{2}A_3^0(l_g;1_Q,\two)\left( |\cm((\wt{12})_Q,i_{\gamma},j_{\gamma},\hat{k}_{\gamma},\hatbl_Q)|^2\right.\nonumber\\
&&\:\:\:\:\:\:\:\:\:\:\:\:\:\:\:\:\:\:\:\:\:\:\:\:\:\:\:\:\:\left. +|\cm(\hatbl_{\bar{Q}},i_{\gamma},j_{\gamma},\hat{k}_{\gamma},(\wt{12})_{\bar{Q}})|^2 \right)J_3^{(3)}(k_{\wt{12}},k_i,k_j)\bigg]\bigg\}\nonumber
\end{eqnarray}

For this subtraction term, we have also checked that in all its collinear and  
quasi-collinear limits, it reduces to the product of a 
Casimir factor multiplied by a splitting function 
corresponding to the given limit and the appropriate non-colour 
ordered matrix-element squared as explained in Section 7.4.1, 
providing us with a powerful check on the correctness of our result.

\section{Conclusions}
We have presented the extension of the antenna formalism required
for the calculation of hadronic processes involving massive final
states in association with jets at the NLO level.
The construction of massive subtraction terms 
with all its mass-dependent ingredients is presented in all required
configurations (final-final, initial-final, initial-initial).
The unknown massive antenna functions are derived, their limiting
behaviour is presented and those are finally integrated
over a factorised form of the massive phase space. 
Besides the massive extension of the flavour-conserving antennae, 
new massive flavour-violating antennae were derived in unintegrated and 
integrated forms.
One of the integrated massive initial-final antenna,
${\cal A}_{g;Q\bar{Q}}$, can be directly related to the well-known 
heavy quark coefficient function. A special section is dedicated 
to this comparison and full agreement is found.
In Section 5, when all antennae are integrated over the appropriate
factorised massive phase space,  we showed that we can capture   
all poles of the massive integrated antennae in  universal factors.
Those poles are related either to massive colour 
ordered ${\bf I}^{(1)}_{ij}$-type operators or well-known splitting 
kernels $p_{(ij)}(x)$ associated to initial-final massless 
collinear singularities.
The colour-ordered massive ${\bf I}^{(1)}_{ij}$-type operators are 
here presented for the first time. 
As a first application of our massive extension of the antenna
formalism, we constructed the colour ordered 
real contributions and subtraction terms for the production of a top quark pair
and for the production of a top quark pair and a jet at NLO.
In the second case, the presence of interference terms in the colour
decomposition of the real matrix element squared renders 
the construction of the subtraction terms more involved. 
The treatment of those terms is explained in detail in the paper in
Section 7.2.1.
All colour-ordered subtraction terms constructed have been checked in
all soft, collinear and quasi-collinear limits of the real matrix
element squared. Furthermore, for each subtraction term, it has been verified
that in all collinear and/or quasi-collinear limits present in it, 
the sum of all terms contributing to a given limit add up to reproduce
the product of the required splitting function with the corresponding
Casimir factor multiplied with the non colour-ordered matrix element
squared. This check is explicitly derived for the subtraction term related to
the process $gg \rightarrow Q\bar{Q}g$ contributing to the process 
$pp \rightarrow t\bar{t}$ at NLO.
This verification provides us with an extremely powerful test on
the correctness of our results for the subtraction terms  
for $t\bar{t}$ and $t\bar{t}$ +1 jet production at NLO.

The results presented in this paper represent a substantial step
towards the calculation of the NNLO corrections to top quark pair
production within the antenna subtraction formalism.
The decomposition of the real matrix elements into colour-ordered  
amplitudes squared and the identification of the different leading and
subleading colour structures, including the treatment of interference
terms, described in Section 7.2.1, will allow the application of the
NNLO antenna subtraction method to compute the double real radiation
contributions to $p p \rightarrow t \bar{t}$ at NNLO. The NLO antenna
subtraction terms for $p p \rightarrow t \bar{t}$ +jet, provided in
this paper, are already part of the NNLO corrections to $t \bar{t}$ 
production at hadron colliders.  
 
\section{Acknowledgements}
This research was supported by the Swiss National Science Foundation
(SNF) under contract PP0022-118864 and in part by the European 
Commission through the 'LHCPhenoNet' Initial Training 
Network PITN-GA-2010-264564', which are hereby aknowledged.

\bibliographystyle{JHEP-2}

\providecommand{\href}[2]{#2}\begingroup\raggedright
\endgroup

\end{document}